\documentclass[11pt]{article}
\usepackage[utf8]{inputenc}

\usepackage[shortlabels]{enumitem}
\usepackage[margin=1in]{geometry}

\usepackage{dgleich-math}
\usepackage{todonotes,hyperref,graphicx}
\hypersetup{colorlinks=true,
    linkcolor=blue,citecolor=black}

\usepackage{booktabs}
\usepackage{tabularx}
\usepackage{longtable}
\usepackage{caption}
\usepackage{subcaption}
\newcommand{\ER}{Erd\H{o}s–R\'{e}nyi}

\usepackage{minitoc}

\title{Multi-scale Local Network Structure Critically Impacts \\ Epidemic Spread and Interventions }
\author{Omar Eldaghar\thanks{Department of Mathematics, Purdue University,              \texttt{oeldagha@purdue.edu}}, 
        Michael W.~Mahoney\thanks{ICSI, LBNL, and Department of Statistics, University of California at Berkeley, \texttt{mmahoney@stat.berkeley.edu}}, 
        David F.~Gleich\thanks{Department of Computer Science, Purdue University, \texttt{dgleich@purdue.edu}}}
\date{}

\begin{document}

\doparttoc 
\faketableofcontents 

\maketitle

\section*{Abstract}
Network epidemic simulation holds the promise of enabling fine-grained understanding of epidemic behavior, beyond that which is possible with coarse-grained compartmental models. 
Key inputs to these epidemic simulations are the networks themselves. 
However, empirical measurements and samples of realistic interaction networks typically display properties that are challenging to capture with popular synthetic models of networks. 
Our empirical results show that epidemic spread behavior is very sensitive to a subtle but ubiquitous form of multi-scale local structure that is not present in common baseline models, including (but not limited to) uniform random graph models (\ER), random configuration models (Chung-Lu), etc. 
Such structure is not necessary to reproduce very simple network statistics, such as degree distributions or triangle closing probabilities.
However, we show that this multi-scale local structure impacts, critically, the behavior of more complex network properties, in particular the effect of interventions such as quarantining; and it enables epidemic spread to be halted in realistic interaction networks, even when it cannot be halted in simple synthetic models. 
Key insights from our analysis include how epidemics on networks with widespread multi-scale local structure are easier to mitigate, as well as characterizing which nodes are ultimately not likely to be infected. 
We demonstrate that this structure results from more than just local triangle structure in the network, and we illustrate processes based on homophily or social influence and random walks that suggest how this multi-scale local structure~arises.
\section{Introduction and Central Findings}

Understanding epidemic spread is fundamental to controlling, containing, and mitigating epidemic spread when it is unwanted, as in the case of a harmful pathogen. 
Controlled experiments with epidemic spread are limited, inherently, to virtual pathogens, and so computational models are key to our understanding. 
Modeling approaches can vary over an enormous range, from simple mathematical models based on dynamical systems~\cite{kermack1927contribution,apolloni2014metapopulation,allen2017primer}, to more complicated dynamical systems, to models fit to data~\cite{zhang2021integrated,chang2021mobility}, and to fine-grained event simulators that provide intricate possibilities for interventions~\cite{le2020impact,hunter2018comparison,bruch2015agent,truszkowska2021high}.
A key component that is often explicit in event simulators but implicit in models based on dynamical systems is the underlying \emph{network substrate} over which the epidemic spreads. 
For instance, many models (e.g.,~meta-population or structured dynamical systems) make use of a homogeneous mixing assumption.
This amounts to assuming the underlying contact network is either an overly simplistic fully connected network (i.e.,~a clique) or consists of uniformly random interactions~\cite{wilson1945law,keeling2005networks}.
These assumptions are contrary to empirical work showing that these properties are not present in empirical networks~\cite{Costa_2007,leskovec2009community,pastor2016distinct,freitas2022graph}.
Put plainly: network structure matters to epidemic spread; and the properties of underlying network, from simple to complex, are critical to understanding epidemic spread in empirical networks~\cite{allard2020role,hebert2020beyond,grossmann2021heterogeneity}.

\emph{Our findings show that local structure at multiple size scales within a network is critical to the success or failure of local quarantine interventions}.
Surprisingly, this observation on the interaction of network structure, epidemics, and interventions has seemingly fallen between the cracks in disciplines that study epidemics and networks. 
For instance, meta-population models often use empirical mobility patterns among different regions~\cite{wang2014spatial,soriano2022modeling,lang2018analytic,gross2020epidemic,butler2023optimal,liu2018measurability,hota2021closed}.
Such models are often used by epidemiologists, 
where the research goals are large-scale population level epidemic forecasts and intervention strategies for widely spread pathogens that are consistent with empirical data, rather than a detailed understanding of \emph{what} properties of the network might give rise to these findings.
While meta-populations are useful for macroscopic modeling, they are known to overestimate epidemic prevalence, compared to more fine-grained approaches that take in account low-level network structure~\cite{keeling2010individual,ajelli2010comparing,zachreson2022effects}.
On the other hand, studies of the network structure itself (often by computer scientists or physicists) typically show that some structural network characteristic leads to qualitative differences in epidemic spread.
Examples in the literature range from heterogeneity in contact patterns \cite{allard2020role,hebert2020beyond,grossmann2021heterogeneity} and community structure \cite{salathe2010dynamics,liu2016community} altering final epidemic size to, on the policy side, details of how the spatial network and interactions result in burstiness and possibly overcrowding treatment facilities~\cite{thomas2020spatial}.
There are also elegant mathematical analyses of how graph properties might impact epidemic properties, in cases that are simple enough to admit closed-form or nearly closed-form solutions~\cite{kiss2017mathematics}.
Capturing realistic heterogeneity present in empirically-measured networks is an ongoing challenge for these analytically tractable models. 
Finally, another community within epidemic modeling (often by researchers in control theory) makes use of networked data to formulate epidemic interventions as a problem of optimal control (see the survey by Nowzari et al., 2016~\cite{nowzari2016analysis} as well as related references~\cite{preciado2014optimal,hota2021closed,butler2023optimal}). While these methods make use of networked data to devise interventions, they do not study what network structures drive or mitigate spreading.

\begin{figure}
    \centering
    \includegraphics[width=0.75\textwidth]{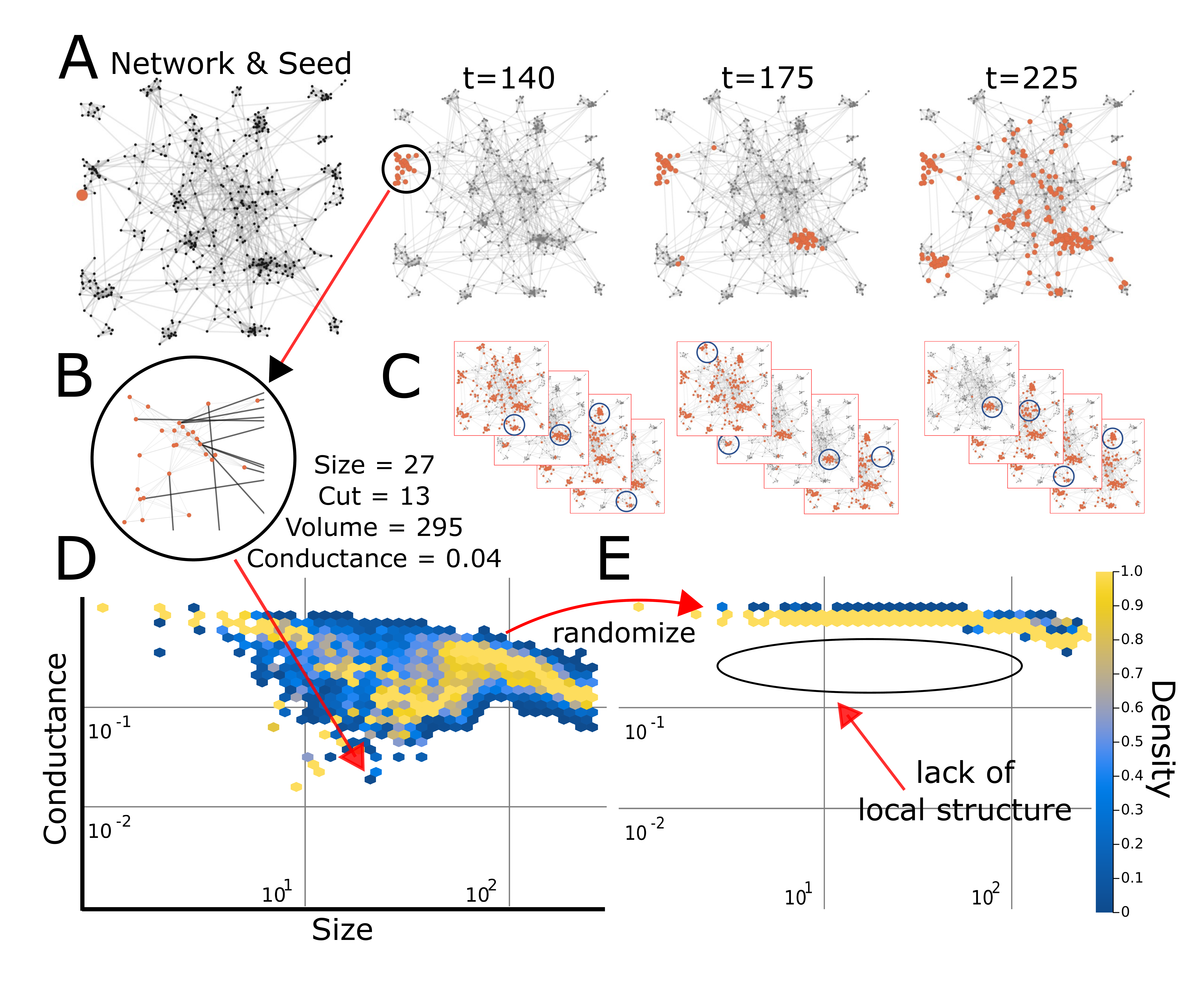}
    \caption{The Network Community Profile (NCP) characterizes local structure in networks by showing the distribution of local conductance bottlenecks at all size-scales.
    Here, we use it to understand epidemic spreading preferences by plotting conductance bottlenecks derived from epidemic spreading. The resulting epidemic NCP plots in (D) and (E) show a characteristic difference between networks with rich local structure (A and D) compared with a randomized network with the same degree distribution (E).
    (A) We initiate an SEIR process on a network stating from a given node and show a few steps of the epidemic spreading  where orange represents an infected or recovered node.
    (B) We zoom into a single group of infected nodes at time $t=140$ and assess the conductance of the set (the ratio of cut to volume is 13/295 = 0.04 for this set).
    (C) This is repeated for 10 or 100 thousand times to get multiple epidemics from multiple start points and evaluated with multiple sets of infected nodes from each epidemic.
    (D) Each evaluated set yields a single point in the size-vs-conductance space, and the NCP is a 2d histogram of these points. 
    The NCP of the network in (A) shows wide variation in the size-vs-conductance space and indicates more local structure is present than in (E). 
    (E) We compare this to the epidemic NCP obtained by randomizing edges from the network in (A) with a configuration model.
    In this case, conductance has minimal variability as the size-scale changes, indicating a lack of local structure. 
    }
    \label{fig:ncp-explaination}
\end{figure}

The specific gap that we address is to observe that empirically-measured networks\footnote{We are using the term empirically-measured networks to denote networks based on datasets relevant to epidemic spreading. While we believe that some of these networks are reasonably realistic substrates for epidemic spread, there remains a gap between the data that exists for analysis and the idealized realistic or ground truth networks for studying epidemic spread.} 
have a rich multi-scale structure, and that this structure has important implications for interventions in epidemic spread. 
This structure arises empirically because individuals interact repeatedly, and so triangles (as well as higher-order local structures) naturally close in such repeated interactions~\cite{granovetter1973strength,watts1998-dynamics}. 
Interactions on the order or 5-10 people frequently occur, and these give rise to small-scale group structure.
Group-level interactions continue up to create medium-sized meso-scale structure, where modular clusters of nodes of many sizes occur.
All of these interactions occur in a background of weak-ties, long range edges, or global quasi-random interactions, with their own implications for epidemic spreading~\cite{eckles2018long,wren2021local}. 
This type of local structure was initially identified in the analysis of community structure in social and information networks, where it was assessed with a network community profile (NCP)~\cite{leskovec2009community}.
The NCP plot was introduced as the minimum conductance\footnote{The conductance of a set in a graph is a surface area to volume ratio, computed as the ratio of how many connections leave a set (surface area) compared with the total number of edges incident on all vertices in the set (volume). It is the probability that a random walk at stationary switches between the set and the rest of the~graph.}
as a function of size-scale, in order to qualitatively study purely structural bottlenecks in interaction networks.
Qualitatively, the distribution of size-vs-conductance scores in empirical networks displays both high variance as well as a characteristic initially-downward-sloping, then upward-sloping, lower-envelope, as the cluster size increases~\cite{leskovec2009community,jeub2015think}.
We adapt this structural measure for epidemics to create an \emph{epidemic NCP} by using sets based on infection times for nodes over many epidemics.
Figure~\ref{fig:ncp-explaination} contains a detailed example for the computation of an epidemic NCP based on an SEIR epidemic model. 
These plots highlight multi-scale local structure in the wide dispersion of sets visited by epidemics, compared with random networks (Figure~\ref{fig:ncp-explaination}, D vs. E).

\begin{figure}
    \centering
    \includegraphics[width=0.7\textwidth]{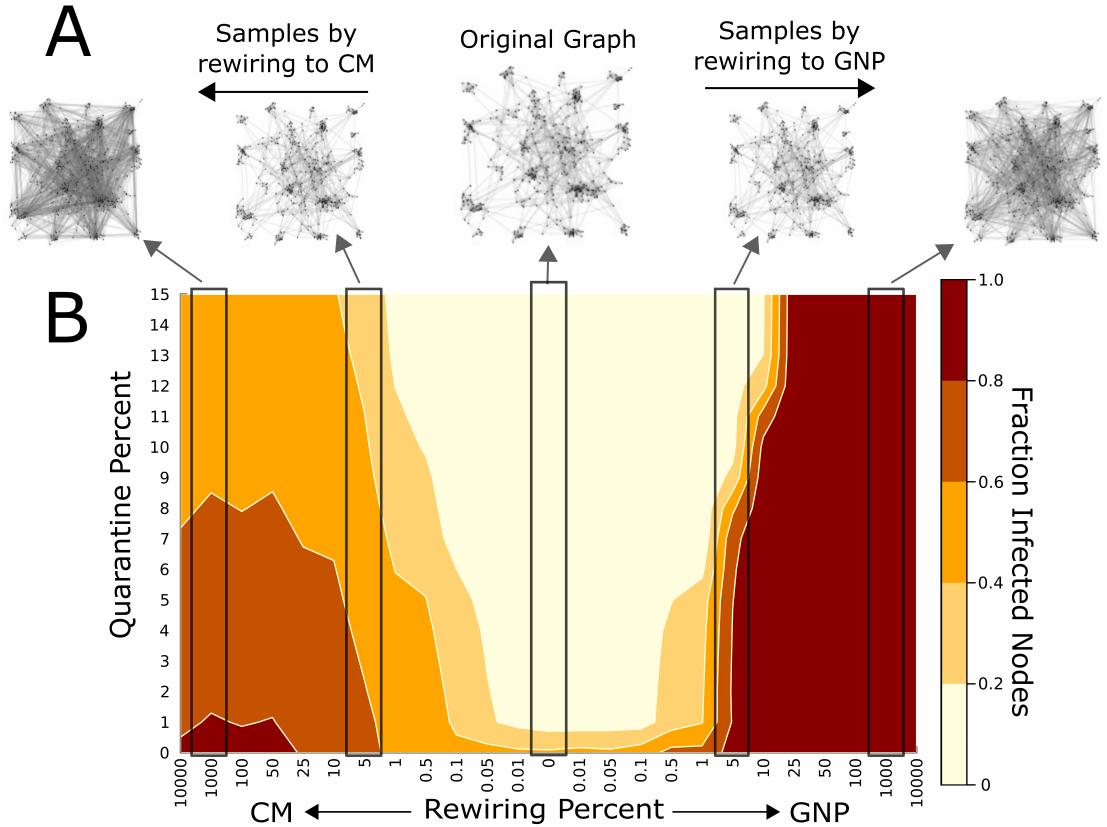}
    \caption{Networks containing realistic multi-scale local structure are more controllable under local interventions than networks without local structure, due to the widespread prevalence of conductance bottlenecks. (A) Shows a network with local structure that is rewired to two different synthetic variants lacking local structure. For each rewired network sample, 800 simulations - split among differing quarantine capacitites - for an SEIR model were performed. (B) Shows the fraction of total infections as the color (darker is a higher fraction) when the original network is a mobility network based on contacts within Mexico City\protect ~\cite{de2020contact}. The vertical axis represents the maximum level of quarantine as a fraction of the total number network size.  The epidemic uses a fixed infection and recovery probability (0.02 and 0.05 respectively). As we move from the center to the left, we are increasingly rewiring to a random configuration model (CM) network that preserves node degrees in expectation; as we move from the center to the right we rewrite to a uniform random graph ($G_{n,p}$) that preserves average degree only. The network depicted in (A) is used for illustrative purposes, as the Mexico City network does not admit a simple visual depiction.}
    \label{fig:uplot-explanation}
\end{figure}

To characterize the impact of multi-scale local structure on epidemic spreading in empirically-measured interaction networks, we compare the behavior of epidemics on a network model with multi-scale local structure to the behavior of the epidemic on networks where this structure has been removed by rewiring the network. 
We do this as we vary the size of a local quarantine intervention from absent (no quarantine) to allowing a large portion of the network to quarantine if needed. 
(See methods section and supplement for precise details and the quarantine model.) 
Our results are illustrated in Figure~\ref{fig:uplot-explanation}.
An important observation is that the presence of multi-scale local structure in networks causes such interventions to work far better than in the rewired random networks that are commonly studied and that lack such structure.  
Figure~\ref{fig:uplot-explanation} displays the total fraction of infected nodes under various levels of the quarantine policy, as we perturb the network to two extremes that lack any local structure. 
The first (shown as the perturbation goes to the left) is a rewiring perturbation going towards a random configuration model, which preserves individual degrees in expectation; and the second (shown as the perturbation goes to the right) is rewiring perturbation going towards a uniform random network, which preserves the total number of edges in expectation. 
In these two extremes, even with high quarantine levels, the quarantine policy is unable to prevent a large fraction of the network from becoming infected.
This should be contrasted with the original network which exhibits multi-scale local structure (the data illustrated here are from the Mexico City contact network~\cite{de2020contact}), where even small quarantine levels are able to quell an epidemic. 
Among other things, this result illustrates that just because one reproduces a coarse summary statistic (e.g.,~average degree or degree distribution) does not mean that more subtle results such as doing an intervention will lead to similar results as doing so in the corresponding empirical network (see Appendix~\ref{sec:app-community-heterogeneity} for an example of this with communities).
That is, interventions are very sensitive to the details of multi-scale local structure that are absent in popular baseline models.

\paragraph{Central findings.}
Here is a summary of our main findings.
\begin{itemize}
    \item 
    We find that epidemics in empirically-measured interaction networks are much more controllable in terms of their response to a quarantine intervention, compared with widely-used synthetic network models of those same networks.
    We also observe larger dominant eigenvalues in the non-random networks --- even though, based on popular baseline models, this would predict \emph{stronger} and less controllable epidemics. 
    (See Section~\ref{sec:epidemic-thresholds-local-structure}.) 
    These results suggest that characterizing multi-scale local structure is important to understanding epidemic spread in interaction networks, going well beyond simple summaries such as average degree or dominant eigenvalue. 
    \item 
    We provide a measure called  Area Above the NCP (AANCP, see Equation~\ref{eq:aancp}) to quantify the amount of multi-scale local structure in a network. 
    We find that this measure has a strong positive association with quarantine impact across a variety of both real and synthetic models (Section~\ref{sec:local-structure-measure}). 
    We cannot establish such a relationship for popular metrics such as the average degree or $\lambda_1$. 
    \item 
    We further study the impact of multi-scale local structure in terms of understanding what sets are missed, i.e., not entirely infected, by the epidemic (Section~\ref{sec:missed-sets}).
    This highlights how sets with low conductance are often protected from infection.  
    This further supports the idea of associating a wide distribution of size-vs-conductance tradeoffs with this multi-scale local structure, because these missed sets occur at many different size scales. 
    \item
    We develop two synthetic models to demonstrate how multi-scale local structure can emerge, as well as its impacts on epidemic spreading (Section~\ref{sec:synthetic}). 
    The first uses a geometry to produce local structure combined with plausible interaction and geometric preferential attachment models. 
    The second uses a random walk process as a surrogate for local geometry (recall that random walks diffuse slowly). 
    These models illustrate compatibility between the rich multi-scale local structure and global quasi-randomness typically found in empirically measured networks. 
    \item 
    We conduct a separate study of epidemic spreading on hypergraph models to justify that the impact of local structure extends beyond simply using triangles as local structure (Section~\ref{sec:hypergraph-diffusions}). This involves rewiring the triangles of a network. 
\end{itemize}

\paragraph{Implications and relationships with existing literature.}

To aid interpretation by the many research areas studying epidemics, we wish to highlight relationships between our results and studies within these areas.

Meta-population models seek macro-level simulation fidelity by modeling spread among sub-populations~\cite{wang2014spatial,soriano2022modeling}. 
These sub-populations are split based on spatial distance, and mobility data among regions is represented as a network of aggregated flow data~\cite{kang2020multiscale}. 
Epidemic models are implemented on top of this mobility data taking into account characteristics of different regions as well as other patterns of human behavior, and they can include some level of heterogeneity~\cite{apolloni2014metapopulation,wang2020network}. 
Within a region and between regions, these models use mass-action or reaction-diffusion dynamics. 
This choice implicitly makes a local homogeneity assumption that differs from the types of local structure we observe empirically. 
These assumptions persist in various ways down into fine-grained event simulator models of epidemic spread, and they often result in overstating epidemic severity, e.g., erasing individual node identity~\cite{keeling2010individual} and related homogeneity assumptions~\cite{ajelli2010comparing}. 
A recent attempt has been made to address this overestimation~\cite{zachreson2022effects} in total infections, but correctly calibrating parameters to match finer epidemic information remains a challenge.
We make use of networks derived from human mobility data, not following the reaction-diffusion approach more common in meta-population models.
Instead, we focus on structural bottlenecks that have implications for quarantining.

The research that arose from the network science community quickly focused on understanding the implications of high level graph structure such as the dominant eigenvalue of the adjacency matrix, $\lambda_1(\mA)$,  or degree fluctuations. 
This has been done in the mean-field or following other moment-closure based approaches~\cite{chakrabarti2008epidemic,prakash2012threshold}. 
Moment-closure approaches like mean-field approximations allow for idealized theory by reducing the complexity of the system of state update equations for node states. 
Perhaps the most well-known of these approaches quantifies the epidemic threshold in the mean-field in terms of $\lambda_1(\mA)$~\cite{chakrabarti2008epidemic,prakash2012threshold}.
For the SIR and SEIR model, these approaches arrive at the following mean-field threshold for epidemics $s=\lambda_1(\mA)\frac{\beta}{\gamma}$. 
This in turn spurred follow up work that sought to mitigate epidemic spread by editing the graph a priori in order to minimize $\lambda_1$ and increase the parameter regime for which these for which epidemics vanish~\cite{tong2010vulnerability,saha2015approximation,chen2015node}. 
However, $\lambda_1$ is known to be localized for empirical data~\cite{cucuringu2011localization,pastor2016distinct}, and this can distort the use of $s$ to localized regions of the network.
Our findings highlight common scenarios where using this threshold to quantify epidemic strength and intervention effectiveness fails and networks with larger values of $\lambda_1$ have more effective interventions.  

A separate line of research is concerned with designing interventions on networks via optimal control methods. These methods make use of networked data but don't \emph{explicitly} probe the impact of that structure on spreading or interventions (see the survey by Nowzari et al., 2016~\cite{nowzari2016analysis}).  
These are often formulated as a continuous-time Markov process and make use of tools from optimal control theory~\cite{preciado2014optimal,hota2021closed,mubarak2022individual,butler2023optimal}. 
Two prominent formulations for epidemic interventions in the optimal control community are budget constrained and minimum-budget. 
In budget-constrained formulations, one minimizes $\lambda_1(\mB - \mD)$ subject to cost constraints, where $\mB$ is the matrix with node-to-node infection rates while $\mD$ is the diagonal matrix of node recovery rates. 
In the minimum-budget formulation, the cost of interventions that decrease infection rates, $f_{i,j}(\beta_{i,j})$ or increase recovery rates $g_k(\gamma_k)$ are used to constrain solutions. 
These methods were global and apriori in nature. 
However, there are optimal control studies that seek to localize interventions at a node level~\cite{mubarak2022agent,mubarak2022individual} or perform interventions in an online rather than offline fashion~\cite{eshghi2014optimal,lorch2018stochastic,hota2021closed}.
While such formulations seek to constrain or contain spreading, neither probes what specific structures lead to more or less controllability.

Finally, our research is most closely related to the study of community structures and epidemic spreading. 
In particular, it was shown that networks with larger modularity~\cite{salathe2010dynamics} were found to have smaller total infections. 
This spurred follow-up work that attempted to mitigate epidemic spread by targeting nodes and edges based on community membership as well as other structural factors in a local~\cite{gong2013efficient,taghavian2017local} and global~\cite{hebert2013global,gupta2016centrality,kumar2018efficient} setting. 
However, there are many quantifications of what ``community structure'' entails. 
This results in ambiguity within the literature and mixed effects from community-targeted interventions~\cite{sah2017unraveling,gosak2021community}, and it calls into question the impact of exact intra-community structures~\cite{stegehuis2016epidemic} in dense communities. 
Broadly, this ambiguity lies in both the size and mixing patterns of communities.
For example, one study~\cite{min2013role} showed that community mixing patterns in networks with identical values of modularity can significantly influence total infections and that this effect is amplified with more communities; while
another study~\cite{liu2016community} showed that the infected fraction within communities is smaller in communities with less nodes.
Our quantification of local structure as size-vs-conductance tradeoffs allows for analysis at all size regimes to resolve such ambiguities. 
In particular, our multi-scale size-resolved approach is calibrated to avoid the ambiguity in community structure, and it shows that networks with modular portions at many size scales admit conductance bottlenecks that can be exploited to mitigate epidemics.

\section{Methods}

\subsection{Data}
Our findings are based on multiple empirically-measured networks that range in size from $10^3$ nodes to over $10^6$ nodes and that are plausible realistic substrates for epidemic behavior. 
Properties of these networks are summarized in Supplemental Table~\ref{table:graph-stats}. 
These networks include a contact network for Mexico City \cite{de2020contact}, aggregated US commuting data~\cite{Nelson-2016-us-commutes}, cellular mobility data~\cite{kang2020multiscale}, friendship networks from colleges \cite{traud2012social}, citation and collaboration networks \cite{leskovec2005graphs,gehrke2003overview,Boldi-2011-layered}, location based social networks \cite{cho2011friendship,snapnets}, among others~\cite{klimt2004introducing,leskovec2009community,leskovec2007graph,richardson2003trust,Wilson-2009-social-networks,Mislove-2007-measurement}. 
The networks were chosen to represent a diversity of possible realistic phenomenon and behaviors that go well beyond popular idealized~models.  

The raw data from these networks were processed to remove structural artifacts. In particular, small conductance sets at large size scales were removed (see Supplemental Section~\ref{sec:large-set-removal} for more information). As an example, in the college networks, the incoming class of freshman formed many connections among other freshman due to the timeframe when the networks were collected. This reflects a different process from students in other class years~\cite{Veldt-2019-resolution}. 
In order to more readily observe effects of local structure, the college friendship networks were further sparsified to highlight local heterogeneity. We used a simple method of sparsification where each node ranks its neighbors based on normalized common neighbors and elects to keep some fraction of edges based on that ranking. This is designed to model the most \emph{likely} interactions to simulate the highest probability of epidemic spread. 
See Supplemental Section \ref{sec:sparsified-data} for additional detail on this sparsification. 
 
There are 15 networks we study throughout in the main text: 

\begin{center}
     \textbf{Human Mobility} \\
     US Commutes, Mexico City, US Flows\\[1ex]
     \textbf{Sparsified College Friendships} \\  Illinois, Penn, Wisconsin\\[1ex]
    \textbf{Electronic Social Interactions} \\ Collaboration, Email, Facebook Interactions\\[1ex]
    \textbf{Distant Networked Interactions} \\ Citation, Slashdot, Flickr\\[1ex]
    \textbf{Synthetic} \\ Local Geometric, GeometricCommunities, RandomWalkCommunities 
\end{center}
These networks are loosely ordered in terms of relevance to epidemic spreading, with the human mobility data most relevant, and the distant networked interactions least relevant. The synthetic network \emph{Local Geometric} is meant to represent a purely local connections in a planar geometric (details in Appendix~\ref{sec:local-geometric-alldetails}). 
In comparison, the synthetic networks \emph{GeometricCommunities} and \emph{RandomWalkCommunities} are meant to emulate more plausible combinations of local structure as well as large scale randomness. We discuss both of these further in Section~\ref{sec:synthetic}. 

\subsection{Epidemic Methods}
    We make use of the standard SIR and SEIR compartmental models with local quarantining. Nodes are partitioned by current infection state (state S -- susceptible, state E -- exposed, state I -- infected, state R -- recovered) along with an overlapping class Q for quarantining. Nodes in S that undergo quarantining are returned to S after the conclusion of quarantining. Nodes in E or I that are quarantined are moved to R at the end of the quarantine period. 
    Additional details on the model and a schematic figure is given in Appendix~\ref{sec:epidemic-models-alldetails}. 
    Epidemic spread is simulated while varying contact structure, intervention strength, and infection probability. 
    Intervention strength and infection probability are purely epidemic parameters, while network structure is topological.
    For each parameter set, we start the epidemic from a single node. 
    We repeat this from 50 random nodes and record epidemic information including total infections and explored local structure for that set of parameters. 
    For each network, we also record the dominant eigenvalues of the adjacency matrix, which has been implicated in epidemic spread~\cite{chakrabarti2008epidemic,prakash2012threshold}. 

    When selecting epidemic parameters, one often chooses parameters relative to the epidemic threshold \cite{chakrabarti2008epidemic,prakash2012threshold}, defined as $\frac{\beta}{\gamma}\lambda_1(\mA)$, where $\beta,\gamma,\lambda_1(\mA)$ respectively denote the infection probability, recovery probability, and dominant eigenvalue of the adjacency matrix $\mA$. However, we want to make direct comparisons across synthetic variants of the same network. The dominant eigenvalues of these networks can differ dramatically -- a point we return to in Section~\ref{sec:epidemic-thresholds-local-structure} (and see Figure~\ref{fig:eigenvalue-fig}) -- which would change epidemic parameters. Since we want to simulate the spread of the same pathogen across rewired versions of the same graph, we often fix epidemic parameters ($\beta,\gamma$) when simulating spread on a network and it's rewired variants (see Appendix~\ref{sec:epidemic-params-app} for details). In instances where we compare one network to another network that differs in average degree (see Figure~\ref{fig:spatial-graph-fig}), we make these comparisons based on a fixed value of $\frac{\beta}{\gamma}\lambda_1(\mA)$.

\subsection{Quarantine Procedure}
    We perform local quarantining by moving a node and and its immediate neighbors to state Q.\footnote{This is similar to, but distinct from, acquaintance quarantining~\cite{cohen2003efficient} since we quarantine all neighbors rather than a randomly selected neighbor.} 
    A maximum quarantine capacity is maintained at each time step where nodes are able to be quarantined so long as we have not reached the maximum capacity to do so. 
    This capacity is given as a percentage of total nodes in the base graph. 
        Several time delays are used at several portions of the model in order to avoid trivially ending the epidemic under this local quarantine policy. 
    These are fixed in the model for all epidemic simulations.
    The first time delay is given as a threshold for disease detectability in the population, which is fixed at 100 nodes. 
    The second time delay is between node exposure (node in compartment E or I) and detectability of the pathogen in a node.
    This is fixed at a single time step for all nodes. 
    Lastly, there is a time delay from when a node is quarantined to when it's neighbors are quarantined. 
    This is also fixed at a single time step for all~nodes.

\subsection{Network Community Profiles}
    The Network Community Profile (NCP) displays the distribution of size vs conductance for many sets of nodes in a network. 
    Originally, the NCP was designed to highlight structural properties of social and information networks (in the minimum envelope of the samples for a conductance minimizing procedure).
    However, it has since been used to study the behavior of processes on networks more generally~\cite{Gleich-2012-neighborhoods,jeub2015think,Fountoulakis-2023-flow}. 
    Consequently, there are many distributions of node subsets that can be defined. 
    For the epidemic NCPs we use here, the sets are derived from epidemic spread itself. 
    In particular, we sample 50,000 seed nodes; and, for each trial, we simulate epidemic spread for an SEIR model from each node 20 times. 
    We combine the diffusions from each trial, and we use aggregated infection times for each node as a node ranking. 
    Using this node ranking $\vs$, we sequentially compute the conductance of the sequence of sets defined by the top ranked nodes. 
    Explicitly, we are forming sets, $S_k$, consisting of the top-$k$ ranked nodes and then finding their conductances.\footnote{Forming sets of top-$k$ ranked nodes, $S_k$, and computing conductance values of this sequence is known as a sweepcut \cite{andersen2006local} in the local community detection literature.} 
    From there, we sub-sample several sets on a uniform-log scale using several bins in order to have size-vs-conductance information at all size scales. See Appendix~\ref{sec:epidemic-ncps-alldetails} for full details and modeling rationale on epidemic NCPs. 

    There is another type of NCPs used in this paper. This one uses a conductance minimizing procedure based on seeded PageRank vectors to form node rankings before computing conductance values via a sweepcut via the Andersen-Chung-Lang procedure~\cite{andersen2006local,leskovec2009community}. 
    In this instance, we can tune the locality directly via parameters from the underlying algorithm. In addition to storing information about sets, we explicitly store these sets. Storing the sets allows us to compare conductance minimizing NCPs with the epidemic NCPs. This allows us to display an NCP-like plot of the difference to highlight the nodes from the seeded PageRank NCP sets that were not infected during the epidemic. 
    In this instance, each set from the seeded PageRank NCP is weighed based on the proportion of infected nodes from epidemic simulations. 
    Thus each bin is weighed by the fraction of nodes in that bin that were in the S class at the end of several epidemic simulations without quarantining. Put simply, each bin reports the fraction of susceptible nodes in that bin. Values close to 1 indicate that nodes among sets in that bin were often avoided by the epidemic. Values close to 0 indicate most nodes among sets in that bin tended to become infected.
\section{Main Results}

\subsection{Empirical Networks have Worse Epidemic Thresholds but Better Interventions}
\label{sec:epidemic-thresholds-local-structure}

    Total infections often serves as a measure of the severity and strength of an epidemic. Recall that Figure~\ref{fig:uplot-explanation} shows total infections as the fraction of infected nodes (color) as a function of intervention strength (vertical axis) and network structure (horizontal axis) for an SEIR model on a mobility network of Mexico City \cite{de2020contact}. 
    For the empirical network (the middle), relatively little intervention (quarantine) is needed to halt the epidemic. 
    On the other hand, the randomly-rewired models, even with extensive quarantine levels, still result in a higher fraction of total infections. This results in a ``U'' shape in the plot.
    
\begin{figure}
    \centering
    \includegraphics[width=0.8\textwidth]{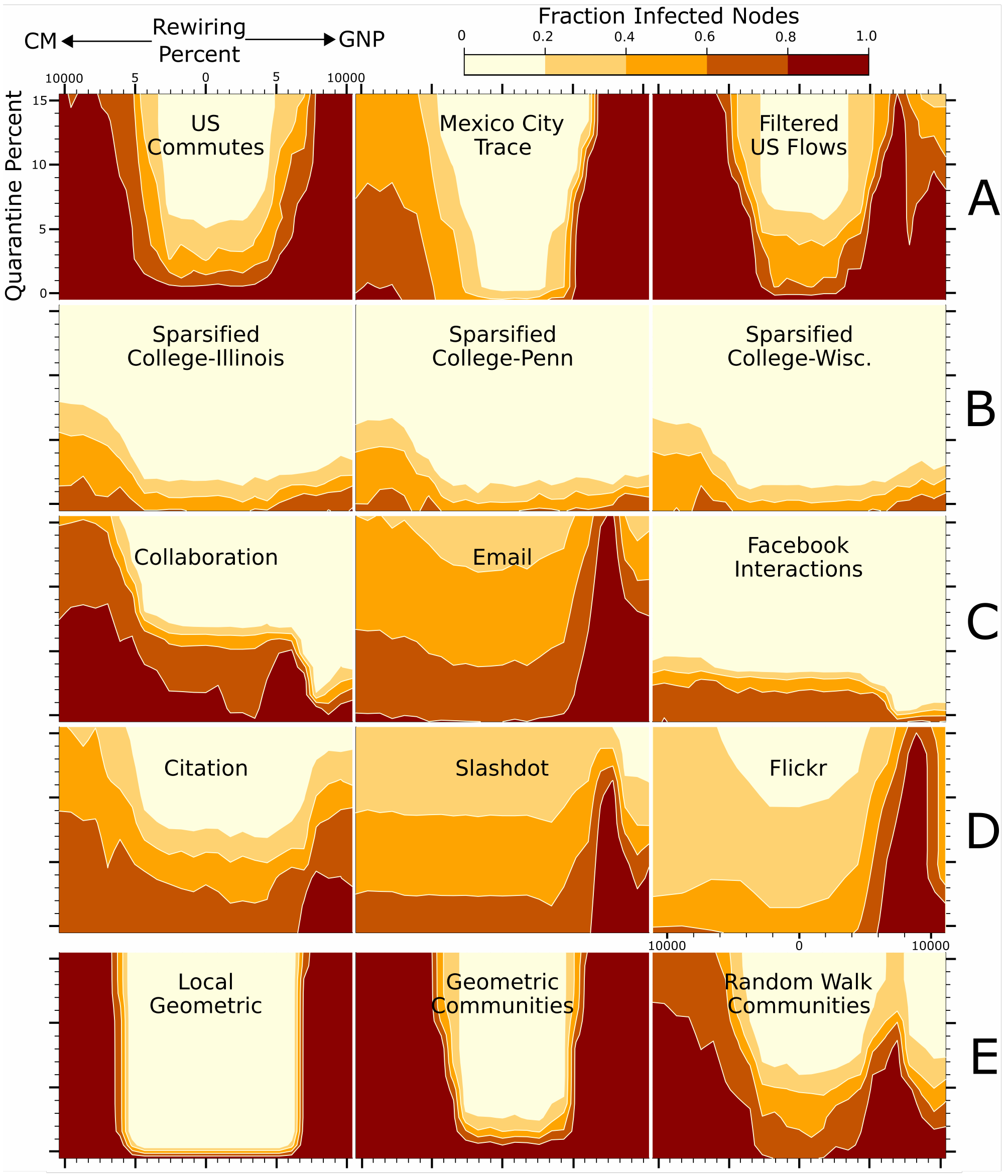}
    \caption{
    The presence of multi-scale local structure dramatically improves the impact of a quarantine intervention in an epidemic. Rows (A)-(E) show total infections as a function for rewiring (x-axis) and quarantine capacity (y-axis), as in Figure~\ref{fig:uplot-explanation}, for 15 networks. As we increasingly rewire networks (move away from the center), total infections deviates from those on empirical networks under local interventions. This effect is more pronounced in some networks (Row A) while absent in others (Facebook Interactions, etc). 
    These results were produced using SEIR simulations on 50 node samples for each parameter set (quarantine percent and rewiring). See Appendix~\ref{sec:app-uplots-all} for details concerning choice of epidemic parameters. }
    \label{fig:uplots-all}
\end{figure}

    The inability of random models without local structure to model epidemic behavior is not an artifact of this one network. Figure~\ref{fig:uplots-all} shows similar effects for a large variety of networks (see Table~\ref{table:graph-stats} for properties of each network). 
    These were deliberately chosen to span a wide range of different network types where the effects we describe are more or less pronounced. 
    For instance, row (B) corresponds to sparsified college friendship networks, whereas row (E) corresponds to synthetic networks designed to emulate and possess local structure. 
    The dramatic difference under quarantining observed in Mexico City is present to differing degrees for many of the networks (US Commutes, Filtered US Flows, Citation, Flickr). On the other hand, networks where links have only a small basis in physical space display a range of effects (Collaboration, Email, Facebook Interactions, Slashdot). For two of the networks, Collaboration and Facebook Interactions, rewiring to $G_{n,p}$ produces a more controllable epidemic. Both of these have an extremely low average degree (6.3 and 3.5, respectively), which we believe is responsible for the difference in behavior. However, there is no straightforward relationship between the average degree and any of the effects beyond this. For example, all of the sparsified college networks have average degree around 3, and yet we see a different result there where rewiring to $G_{n,p}$ produces a stronger epidemic.

    Beyond average degree, epidemics in networks are often studied by their epidemic threshold: the critical value where the epidemic either explodes or dies off.
    This criticality threshold for epidemic spread under idealized conditions was shown 
    to be proportional to the largest eigenvalue of the adjacency matrix, i.e.,~$\lambda_1(\mA)$, for various compartmental models~\cite{chakrabarti2008epidemic,prakash2012threshold} via mean-field approximations. 
    The classical result is that the disease free state is asymptotically stable for an SIS model if $\frac{\beta}{\gamma}\lambda_1(\mA)<1$ and unstable when $\frac{\beta}{\gamma}\lambda_1(\mA)>1$. 
    Hence, for these models, a larger value of $\lambda_1(\mA)$ indicates \emph{stronger epidemic spread}. Moreover, the speed of extinction in the asymptotically stable case $\left(\frac{\beta}{\gamma}\lambda_1(\mA)<1\right)$ was shown to be exponential~\cite{chakrabarti2008epidemic} ($Pr(v\text{ infected at time }t)\le  C\left(\lambda_1(\mA)\right)^t$). 
    These results have spurred research into methods for mitigating disease spread by reducing the dominant eigenvalue either explicitly or via an approximation \cite{tong2010vulnerability,saha2015approximation,chen2015node}. The main avenue of study is formulated as finding a subset of nodes/edges to remove from the graph so as to minimize the spectral radius of the resulting graph and reduce the parameter regime for which the disease free state is unstable. In other words, reducing the spectral radius increases the extinction regime.

\begin{figure}
    \centering
    \includegraphics[width=0.75\textwidth]{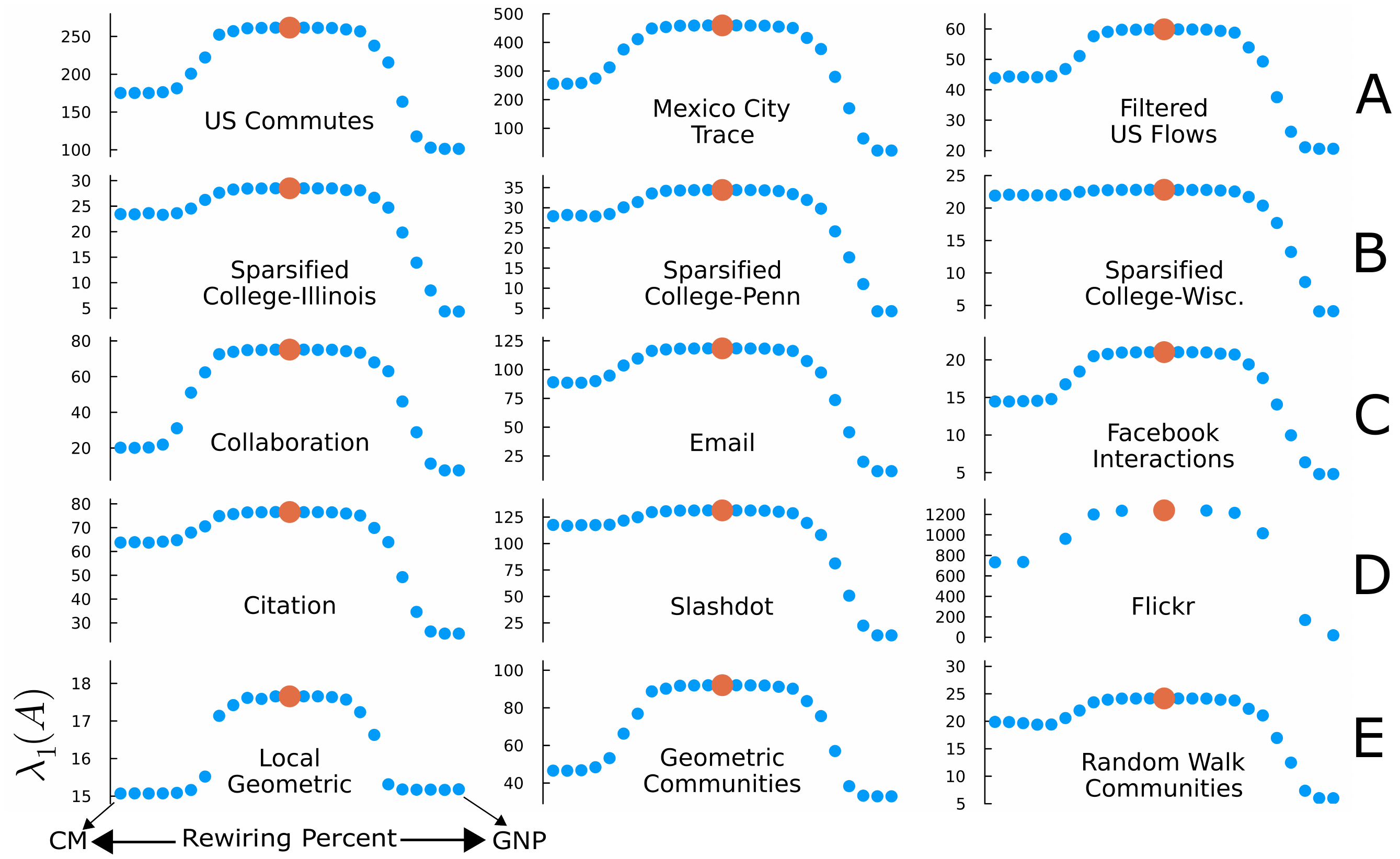}
    \caption{Using the epidemic threshold, $\frac{\beta}{\gamma}\lambda_1(\mA)$, as a measure of epidemic strength would predict that higher values of $\lambda_1$ produce infections that are harder to control. However, the opposite result holds for these empirically measured networks. We display the dominant eigenvalue $\lambda_1(\mA)$ of the adjacency matrix ($\mA$) vs the same $\text{CM} \leftarrow \text{Original} \rightarrow \text{GNP}$ rewiring on the horizontal axis (omitted for clarity). As we move from the center to the left (right) we are increasingly rewiring to CM (GNP). The original networks have larger eigenvalues, yet, comparing with Figure~\ref{fig:uplots-all} shows that almost all of these epidemics require smaller interventions than synthetic variants to impede or halt spreading, contrary to what would be expected from eigenvalues alone.}
    \label{fig:eigenvalue-fig}
\end{figure}

To understand whether these theoretical results apply to realistic networks, we study values of $\lambda_1(A)$ as the networks are rewired in Figure~\ref{fig:eigenvalue-fig}.  The original networks (red marker in subplots) all have larger values of $\lambda_1$. 
Contrary to what might be expected given the preceding discussion, $\lambda_1$ is \emph{not} predictive of intervention effectiveness. 
Despite larger eigenvalues, many of the epidemics on the original networks are easier to disrupt and hence are more controllable. 
Referring back to Figure~\ref{fig:uplots-all}, most of the original networks exhibit fewer total infections compared to rewired variants, despite having larger spectral radius and hence a smaller extinction regime. 
Moreover, as networks are rewired, most eigenvalues decrease monotonically. 
Despite this, total infections can exhibit a spike when rewiring to GNP under quarantining (Email, Slashdot, Flickr, RandomWalkCommunities).

These results suggest that minimizing the spectral radius is not appropriate for empirical data containing multi-scale local structure. 
The idea behind minimizing spectral radius is to extend the extinction regime in the mean-field, under the assumption that this is a good heuristic for mitigating spread.
However, this provides no information about epidemic indicators such as total infections, maximum infections, etc, when one is far from that threshold.  
Moreover, the spectral radius is fundamentally a global metric that can be locally distorted.  
One reason for this lies in the empirical localization of the dominant eigenvalue in empirical networks~\cite{pastor2016distinct}. 
A previous line of work~\cite{cucuringu2011localization,goltsev2012localization,pastor2016distinct,pastor2018eigenvector} has demonstrated the impacts of such localization on epidemic spreading for targeted intervention. 
By way of analogy, the spectral radius measures how easy a material is to ignite rather than measure how difficult it is to extinguish.
Consequently, a large spectral radius indicates how easy it is to start an epidemic on the network rather than a forecast of an ongoing epidemic.

\subsection{Local Structure Modulates Epidemic Spread Under Local Intervention} 
\label{sec:local-structure-measure}
\begin{figure}
    \centering
    \includegraphics[width=0.8\textwidth]{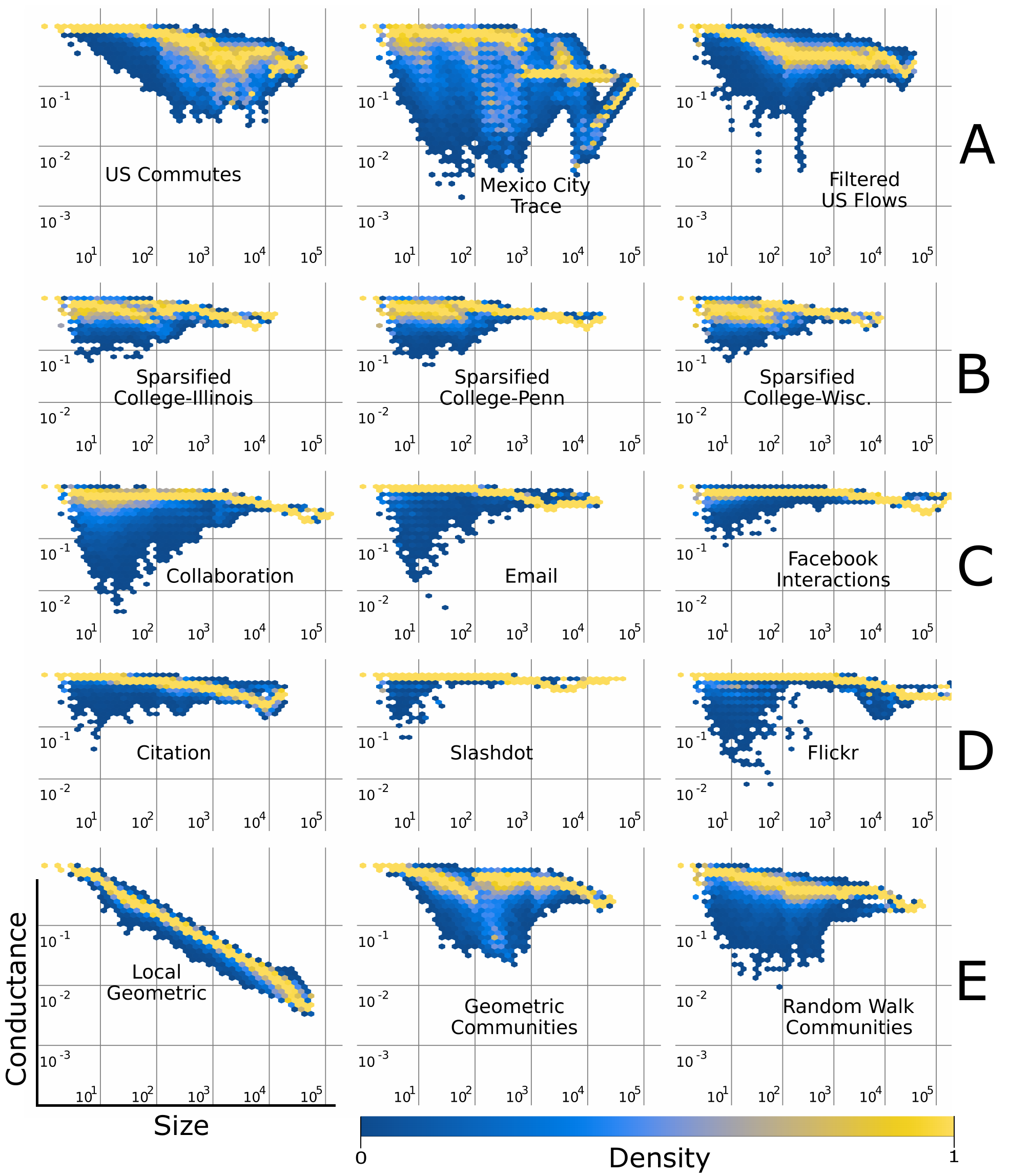}
        \caption{Local set structure explored by epidemics in these conductance-vs-size plots for graphs derived from the 15 networks. Mobility networks (row A) have some of the highest diversity in sets explored, whereas networked interactions (rows C and D) have lower diversity in this space. Compare with Figure~\ref{fig:uplots-all}. 
        }
    \label{fig:fig-ncp-trajectory}
\end{figure}

    The relationship between controllability and epidemic behavior can be understood through the degree to which local structure manifests in these networks.
    Intuitively, we expect that smaller conductance values at multiple size-scales should correspond to more successful epidemic interventions.
    This is due to local interventions being able to exploit near-by bottlenecks.
    Figure~\ref{fig:fig-ncp-trajectory} shows the epidemic NCP for the same set of networks as in Figure~\ref{fig:uplots-all}.
    Recalling our analogy with local structure as high diversity of sets visited by epidemics, we note that networks where interventions are better able to curb spreading tend to have larger support in the size-vs-conductance space.
    
    This led us to take an \emph{area-based} measure of the support of the NCP plot as a quantitative measure of local structure at multiple size scales. 
    The specific measure we compute is a normalized version of the Area Above the NCP (AANCP). 
    If $\vx$ is a normalized measure of set-size (maximum on x-axis in Figure~\ref{fig:fig-ncp-trajectory} normalized to 1), and $\vy$ are the minimum values  
    from the NCP at the corresponding values of set-size, then we use 
    \begin{equation}
        \text{AANCP}(\vx,\vy) = \text{AUC}(\log_{10}(\vx),\ -\log_{10}(\vy)),
        \label{eq:aancp}
    \end{equation}
    where AUC denotes the traditional Area Under the Curve metric.
    Equation~\ref{eq:aancp} is simply the normalized area from maximum possible conductance $\phi = 1$ to the observed minimum in empirical sampling (further details in Appendix~\ref{sec:area-vs-local-structure-details}). 
    This measure is closely related to the notion of the minimum envelop used in the initial literature on the NCP~\cite{leskovec2009community}.
    While the authors used the minimum conductance to define the NCP, we use normalized area above the minimum to capture the size of the support.

\begin{figure}
    \centering
    \includegraphics[width=\textwidth]{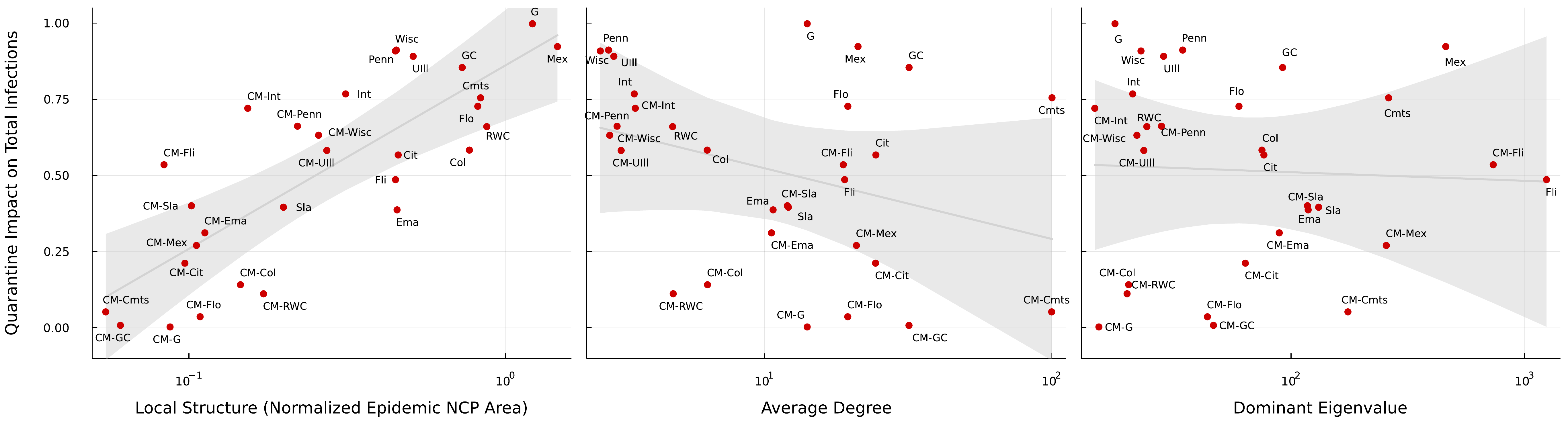}
    \caption{Local structure has a positive association with effectiveness of local interventions while measures such as $\lambda_1(\mA)$ and average degree have little or  no correlation. 
    When we use normalized area above NCP, AANCP (see discussion in text), of the epidemic NCP as a measure of local structure (horizontal axis in leftmost plot) and quantify the impact of quarantine as: $1-\frac{\text{average infections under quarantining}}{\text{average infections without quarantining}}$, we then find a positive correlation (Spearman $\rho$ correlation coefficient $0.78$ with 95\% CI $[0.58,0.89]$). 
    Values close to 1 (the maximum impact) indicate that quarantining was more effective.
    Epidemic parameters used for the data shown are the same as those in Figure~\ref{fig:uplots-all}.
    }
    \label{fig:area-vs-local-structure}
\end{figure}

    Using Equation~\ref{eq:aancp}, Figure~\ref{fig:area-vs-local-structure} shows that, as a measure of local structure, AANCP  has a positive association with quarantine impact, while average degree and the dominant eigenvalue do not (middle and rightmost subplots). 
    One would expect that average degree and dominant eigenvalue have a negative association. 
    However, the presence of multi-scale local structure distorts this quite substantially.
    Figure~\ref{fig:area-vs-local-structure} contains a network as well as it's CM rewired variant, and it shows that the difference in quarantine impact in the presence of local structure can be quite substantial.
    Quarantine impact is quantified as 
    \[1-\frac{\text{average infections under quarantining}}{\text{average infections without quarantining}},\]
    so that values close to 1 indicate that quarantining was very effective while values close to 0 indicate quarantining had no impact.
    (Additional statistics discussions on these results are given in Appendix~\ref{sec:area-vs-local-structure-details}.)

\subsection{Local Structure is Indicative of Who Does Not Get Infected}
\label{sec:missed-sets}

Multiple local~\cite{salathe2010dynamics,gong2013efficient,taghavian2017local} and global~\cite{hebert2013global,gupta2016centrality,kumar2018efficient} immunization strategies have been proposed for networked epidemics that explicitly target nodes or edges based on the modular structure of networks. 
Implicitly, these methods seek to contain epidemic spread by making the networks more modular by targeting nodes and edges that accelerate spreading (e.g., community bridges or hubs). 
Figure~\ref{fig:fig-ncp-trajectory} suggests that epidemic spread is global since for most networks,  higher conductance sets are explored far more often than low conductance sets at all size scales.
Moreover, lower conductance tends be seen at smaller sized sets. 
What this suggests is that very good conductance sets (i.e., communities) may be more indicative of what portions of the graph are \emph{not} explored. 
To support this point, we explore the extent to which an epidemic fails to infect extremal conductance sets. 

\begin{figure}
    \centering
    \includegraphics[width=\textwidth]{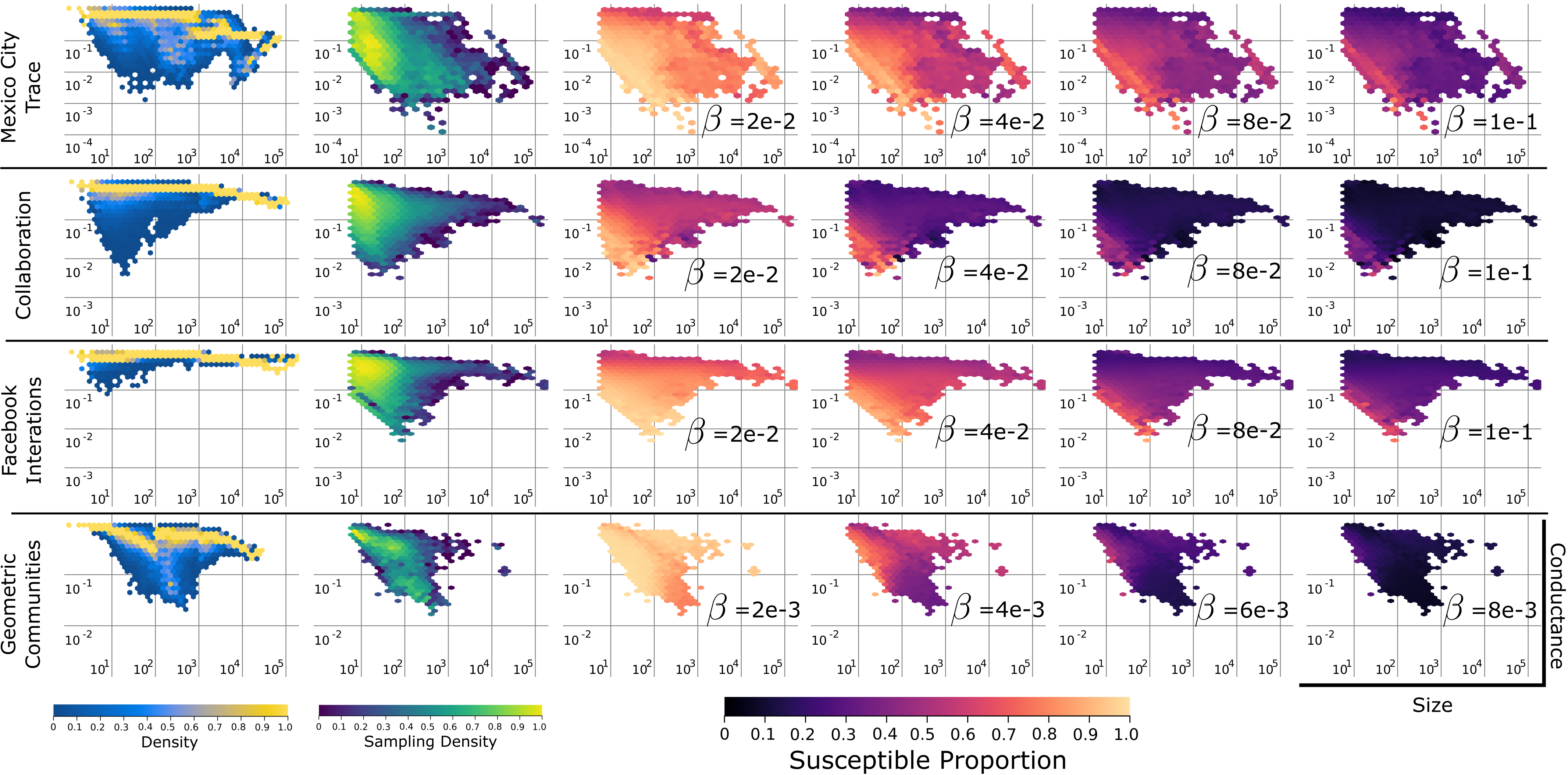}
    \caption{Small conductance sets at small sizes serve as barriers to infection. The first column shows the epidemic NCP. The second column uses PageRank vectors to better sample extremal conductance sets at small size scales. Columns 3-6 uses sets found by PageRank vectors but weighs each set according to the proportion of nodes susceptible at the end of 50 SEIR epidemics. Nodes in low conductance sets at small size scales are more resistant to infection and require larger infection probabilities ($\beta$) for epidemics to reach those nodes. This suggests that these sets serve as local mixing bottlenecks and provide barriers to infection. Epidemics in this figure do not use any quarantining. The recovery probability is fixed at $\gamma=0.05$ while the infection probability, $\beta$, is displayed in columns 3-6.}
    \label{fig:missed-sets-fig}
\end{figure}

Figure~\ref{fig:missed-sets-fig} shows that sets at small sizes with smaller conductance values resist infection and serve as barriers to infection (Columns 3-6). 
Columns 3-6 shows the evolution of node susceptibility in the size-vs-conductance space.
As the infection strength ($\beta$) increases, the portion of these NCPs more likely to remain susceptible are sets that have low conductance and small size (bottom left of individual plots in columns 3-6 of Figure~\ref{fig:missed-sets-fig}). 
Overcoming these conductance bottlenecks requires even stronger epidemics.
Sets in columns 3-6 are derived from a different NCP that uses seeded PageRank information to retrieve sets with good conductance scores. 
This approach comes with strong theoretical guarantees~\cite{andersen2006local,gleich2016seeded} (see Appendix~\ref{sec:epidemic-ncps-alldetails} for more details). 
The local structure derived from seeded PageRank is combined with epidemic information to show susceptible nodes in the size-vs-conductance.
While the epidemic NCP in column 1 gives us information about where the epidemic explored, the seeded PageRank NCP is used to better sample extremal sets at the small size regime.
Put simply, the epidemic NCP is biased towards disease spread, while the PageRank NCP is biased towards minimal conductance sets.

While previous research has noted lower epidemic prevalence in smaller sized communities~\cite{liu2016community,ghalmane2019immunization}, Figure~\ref{fig:missed-sets-fig} offers an explanation for \emph{why} this is the case.
These sets are harder to get into during epidemic spreading due to the nearby local mixing bottlenecks in the network.

\subsection{Generative Models of Local Structure }
\label{sec:synthetic}
Empirical results from the realistic networks we have analyzed suggests that multi-scale local structure critically impacts the evolution of epidemics in light of local interventions. 
However, realistic empirical networks are often a confounding mixture of many phenomena. 
Here, we give two generative network models that produce local structure in distinct ways. 
The first uses local geometry to add community structure while maintaining large scale randomness.  
The second starts with planted community structure and adds local structure via local random walks, akin to what was studied in ref.~\cite{leskovec2009community}. 
Both models feature long-range edges or weak ties, the presence of which is know to impact disease spreading~\cite{eckles2018long,onnela2007structure}.
In both, we will see that the effect of quarantine interventions is similar to the most realistic empirical networks formed from human mobility traces.

\paragraph{GeometricCommunities.}
The first synthetic model (GeometricCommunities) seeks to create local pockets of community structure with large scale randomness. 
It also aims to produce networks that have a core-periphery structure~\cite{borgatti2000models} consisting of a large dense ``core'' and a ``periphery'' consisting of relatively small dense pockets that are loosely connected to the core. 
This feature of networks has been observed in a wide range of empirical networks~\cite{leskovec2009community,ASM13,ASM16_IM}.
We begin by assigning each node a random coordinate in a toroidal space (2d box with wrap around). 
Each node picks a degree from a log normal distribution and connects up to the nearest local neighbors. 
Each node also samples a much smaller number of neighbors randomly across the network to simulate weak global ties. 
Distances are computed in toroidal space (allowing wrap-around), while weak ties are selected in a preferential attachment manner, with high degree nodes adding more edges than low degree nodes. 
Node degree is then used as a proxy for node importance. 
Nodes \emph{update} their positions by moving to their most important (highest degree) neighbor as a means to emulate social influence or homophily. 
Local edges are reformed while some fraction of weak-ties is maintained (details in Appendix~\ref{sec:local-geometric-alldetails}). 
After updating positions and keeping weak ties, they reconnect to nearest neighbors to roughly maintain their degree. 
Each step of updating the node positions and reallocating links is called an iteration. 
Nodes that become disconnected are randomly re-initialized to simulate migration. We see the development of similar local structure (as measured by the NCP plots) in about 50-150 iterations. 
The goal is to induce community structure (local accretion to the high degree nodes) with large-scale randomness (weak ties) while maintaining some of the planted 2d geometry (coordinates).
Note that the average degree in the network grows via this process because the connection decisions are made from either side.

\paragraph{RandomWalkCommunities.} 
The second synthetic model (RandomWalkCommunities) used to generate local structure uses a sparse LFR model~\cite{lancichinetti2008benchmark} augmented by local random walks. This model plants community structure and then augments it with local structure. We start with a very sparse LFR model, then connect any disconnected components via a Chung-Lu~\cite{chung2002connected,miller2011efficient} process that only adds an edge if it decreases the number of connected components. Local random walks are then used to augment this base graph. At each iteration, we sample a seed node and begin a random walk from it. At each time-step of the random walk, we first stop the walk with a fixed probability, otherwise we randomly pick whether to mark the current node. Once the walk stops, all marked nodes are connected to the seed node. The main goal is to take planted structure and augment it with rich local structure around a node. Since each iteration is only one random walk, we perform many thousands of steps. And since we are adding edges in each step, the average degree grows in this process as well.

\paragraph{Epidemic Simulation Results for Synthetic Models of Local Structure.}

\begin{figure}
    \centering
    \includegraphics[width=\textwidth]{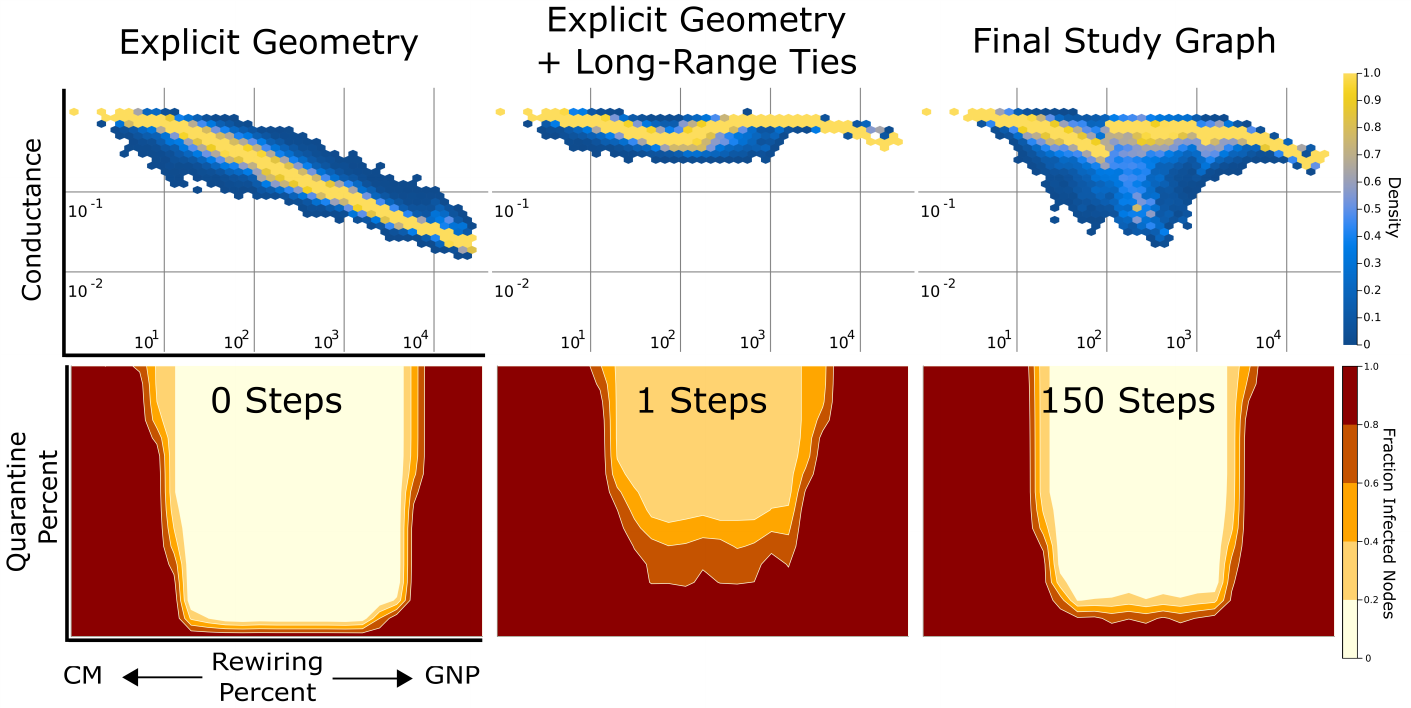}
    \caption{The GeometricCommunities model helps us understand how long-range edges can simultaneously mask multi-scale structure and reduce epidemic controllability.
    The first column shows the epidemic NCP (top) and total infections (color in bottom plot) for a network where edges are derived from explicit geometry.
    Minimal quarantining is needed to halt the epidemic entirely.
    The middle column shows the same information using the network in column 1 but adding a small fraction of long-range ties.
    This destroys local structure and dramatically reduces epidemic controllability, as larger interventions are required to mitigate spreading.
    The last column shows the same information for the network in column 2 after 150 iterations of the GeometricCommunities model in which we have both more local structure and the epidemic is more controllable. 
    Epidemic parameters are chosen based on the epidemic threshold $s=\lambda_1(\mA)\frac{\beta}{\gamma}$ due to large differences in average degree (see Appendix~\ref{sec:app-spatial-graph-fig} for details).
    }
    \label{fig:spatial-graph-fig}
\end{figure}

GeometricCommunities is a generative model that produces rich multi-scale local structure while qualitatively matching observed epidemic behavior of empirical networks.
In Figure~\ref{fig:uplots-all} we see that local interventions on GeometricCommunities can produce the same type of epidemic response as epidemics on human mobility networks by comparing row (A) to the subplot for GeometricCommunities (middle of row (E)). 
Moreover, Figure~\ref{fig:fig-ncp-trajectory} shows that GeometricCommunities has qualitatively similar levels of multi-scale local structure as empirical mobility networks. 
This is further exemplified by Figure~\ref{fig:area-vs-local-structure}, as we compare epidemic response vs $\lambda_1$ and average degree. 
As a generative model with both geometry and long-range edges, we are further able to probe the impact of long-range edges on local structure and spreading.
Long-range edges or weak-ties~\cite{onnela2007structure} are known to accelerate epidemic spreading~\cite{balcan2009multiscale,eckles2018long,mercier2022effective} as well as to seed new local epidemics~\cite{wren2021local}. 
From this alone, we should expect that long-range edges should mask local structure.
Figure~\ref{fig:spatial-graph-fig} shows that is exactly what happens when one adds in long-range edges.
As we move left to right column-wise, the networks on the right are derived from networks on the left by adding long-range edges (middle subplot) and then applying an iterative local homophily or social influence update to node positions and reforming local edges while preserving long-range edges.
In doing so, the long-range edges completely masks the latent geometry in the network, while simultaneously accelerating spreading and reducing controllability.
However, after 150 steps of the iterative node and edge update procedure, we arrive at the final network for GeometricCommunities which behaves like a spatial model (Local Geometric) with regards to local interventions but which has a wider distribution of local structure than a purely geometric network.

\begin{figure}
    \centering
    \includegraphics[width=\textwidth]{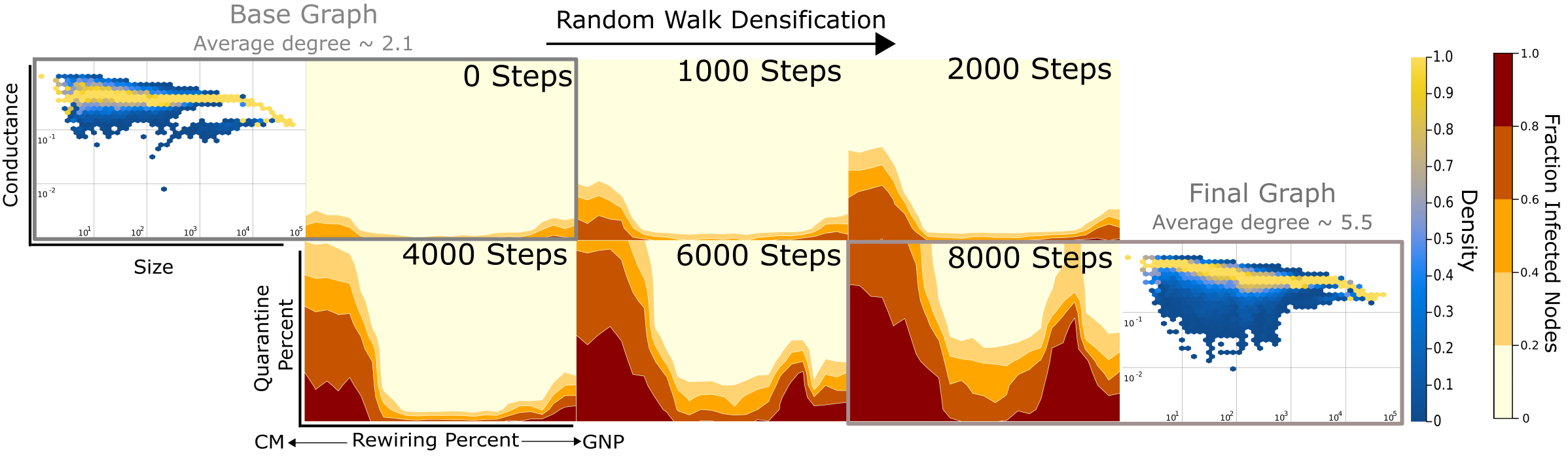}
    \caption{The RandomWalkCommunities model emulates local structure by combining a very sparse LFR graph to plant community structure along with local structure that arises by adding edges with local random walks. This model exhibits qualitatively similar effects as Filtered US Flows under local interventions (see Figure~\ref{fig:uplots-all}). RandomWalkCommunities is generated with planted community structure augmented by local random walks. As the number of random walks increase (number of steps in figure increase), the model exhibits increasing levels of local structure (larger values of AANCP). 
    Moreover, epidemics on these networks exhibit greater deviations from synthetic models lacking local structure under local interventions(contour plots).}
    \label{fig:lfr-fig}
\end{figure}

RandomWalkCommunities gives a separate generative model that is produced in a reverse manner from GeometricCommunities.
It starts with planted communities and augments them with local random walks rather than purely geometric or spatial patterns. 
In Figure~\ref{fig:lfr-fig}, we depict the total infection vs rewiring and quarantining as the number of local random walks increases.
As we increase the number of random walks taken, the difference in total infections between the generated model and rewired variants become more pronounced.
Long-range connections are more prevalent in this model as they emerge from long paths in random walks themselves, rather than having a fixed proportion being planted. 
Looking back to Figure~\ref{fig:uplots-all} (bottom right of that figure), note that RandomWalkCommunities exhibits similar behavior to Filtered US Flows under local interventions, notably so as we rewire to ER.
While less interpretable than GeometricCommunities, RandomWalkCommunities tracks well with empirical networks in producing local structure (Figure~\ref{fig:fig-ncp-trajectory}) and emulates epidemic behavior (Figures~\ref{fig:uplots-all} and~\ref{fig:area-vs-local-structure}).

\subsection{Triangle Weighted Diffusions}

It is common to assume that local structure in networks is rooted in triangles and local clustering coefficients~\cite{watts1998-dynamics}.
To test this hypothesis and determine whether our findings are simply a matter of the number of triangles in a network, we randomized the triangles of the network. 
One challenge with this experiment is that randomizing triangles produces changes to the number of edges, which makes it difficult to use the same epidemic parameters, as they are sensitive to average degree. 
To account for this effect, we simulate an epidemic on the hypergraph formed via triangles and edges. 

\label{sec:hypergraph-diffusions}

    \paragraph{Hypergraph Weighted Diffusions.}  
    We consider a hypergraph model of epidemic spread using triangles or edges as modes of disease spread.
    Each hyperedge represents a joint meeting of all nodes in that hyperedge.
    In particular, we allow triangles to serve as modes for disease transmission with the same edge infection probability, $\beta$.
    We also include pairwise-edges since otherwise spreading would be impossible on the triangle-free portion of the network.
    Fixing $\beta$ for all edges and hyperedges allows this hypergraph model of spread to be expressed as a weighted pairwise model on the original network. 
    (See Appendix~\ref{sec:hypergraph-epidemic-alldetails} for details and exact formulation.) Edges in the original graph, $\mA=(a_{i,j})$, are weighted so that $w_{i,j} = \max(a_{i,j},T_{i,j})$, where $T_{i,j}$ is the number of triangles in which nodes $i$ and $j$ jointly participate.
    The probability of infection along each edge is then given as $\beta w_{i,j}$.
    This hypergraph model is used on networks in which triangles are shuffled around the network so that the average driving force of the infection ($\sum_{i,j} \beta w_{i,j}$) is preserved as triangles move. 

\begin{figure}
    \centering
    \includegraphics[width=0.75\textwidth]{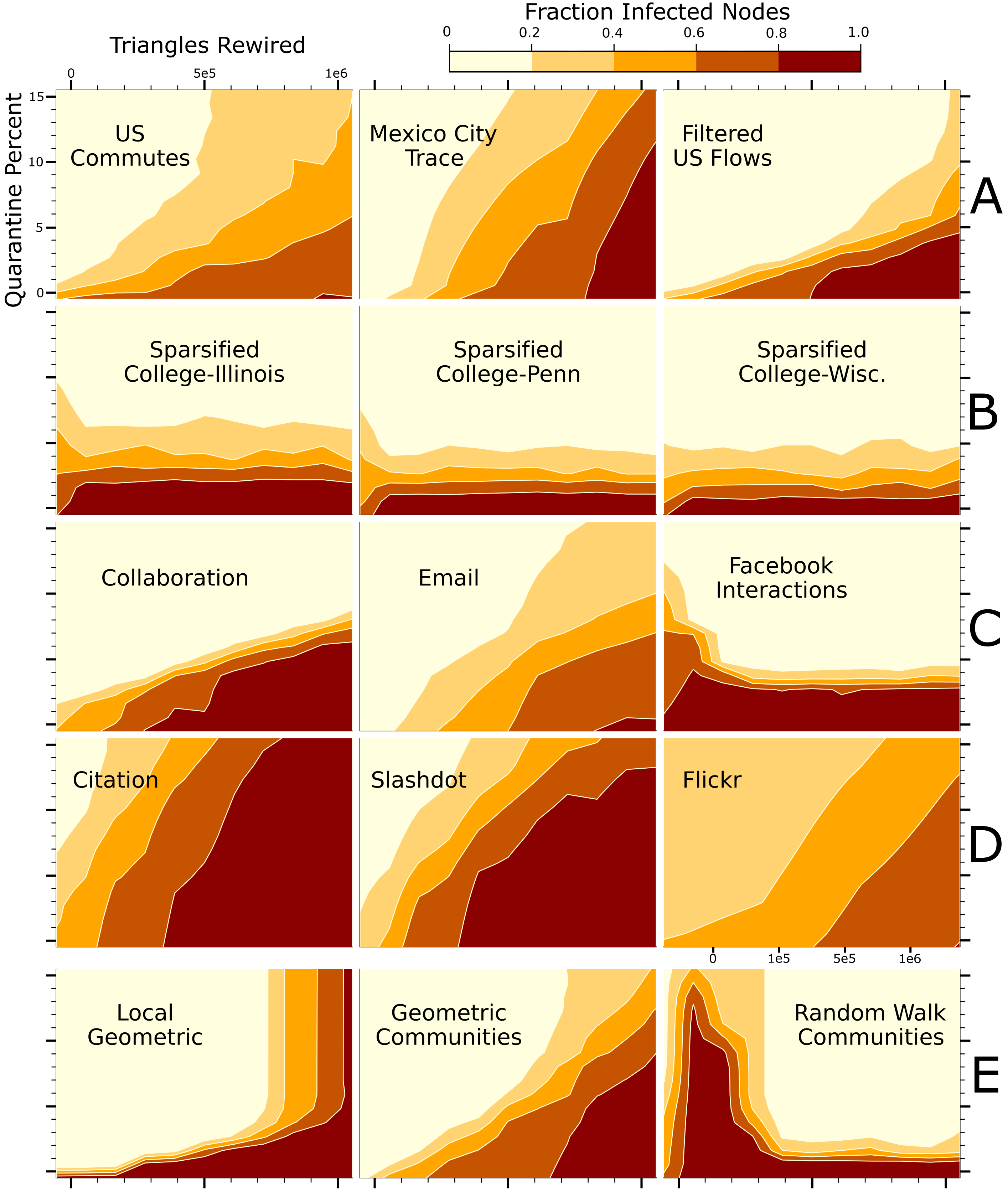}
    
    \caption{Triangles are insufficient to explain the effects of local interventions.
    To test this, we shuffle the location of triangles in the network (horizontal axis) and measure the impact of quarantine interventions (this is similar to one half of Figure~\ref{fig:uplot-explanation}).
    For most networks, randomizing triangles requires more quarantining to mitigate spread (Rows A and D, most of C and E). 
    This shows that the local structure we study exists beyond local triangle structure because epidemic response varies as we shuffle triangles.
    See Appendix~\ref{sec:app-uplots-all} and~\ref{sec:app-uplots-triangles} for details concerning choice of epidemic parameters. 
    } 
    \label{fig:uplot-triangles}
\end{figure}

    \paragraph{The Impact of Randomizing Triangles.}
    To randomize the triangles of the network, we treat each triangle as a hyperedge, and we randomize its endpoints at each step for up to a million steps.
    Figure~\ref{fig:uplot-triangles} shows that for a majority of networks, triangles are insufficient to model local structure under the above hypergraph weighted diffusion using the same quarantine intervention as in the pairwise case. 
    The leftmost side of each subplot represents the original network.
    As we move to the right, the triangles in the network are increasingly shuffled. 
    There are three distinct types of observed phenomena (demonstrated by the rightmost subplots of rows A, B, E). 
    Case 1 (triangles insufficient) is exemplified by Filtered US Flows, Case 2 (triangles sufficient) by Sparsified College-Wisc, and Case 3 (sensitive to parameters) by RandomWalkCommunities). 

    For most networks, \emph{triangles are insufficient to model local structure}.
    This is Case 1, where shuffling of triangles requires more quarantining to stop spreading (e.g.~row A).
    For a smaller minority of networks (row B and Facebook Interactions), triangles are the full extent of local structure.
    For the networks in row B, this makes sense since those networks were sparsified in a biased manner towards edges that participate in many triangles. 
    Put simply: we sparsified them such that triangles are the only remaining local structure.
    Facebook Interactions (rightmost of row C) also exhibits the same behavior, even without this sparsification.
    Case 3 only occurs for RandomWalkCommunities (rightmost plot of row E). 
    Here, triangle shuffling initially causes a large spike in total infections before dramatically reducing and becoming independent of total infections.
    If we ignore the initial spike, then we see behavior similar to case A in that the networks are harder to control after randomization.

\section{Discussion}
\label{sec:discussion}

Our results clearly highlight the significance of multi-scale local structure on the epidemic spread and the effect of interventions on realistic network epidemics. 
We now discuss two related aspects: existing measures of epidemic strength; and community structure. 

The reproduction number $R_0$ is canonically used to characterize the strength or severity of an epidemic~\cite{cushing1994net}.
The value of $R_0$ gives the expected number of infections that an individual will cause in a completely susceptible population. 
In the case of non-networked epidemics (epidemic modeling without access to network structure), 
the critical threshold for epidemic spread is $R_0=1$. 
However, in the SIR and SEIR networked case~\cite{prakash2012threshold}, $s=\lambda_1(\mA)\frac{\beta}{\gamma} = 1$ is the critical threshold in the mean-field. 
Despite this similarity between $R_0$ and $s$, we did not find a good use for $s$ in terms of quantifying epidemics on realistic interaction networks. 
For example, in Figure~\ref{fig:area-vs-local-structure}, two networks can have similar values of $\lambda_1$ and $s$ despite very different behaviors in epidemic spreading. 
Moreover, $s$ is distinct from $R_0$ since $R_0$ is bounded above by average degree in a network while $\lambda_1$ is bounded in terms of the maximum degree of a network. 
The dependence of $s$ on $\lambda_1(\mA)$ has further epidemic consequences due to known issues of localization in the corresponding dominant eigenvector~\cite{pastor2018eigenvector,silva2021dissecting}. 
Our results suggest that $s$ should be interpreted as a best-case threshold in stochastic simulations, rather than a proxy of epidemic strength~\cite{tong2010vulnerability,saha2015approximation,chen2015node}. 

As previously mentioned, it is recognized that community structures are correlated with fewer infections~\cite{salathe2010dynamics} and have several local~\cite{salathe2010dynamics,gong2013efficient,taghavian2017local} and global~\cite{hebert2013global,gupta2016centrality,kumar2018efficient} strategies for targeted interventions. 
However, the use of a metric like modularity can mask heterogeneity in community sizes and mixing patterns.
The variability arises from the diverse nature of what constitutes ``community structure.'' 
Moreover, community mixing patterns~\cite{min2013role} as well as sizes~\cite{liu2016community} can alter the fraction of infected nodes within communities.
For instance, a network composed of many small communities but smaller modularity can require smaller interventions than those composed entirely of medium or large sized communities but with larger modularity (see Supplementary Section~\ref{sec:app-community-heterogeneity} for an example of this). 
The use of the epidemic NCP that we introduce examines all size-scales to overcome this ambiguity.  
Furthermore, the epidemic NCP augments purely structural information on communities with epidemic spreading preferences (section~\ref{sec:local-structure-measure}). 
When combined with purely structural bottlenecks, we find that nodes contained in sets with small conductance and small size are harder to infect; and, in those cases, conductance bottlenecks serve as barriers to infection.

The modeling results show that local structure---in particular multi-scale local structure---can critically impact epidemic spreading. 
Practical success at mitigating epidemics depends on the relevance of the modeling regime in terms of speed with which interventions can be implemented. 
For instance, the speed of viral spread must be slow enough that interventions can meaningfully impact spread. 
For this reason, we assume a mean distribution of 5 time-periods (roughly, days) from exposed to infectious, which is faster than diseases such as measles. 
Thus, our findings reiterates the importance and likely effect of targeted interventions: paid sick leave and rapid tests to effectively implement quarantine procedures along with contact reduction. 
Indeed, our results suggest that sparsifying a network in such a manner will, in general, reduce the speed of viral spread and enhance the role that local bottlenecks play in the network. 
This shows how multiple interventions may amplify the effects of each other in empirical networks. 
In contrast, synthetic models lacking multi-scale local structure often misrepresent how much intervention is required to impact epidemic spreading, let alone halt it entirely. 
This is due to the lack of local conductance bottlenecks in synthetic networks that do not reflect the realistic heterogeneity in human mobility data. 
This disconnect between common synthetic models and empirical data highlights the need for widespread access to human mobility data and synthetic models that contain realistic multi-scale local structure. 
The widespread access of models with local structure and human mobility data will enable policy choices grounded in real-world behavior.

There is also the question of implementations of interventions. While we performed extensive experiments over a number of parameters ($\approx$ 380,000 for each graph in Table~\ref{table:graph-stats}), we assumed 100\% adherence to quarantining. This is an idealized case and speaks to the need for a real-time network for epidemic surveillance and disseminating information about positive tests. As was made apparent during the COVID-19 pandemic, not everyone will adhere to these interventions. A critical aspect to study in this regard is that intervention adherence  may itself be network mediated phenomenon (e.g.,~through information and misinformation spread through both in-person and online social networks), which makes studying and modeling this aspect particularly challenging.  

Finally, in terms of the prospect of future pandemics, our results here also reiterate the critical importance of monitoring, surveillance and early action in such regions prone to viral spillover~\cite{grange2021ranking}, due to the increased potency of early intervention in networks of human mobility with local structure.
For instance, it is estimated that that over 60,000 people are infected with bat-based coronaviruses annually in southeast Asia~\cite{sanchez2022strategy}.
That pandemics are not common may be related to the protective nature of multi-scale local structure in our contact networks.

\subsection*{Funding and Acknowledgements}
Gleich and Eldaghar acknowledges partial NSF support from awards CCF-1909528, IIS-2007481, as well as DOE award DE-SC0023162. 
Mahoney would like to thank the ONR and NSF for partial support of this work.

\bibliographystyle{plain}
\bibliography{refs}

\appendix

\newpage 
\part{Supplementary Information} 
\parttoc 

\section{Epidemic Models}
\label{sec:epidemic-models-alldetails}

\begin{figure}
         \centering
         \includegraphics[width=0.35\textwidth]{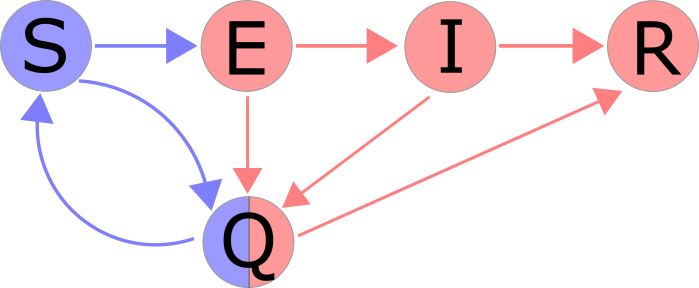}
         \caption{A schematic representation of transitions between disease states. S, E, I, and R form a partition of the nodes. A node can only transition out of Q into either S or R.}
         \label{fig:seir-model-description}
     \end{figure}

 We use discrete time SIR and SEIR models with quarantining. A diagram of state transitions is for the SEIR model with quarantining is shown in Figure~\ref{fig:seir-model-description}. 
     
There is only one infection probability, $\beta$, used for a single simulation. (For the triangle rewiring experiments, we used a hypergraph diffusion which differs slightly, see Appendix~\ref{sec:hypergraph-epidemic-alldetails}). The recovery probability is fixed in all experiments to $\gamma = 0.05$ so that, on average, nodes are infected for 20 time steps. This choice is made to fix the time scale of the epidemic. For SEIR, the standard exponential distribution with parameter $a$ for time between exposed and infectious is used. We fix $a=1/5$ so that, on average, nodes spend $5$ time steps in the exposed class. All epidemics are initialized with a single infected node. All epidemic simulations were repeated from 50 random seeds. Where possible, we avoid averaging over separate epidemics and represent results via distributions.

    \subsection{On the Choice of epidemic parameters}
    \label{sec:epidemic-params-app}

\begin{figure}
        \centering
        \includegraphics[width=0.8\textwidth]{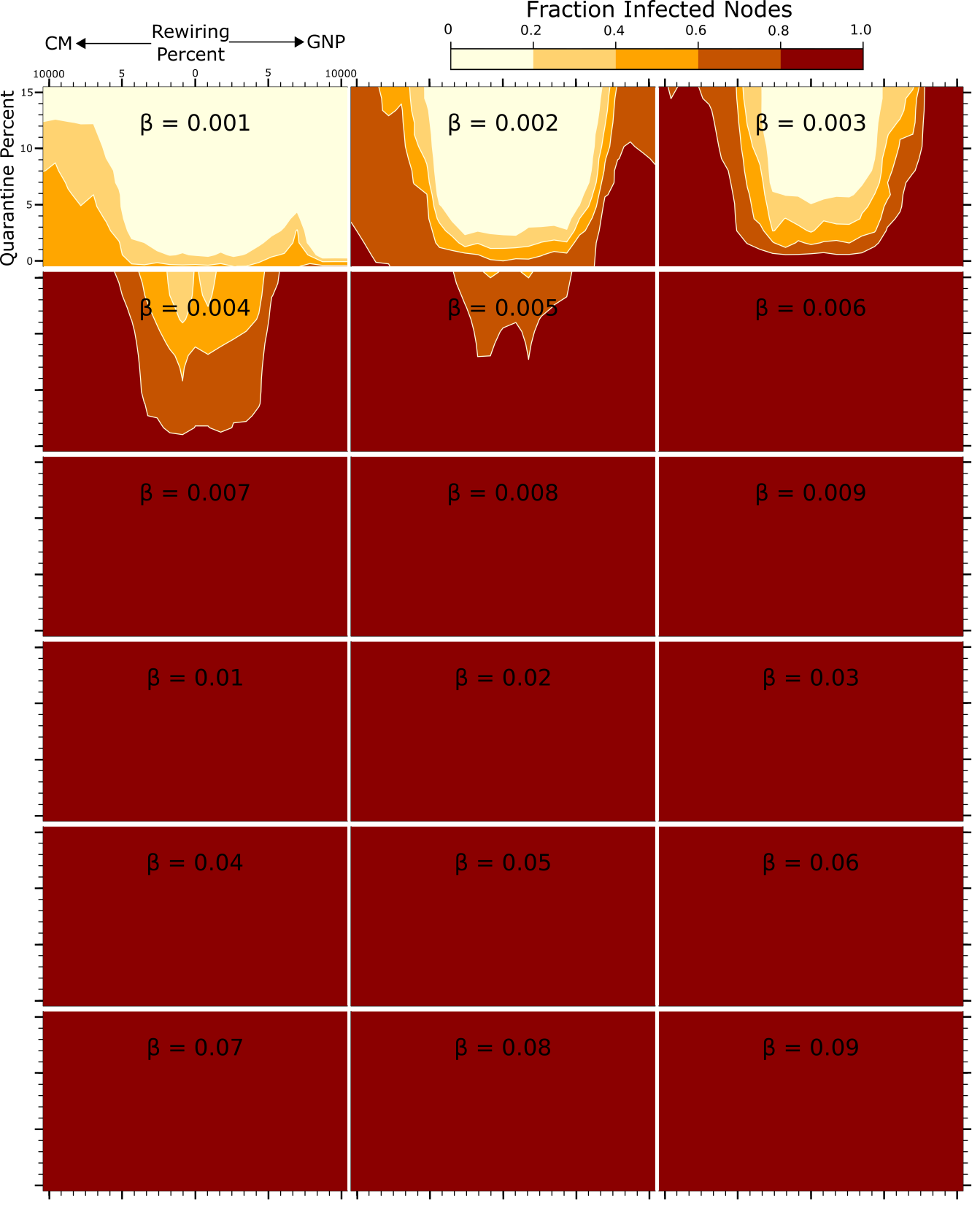}
        \caption{US Commutes diffusion data. Epidemic ``strength'' (original network with no quarantining) varies from $s\approx 5$ to $s\approx 471$.}
        \label{fig:uplot-params-commutes}
    \end{figure}

\begin{figure}
        \centering
        \includegraphics[width=0.8\textwidth]{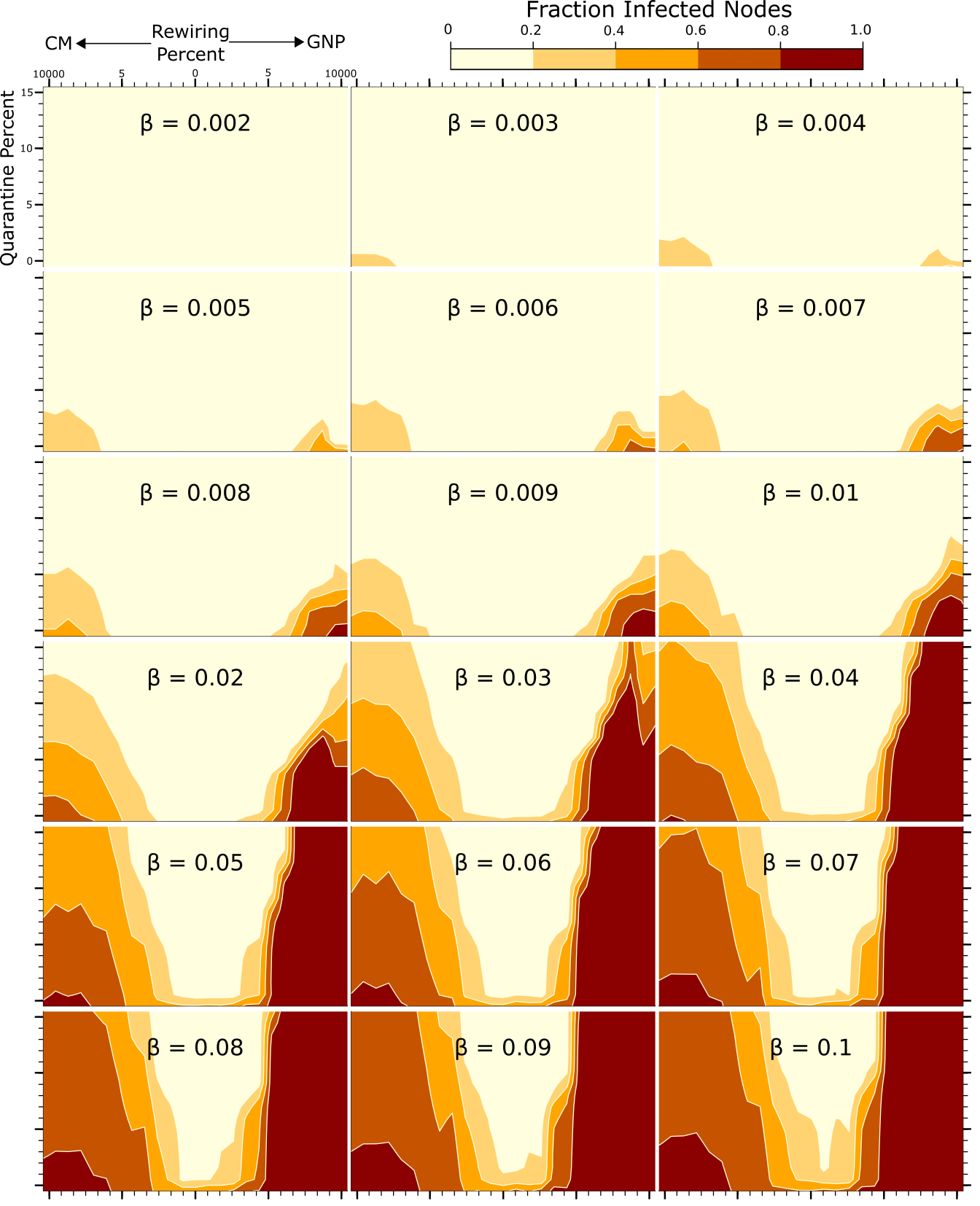}
        \caption{Mexico City Trace diffusion data. Epidemic ``strength'' (original network with no quarantining) varies from $s\approx 18$ to $s\approx 919$.}
        \label{fig:uplot-params-mexico}
    \end{figure}

\begin{figure}
        \centering
        \includegraphics[width=0.8\textwidth]{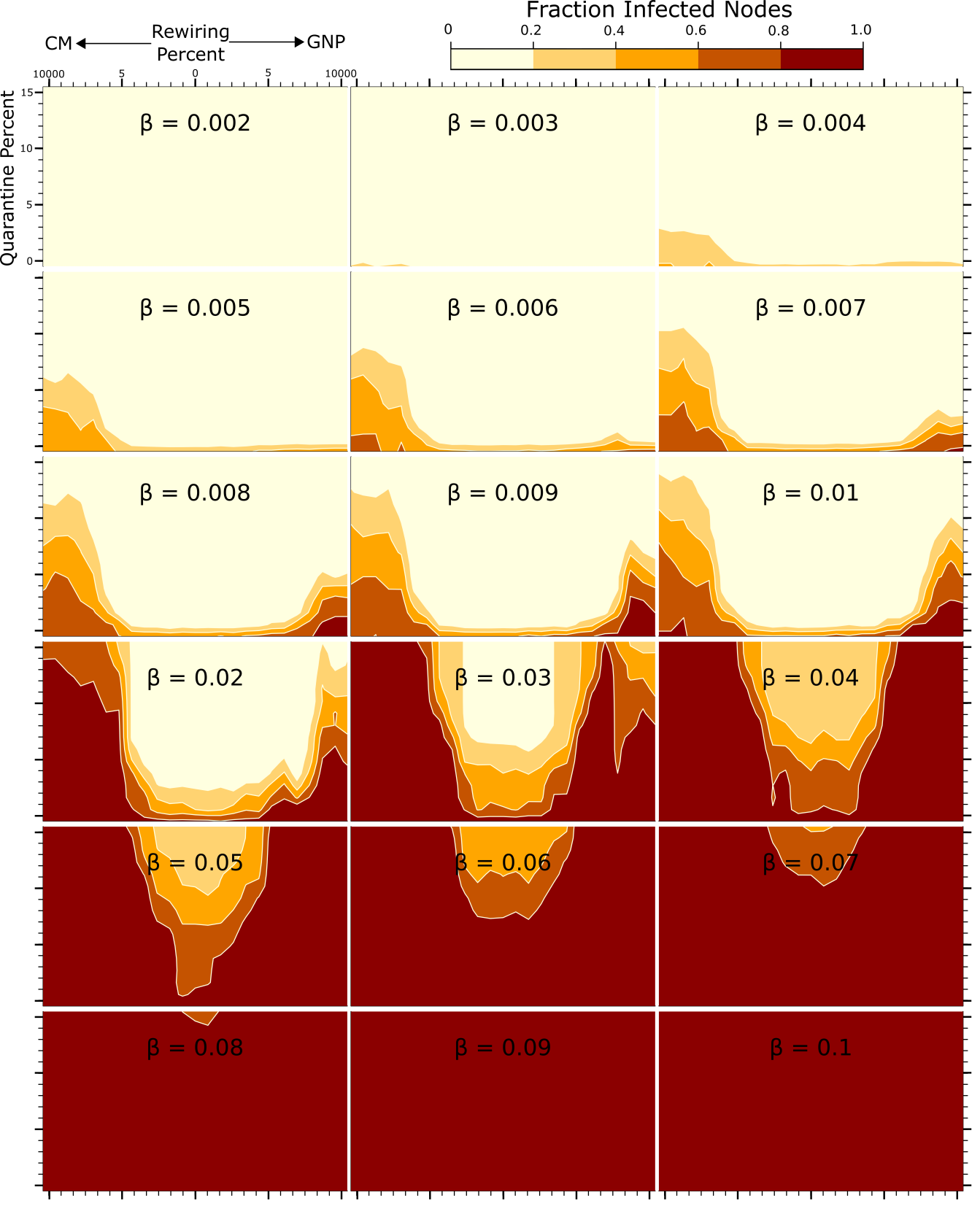}
        \caption{US Flows diffusion data. Epidemic ``strength'' (original network with no quarantining) varies from $s\approx 2$ to $s\approx 120$.}
        \label{fig:uplot-params-us-flows}
    \end{figure}

\begin{figure}
        \centering
        \includegraphics[width=0.8\textwidth]{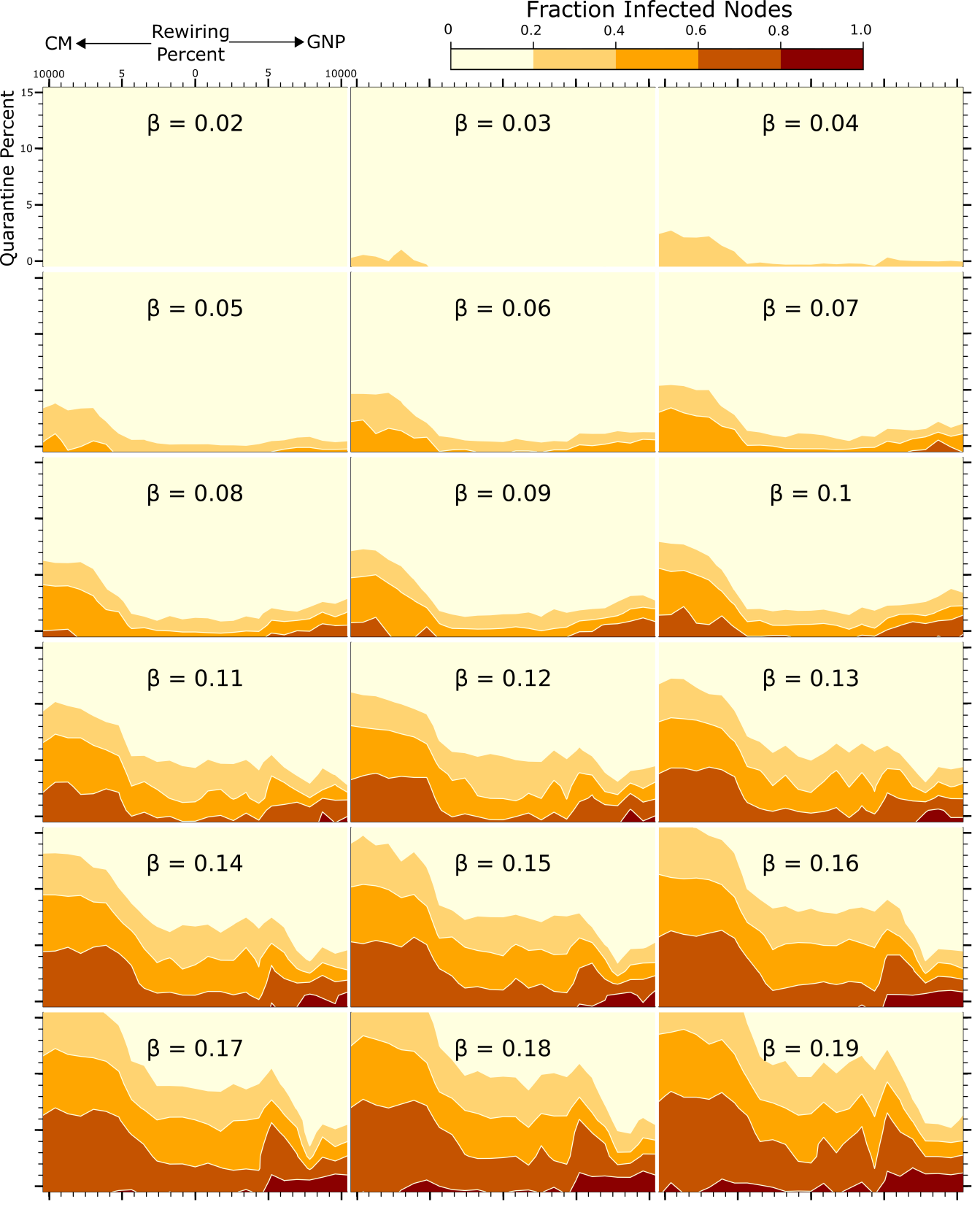}
        \caption{Sparsified College-Illinois diffusion data. Epidemic ``strength'' (original network with no quarantining) varies from $s\approx 11$ to $s\approx 108$.}
        \label{fig:uplot-params-cn-uill}
    \end{figure}

\begin{figure}
        \centering
        \includegraphics[width=0.8\textwidth]{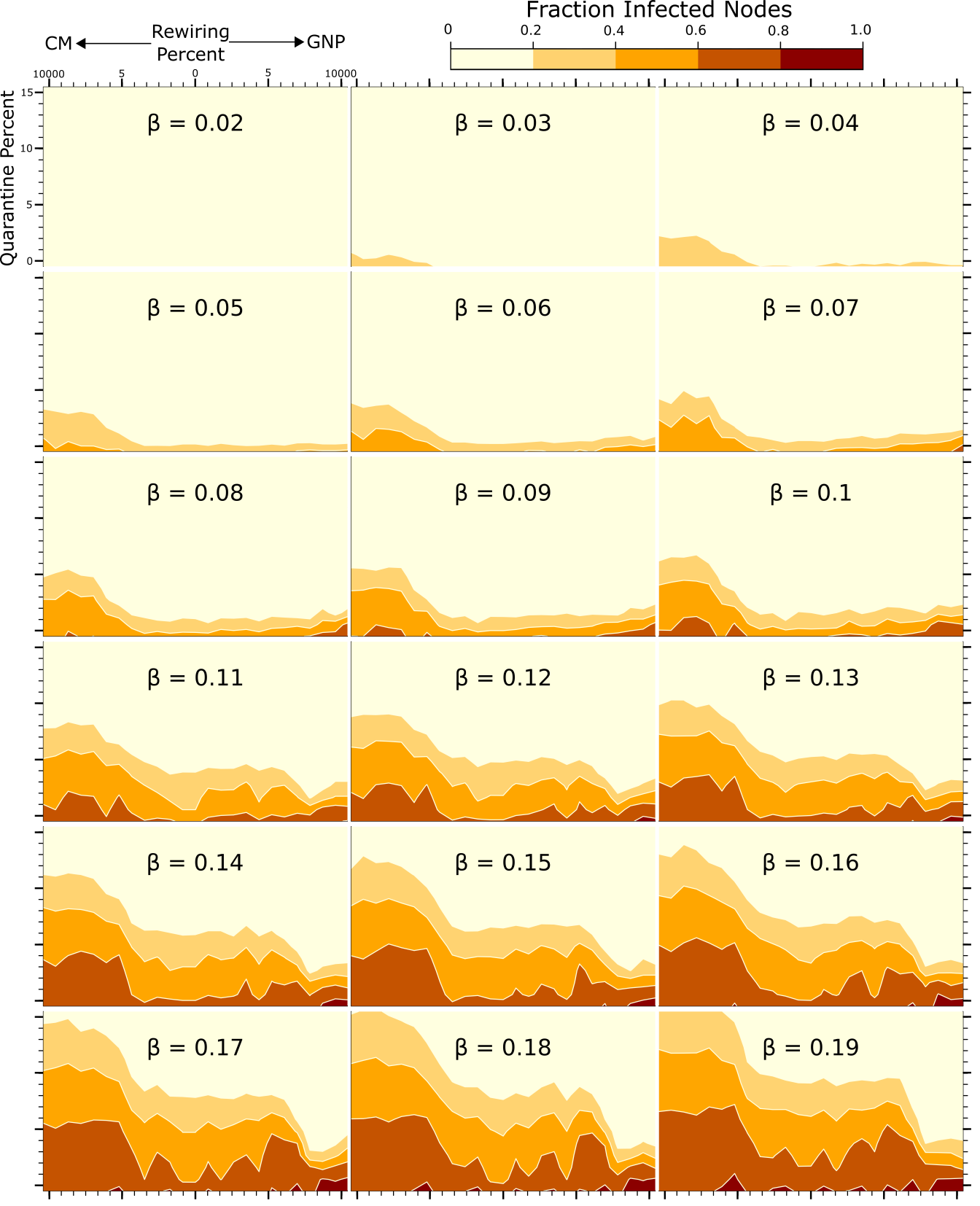}
        \caption{Sparsified College-Penn diffusion data. Epidemic ``strength'' (original network with no quarantining) varies from $s\approx 14$ to $s\approx 131$.}
        \label{fig:uplot-params-cn-penn}
    \end{figure}

\begin{figure}
        \centering
        \includegraphics[width=0.8\textwidth]{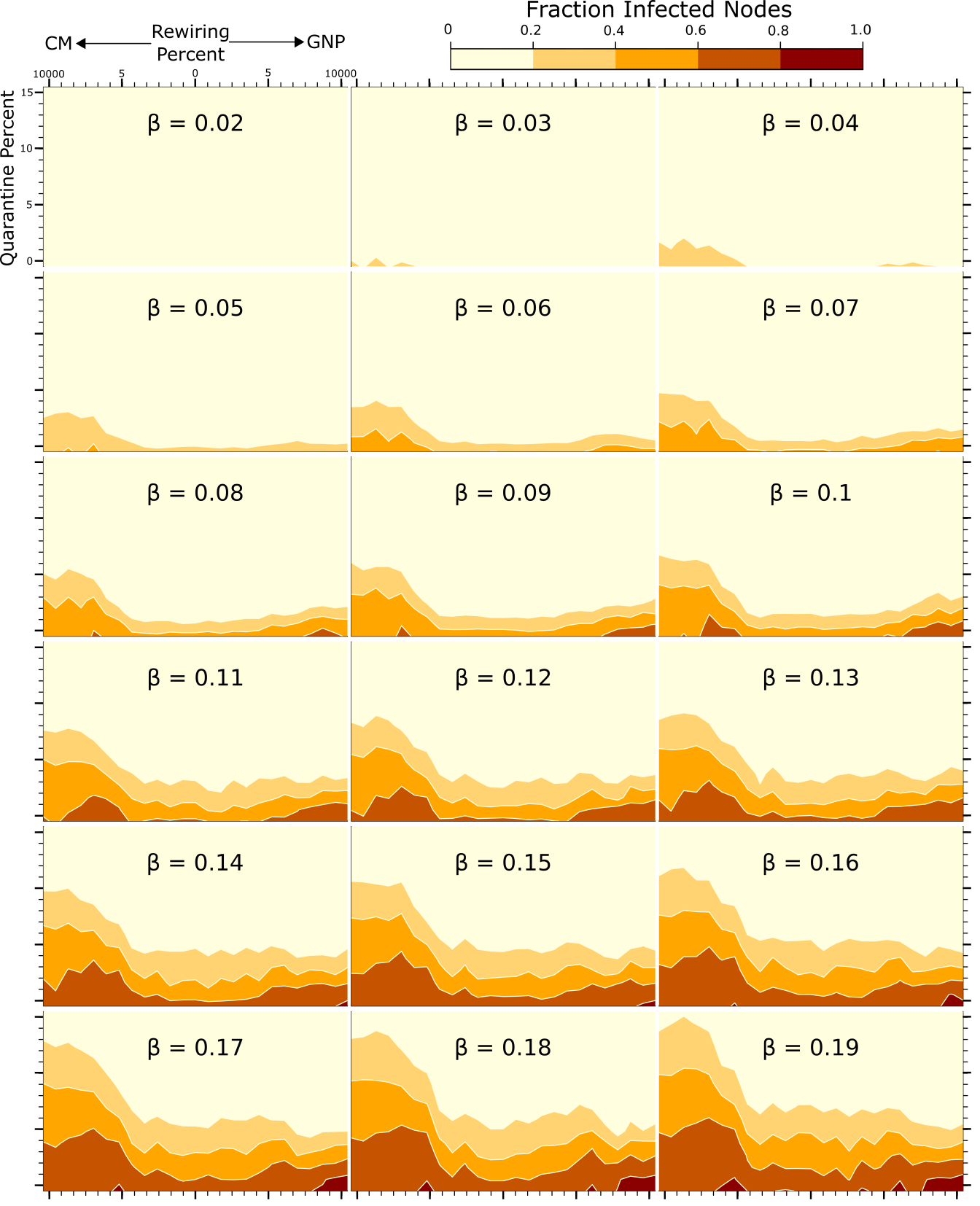}
        \caption{Sparsified College-Wisc diffusion data. Epidemic ``strength'' (original network with no quarantining) varies from $s\approx 9$ to $s\approx 87$.}
        \label{fig:uplot-params-cn-wisc}
    \end{figure}

\begin{figure}
        \centering
        \includegraphics[width=0.8\textwidth]{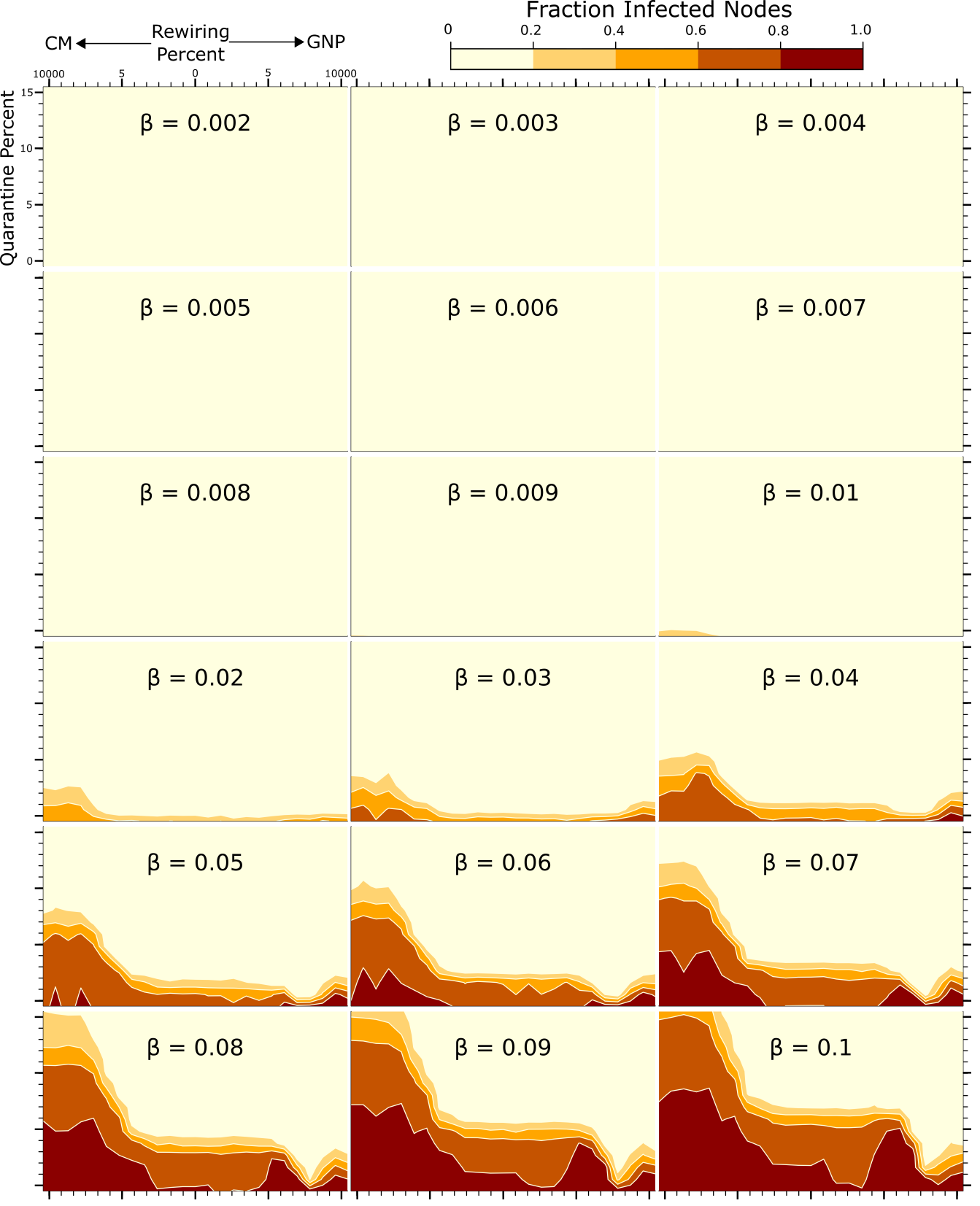}
        \caption{Collaboration diffusion data. Epidemic ``strength'' (original network with no quarantining) varies from $s\approx 3$ to $s\approx 150$.}
        \label{fig:uplot-params-dblp}
    \end{figure}

\begin{figure}
        \centering
        \includegraphics[width=0.8\textwidth]{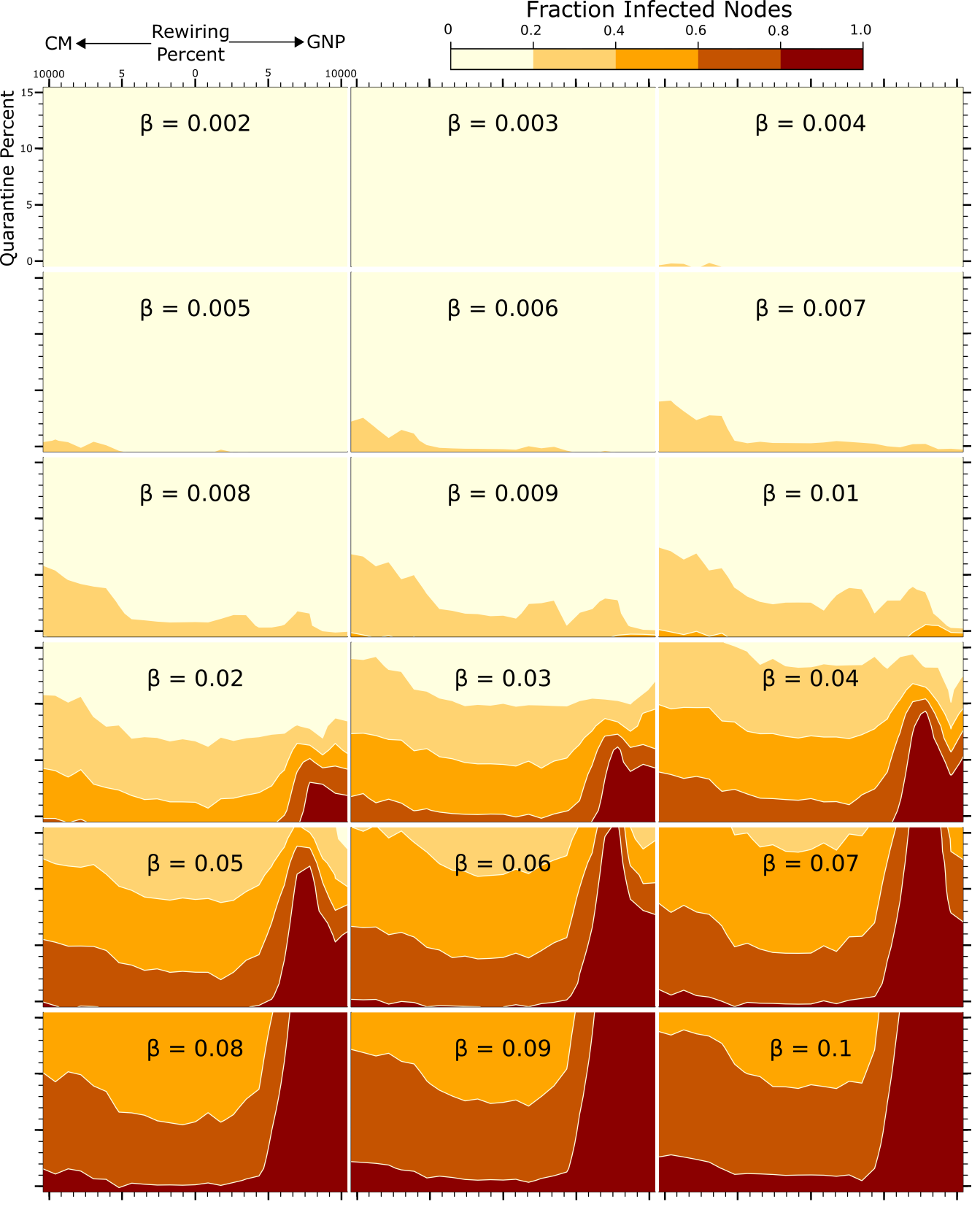}
        \caption{Email diffusion data. Epidemic ``strength'' (original network with no quarantining) varies from $s\approx 5$ to $s\approx 237$.}
        \label{fig:uplot-params-email}
    \end{figure}

\begin{figure}
        \centering
        \includegraphics[width=0.8\textwidth]{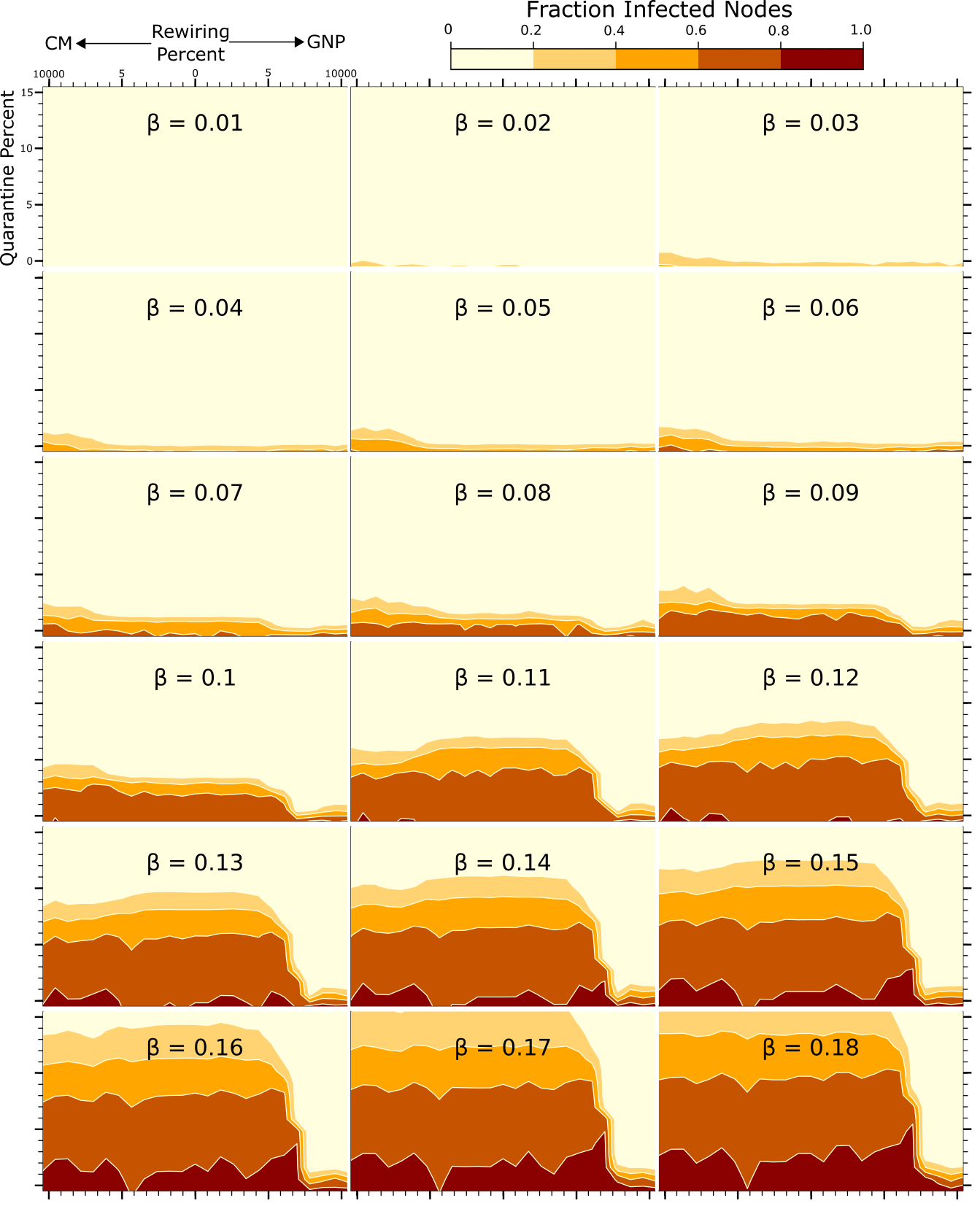}
        \caption{Facebook Interactions diffusion data. Epidemic ``strength'' (original network with no quarantining) varies from $s\approx 4$ to $s\approx 76$.}
        \label{fig:uplot-params-anon}
    \end{figure}

\begin{figure}
        \centering
        \includegraphics[width=0.8\textwidth]{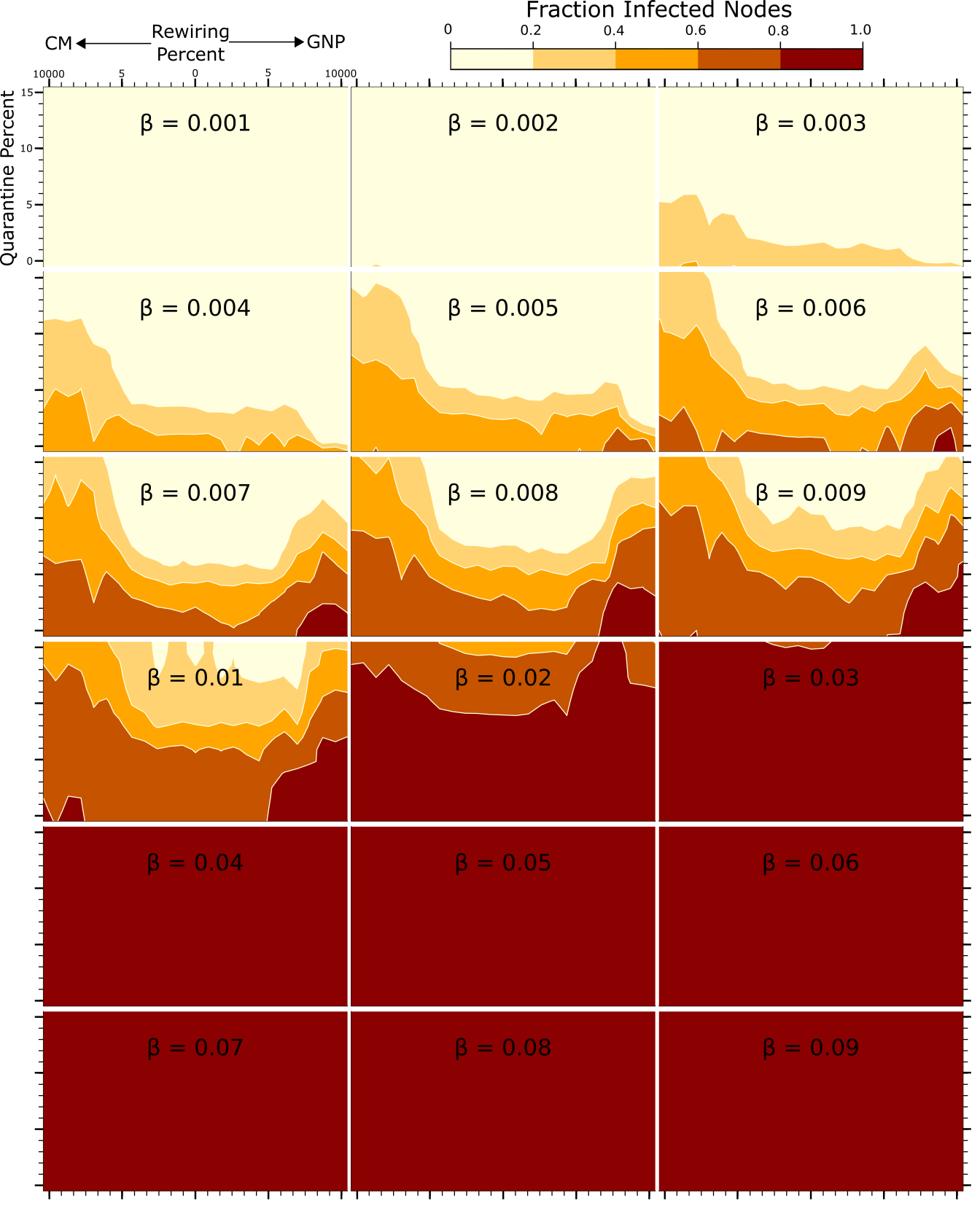}
        \caption{Citation diffusion data. Epidemic ``strength'' (original network with no quarantining) varies from $s\approx 2$ to $s\approx 138$.}
        \label{fig:uplot-params-citation}
    \end{figure}

\begin{figure}
        \centering
        \includegraphics[width=0.8\textwidth]{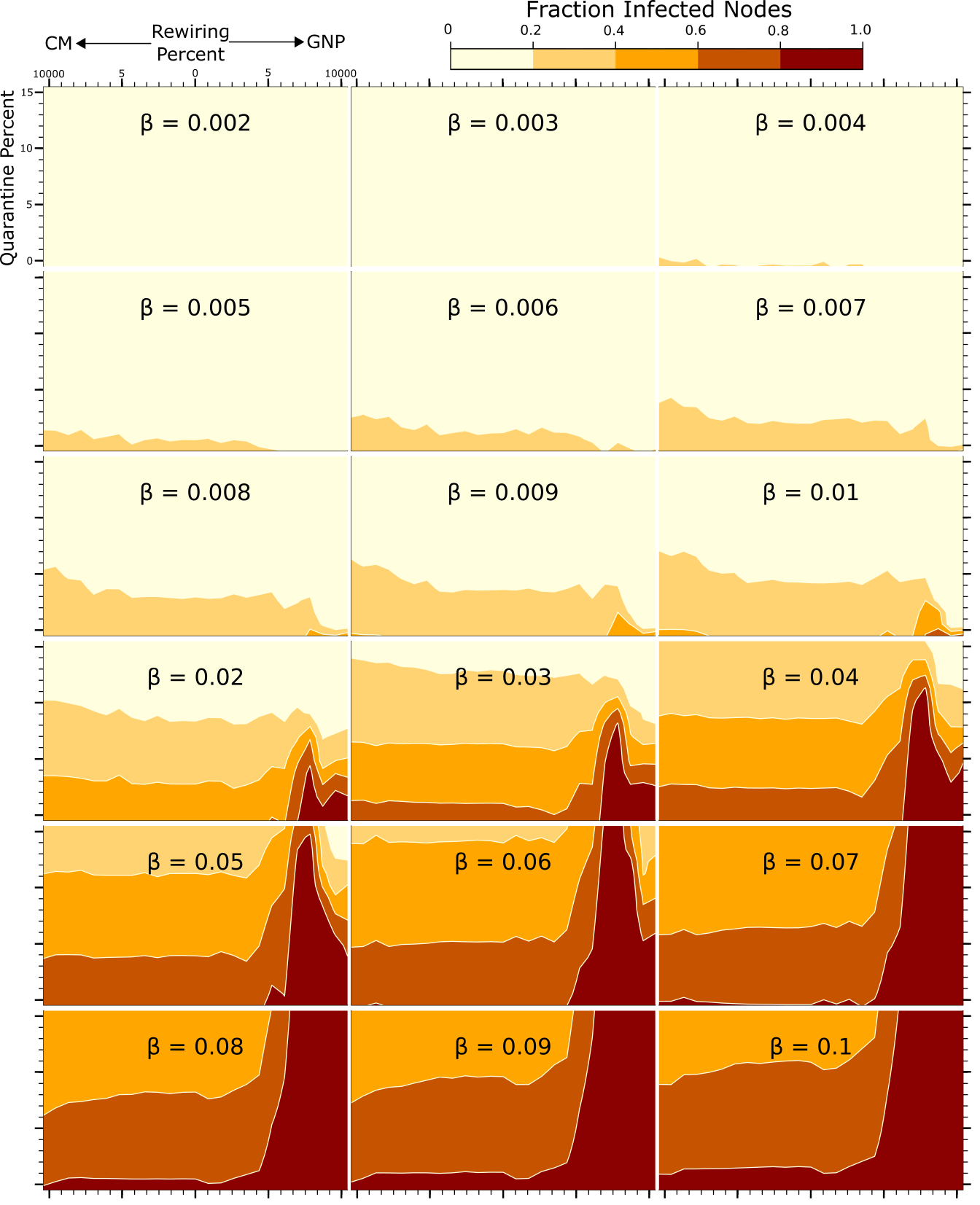}
        \caption{Slashdot diffusion data. Epidemic ``strength'' (original network with no quarantining) varies from $s\approx 5$ to $s\approx 263$.}
        \label{fig:uplot-params-slashdot}
    \end{figure}

\begin{figure}
        \centering
        \includegraphics[width=0.8\textwidth]{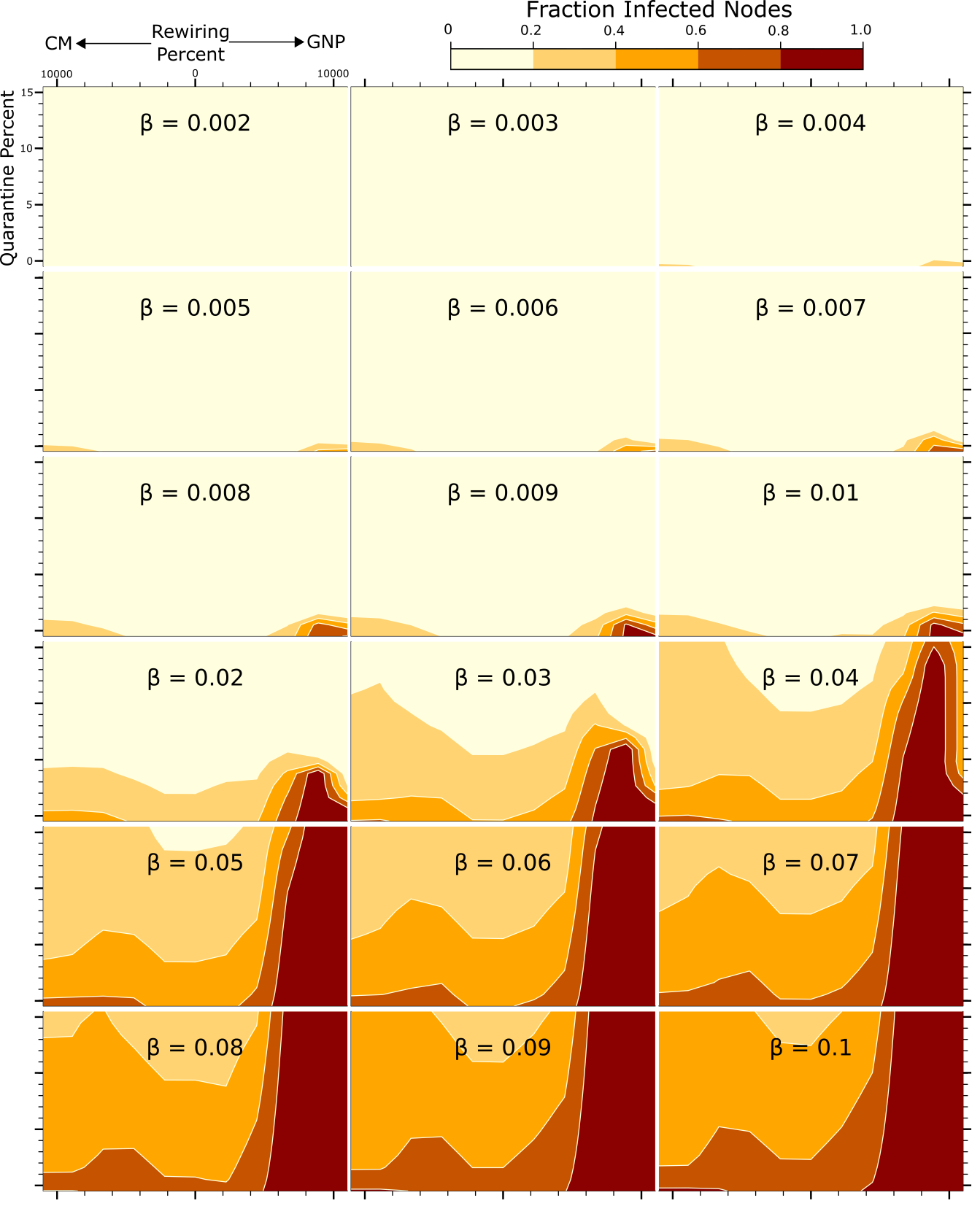}
        \caption{Fickr diffusion data. Epidemic ``strength'' (original network with no quarantining) varies from $s\approx 50$ to $s\approx 2481$.}
        \label{fig:uplot-params-flickr}
    \end{figure}

\begin{figure}
        \centering
        \includegraphics[width=0.8\textwidth]{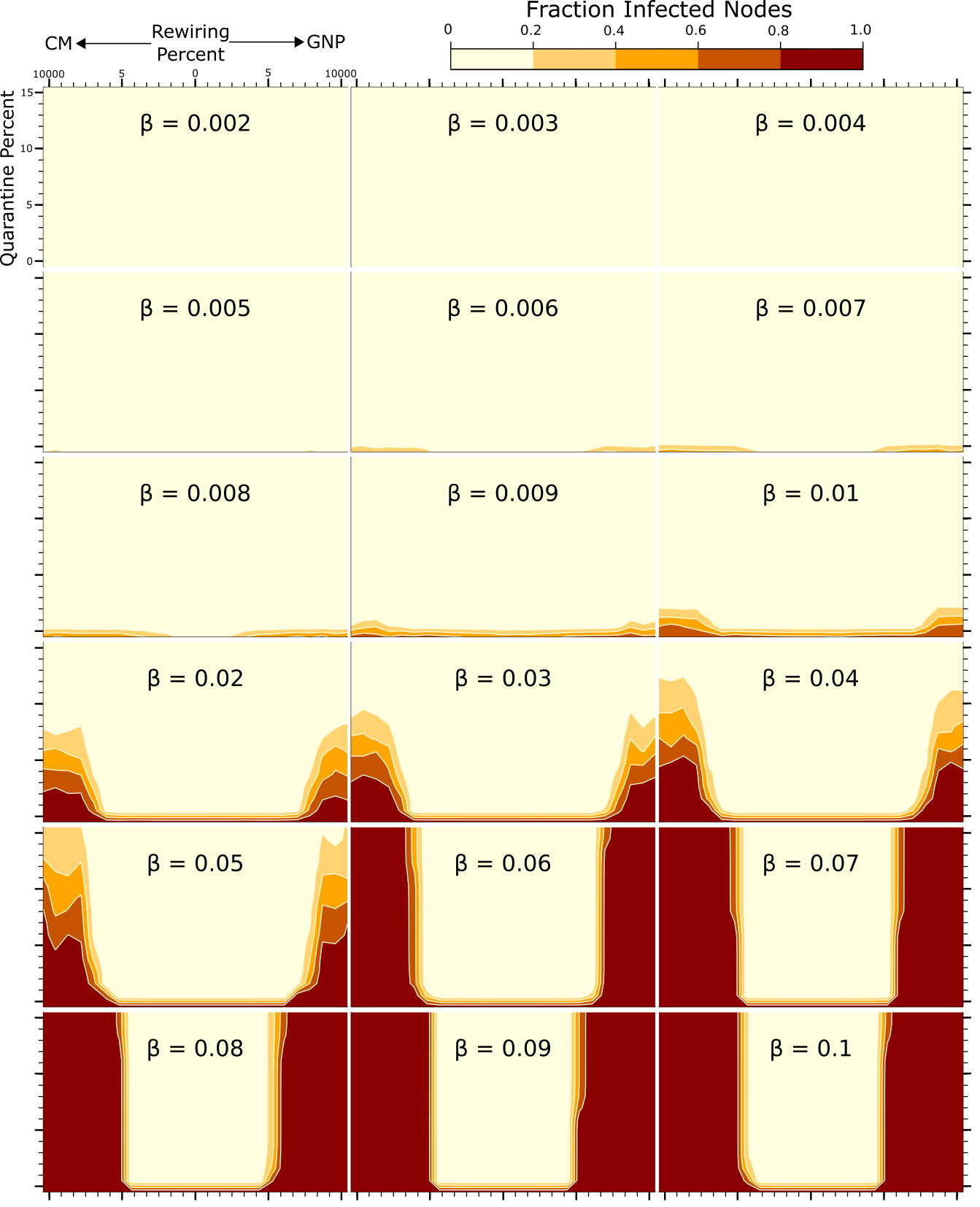}
        \caption{Local Geometric diffusion data. Epidemic ``strength'' (original network with no quarantining) varies from $s\approx 1$ to $s\approx 35$.}
        \label{fig:uplot-params-local-geometric}
    \end{figure}

\begin{figure}
        \centering
        \includegraphics[width=0.8\textwidth]{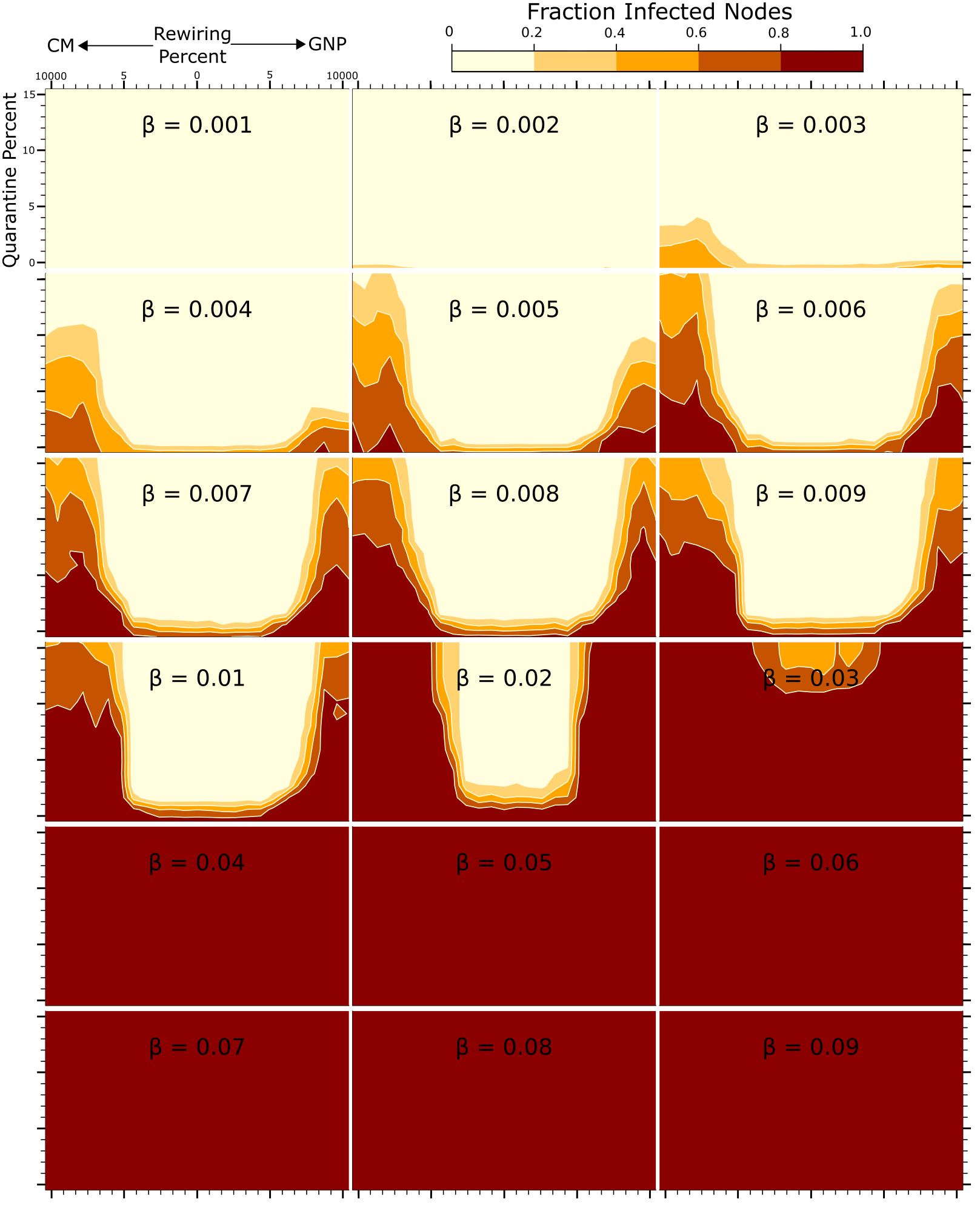}
        \caption{GeometricCommunities diffusion data. Epidemic ``strength'' (original network with no quarantining) varies from $s\approx 2$ to $s\approx 184$.}
        \label{fig:uplot-params-geometric-communities}
    \end{figure}

\begin{figure}
        \centering
        \includegraphics[width=0.8\textwidth]{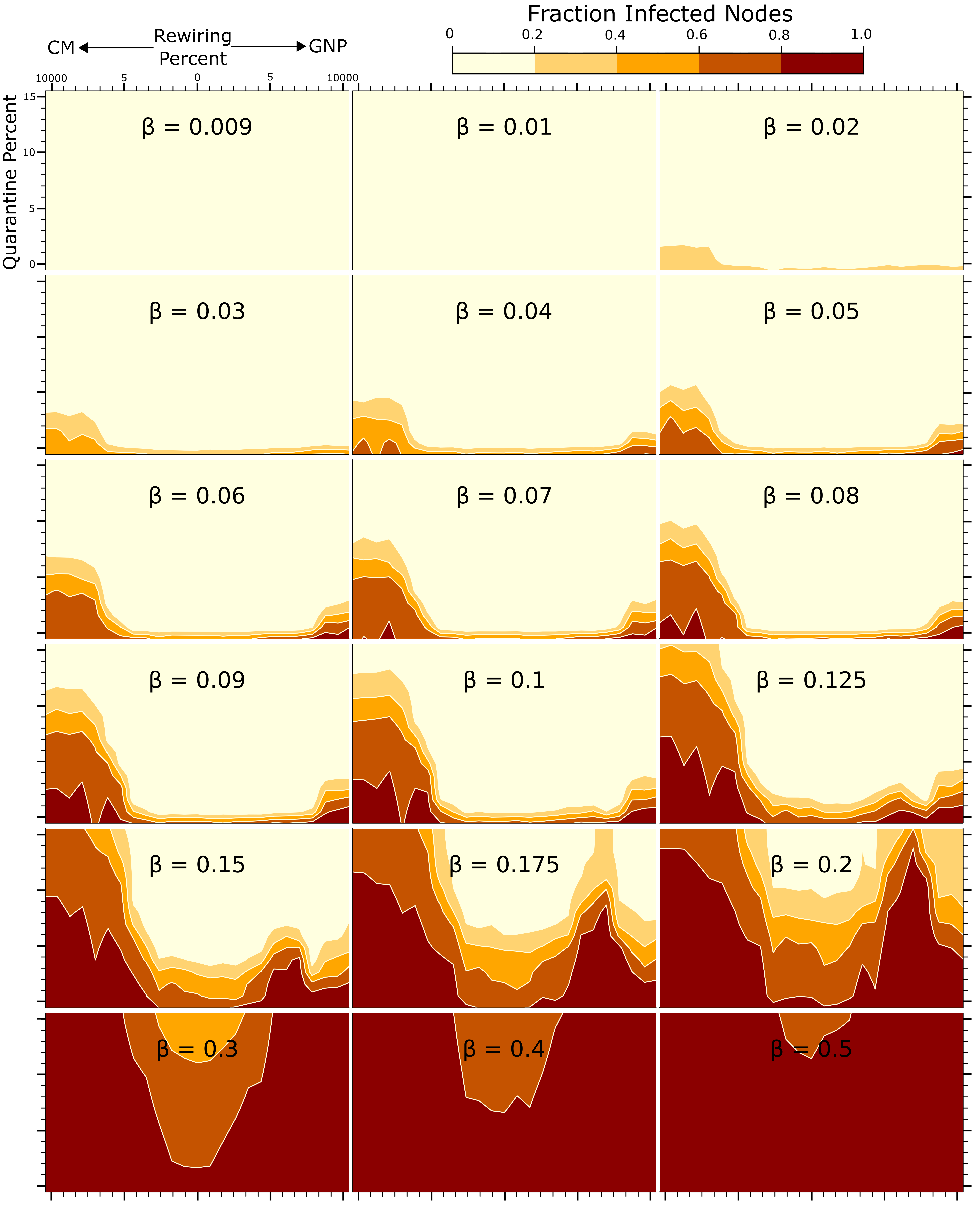}
        \caption{RandomWalkCommunities diffusion data. Epidemic ``strength'' (original network with no quarantining) varies from $s\approx 4$ to $s\approx 242$.}
        \label{fig:uplot-params-rand-walk-communities}
    \end{figure}

    Epidemic parameters are often chosen in relation to the value of the epidemic threshold, $s = \frac{\beta}{\gamma}\lambda_1(\mA)$~\cite{chakrabarti2008epidemic,prakash2012threshold}, where one compares experiments across networks for the same values of $s$. However, our experimental setup is to simulate the spread of the same pathogen across different realizations of the same base network to highlight the impact of network structure. Thus, we fixed the recovery probability $(\gamma=0.05)$ in order to fix the timescale of the epidemic. It was more difficult to choose the infection probability $\beta$.  We did not find the epidemic ``strength'' parameter helpful in regards to choosing this parameter and balancing results across networks.  Consequently, we evaluated results when selecting $\beta$ from among the set $[10^{-3},2\times10^{-3},\dots, 9\times 10^{-3}, 10^{-2},2\times 10^{-2},\dots,9\times 10^{-2},10^{-1},0.1, 0.12, 0.13, 0.14, 0.15, 0.16, 0.17, 0.18, 0.19$ for all graphs. If we already observed that most of the graph is infected by the time $\beta = 0.1$, then we discontinued further evaluations. Then we selected a value of  $\beta$ that was as large as possible such that the infection did not deterministically infect most of the network. We present all of the data used in making this selection in Figures~\ref{fig:uplot-params-commutes}--\ref{fig:uplot-params-rand-walk-communities}. We note that the plots are all qualitatively similar and ``smooth'' in this choice of this parameter, so we simply selected a single representative value to study in the main results. 
      
\subsection{Quarantining}
    We make several choices and assumptions in implementing quarantining.
    The underlying rationale for these choices is to prevent trivialities in quarantining such as immediately halting the epidemic or implementing quarantining too late to notice a meaningful impact. 
    We use a fixed quarantine period of 20 time steps $(1/\gamma)$ to coincide with the average length of the infectious period. 
    Each node that is quarantined must remain quarantined for exactly 20 time steps. There is no partial or early exit from Q.

    To avoid a trivial epidemic, we impose a detectability threshold in the population. This threshold is set to 100 nodes. We count nodes that have passed into the exposed or infected compartment when determining detectability. Note that this implicitly assumes there is a method available for testing for disease exposure. We further impose a time delay from exposure to detectability for each node. This is fixed for all nodes at a single time step. Since the quarantine budgets we often use are much larger than 100 nodes, this local time delay ensures we don't immediately quarantine all exposed and infectious nodes and stop the epidemic.

    There are two ways a node can pass to the Q compartment. The first is by being either exposed or infected and being randomly selected. The second is by being adjacent to a quarantined node that was randomly selected. That is, we quarantine randomly selected nodes and all of their one hop neighbors. We do not recurse. Random testing occurs after we've reached the disease detectability threshold. For random testing, while there is space to quarantine, we iterate through the population and for each node that is currently in the exposed or infected compartment we quarantine them with probability $1/(\text{remaining quarantine capacity}+1)$. Nodes adjacent to these nodes are quarantined during the next time step. That is, there is a time delay of a single time step when quarantining nodes adjacent to a randomly selected node.

    A rolling quarantine capacity is maintained and at each time step, no more than that capacity can be quarantined at any given time. Once that capacity is reached, new nodes are not quarantined until space becomes available. In our experiments, we vary this quantity as a percentage of the number of nodes in the graph. So, this varies from $0\%$ (no quarantining) to $15\%$ of the nodes in the~graph.

\subsection{Higher-Order Network Epidemics}
    \label{sec:hypergraph-epidemic-alldetails}
    
    In standard models of epidemic spread, transmission pathways are constrained to strictly pairwise interactions. A node $v$ becomes infected at each time step, $t$, with probability 
    \[1-\prod_{u\in N(v)} (1-\beta\itr{I}{t}[u]),\]
    where $N(v)$ denotes the neighbors of $v$ and $\itr{I}{t}[u]$ indicates whether the node $u$ is infectious at time $t$. This can be generalized by allowing spreading to occur along hyperedges. One continuous time formulation of this~\cite{higham2021epidemics} uses
    \[
    \beta\sum_{h\in H(v)} \chi(u\in h) f\left(\sum_{w\in h} I^{(t)}[w]\right) 
    \]
    as the rate of infection, where $H(v)$ is the set of hyperedges involving $v$, $\chi$ an indicator function, and $f$ a function defining the contribution of $h$ to the rate of infection. The function $f$ can depend on particular nodes, group information (e.g., proportion of infectious node), or other relevant information~\cite{bodo2016sis,higham2021epidemics,iacopini2019simplicial}.

    We use a discrete time analogue of the above. At each time step $t$, the probability that $j$ infects $i$ (assuming $j$ infectious) is given by 
    \[1-(1-\beta)^{w_{j,i}}\approx \beta w_{j,i},\]
    where $w_{j,i} = \max(T_{j,i},a_{j,i})$ and $T_{j,i}$ is the number of triangles that nodes $j$ and $i$ jointly participate in. 
    This choice for $w_{j,i}$ allows for triangle-free edges to contribute to infections. The final value of $\beta w_{j,i}$ is rounded to lie in the range $[10^{-6},1-10^{-6}]$ for computational convenience in simulations.

\subsection{Implementation Details}
    The epidemic diffusions are implemented in Julia using a calendar queue. The main idea is that we schedule changes in node states and use the current state of a node to determine what state change needs to happen. To facilitate this, we use a data structure that stores all the relevant information needed. See the table below for what was stored and the rationale for storing it. Highlighted design choices include: adjacency network representation for faster traversal and look ups, vectors of scheduled state changes (eg, $S\rightarrow E$ time), quarantine information (current quarantine state for nodes, current capacity, etc), and last recorded infection time for nodes for recomputing times if a node is quarantined.

\begin{table}[h]
    \centering
    \footnotesize 
    \begin{tabularx}{0.8\linewidth}{rX}
        \toprule 
        name & description \\
        \midrule 
        $\beta$ & Infection Probability \\
        $\gamma=0.05$ & Recovery Probability \\
        $a=5$ & Parameter for Exponential Distribution for time from Exposed to Infectious \\ 
        log1beta & cached value of $\log{(1-\beta)}$ for geometric sampling \\ 
        log1gamma & cached value of $\log{(1-\gamma)}$ for geometric sampling\\
        \midrule
        inNeighborsA[v] & in-neighbors of node $v$  \\
        outNeighborsA[v] & out-neighbors of node $v$ \\
        \midrule
        itime[v] & time when node $v$ is infected (exposure time for SEIR model)\\
        ctime[v] & time when node $v$ becomes contagious \\
        rtime[v] & recovery time for node $v$\\
        snodes[v] & indicates whether node $v$ is currently susceptible \\
        etimes[t] & nodes with event time $t$\\
        qcap & rolling quarantine capacity \\
        qcurr[t] & number of nodes currently quarantined at time $t$ \\
        qnodes[v] & indicates whether node $v$ is currently quarantined \\
        qtimes[v] & last recorded quarantine time for node $v$\\
        ihist[v] & last recorded incoming infection times for node $v$ (infection time by in-neighbor)\\
        \bottomrule
    \end{tabularx}
    \caption{Table depicting data structure used to implement epidemic simulation}
    \label{tab:epidemic-data-structure-table}
\end{table}

Due to the choice of discrete time epidemics, ties may arise in scheduling that need to be resolved. There are two potential scenarios where ties may arise. In the first case, a susceptible node is scheduled to become infected and quarantined at the same time. In this case, we attempt quarantining first but still allow the node to become infected. If quarantine is successful, the node is not allowed to pass on the infection. 
In the second potential scheduling tie, a node is scheduled to be quarantined and to recover at the same time. In this instance, we still quarantine the node and update the new recovery time to be the end of the quarantine period.

\begin{figure}
    \centering
    \includegraphics[width=0.5\textwidth]{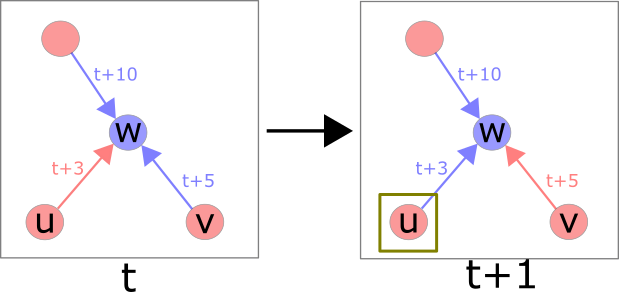}
    \caption{(Best viewed in color.) Example of rescheduling disease transmission time after quarantining an infected node. At time $t$, $u$ is scheduled to infect $w$ at time $t+3$. This is the minimum infection time for $w$. However, when $u$ is quarantined at time $t+1$, we must figure out if the infection time for node $w$ changes. In this example, it does. Node $v$ infects $w$ at $t+5$, and this becomes the new infection time for $w$. }
    \label{fig:adjust-itimes}
\end{figure}

\begin{figure}
    \centering
    \includegraphics[width=0.5\textwidth]{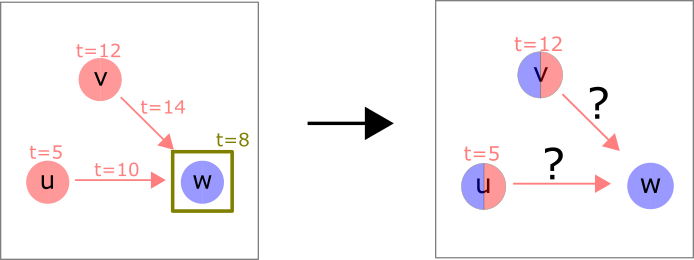}
    \caption{(Best viewed in color.) An example illustrating a rescheduling of infection times due to quarantining.
    Node $w$ is quarantined at time $t=8$ but is susceptible after exiting quarantining.
    If either of $u$ or $v$ are infectious after $w$ leaves quarantine, they may potentially infect $w$.
    }
    \label{fig:post-quarantine-infection}
\end{figure}

Quarantining makes the implementation more difficult, as care must be taken to handle all of the edge cases. In particular, moving nodes from $Q$ back to $S$ if they haven't been exposed or infected requires extra care. There are two main cases that need to be handled with care. The first is that when we quarantine an infected node, $u$, we must potentially recompute exposure times for all $v\in N(u)$. This is to account for cases where $u$ was scheduled to infect a node $v\in N(u)$. However, due to the quarantining, we must recompute exposure times. The other case involves when a node, $w$, moves from $Q$ to $S$. In this instance, we need to handle the case where $w$ exits quarantine but has a neighbors that is infectious. These two cases are depicted in Figures~\ref{fig:adjust-itimes} and~\ref{fig:post-quarantine-infection}, respectively.

\subsection{Parallel Implementation Details}
    The above describes how to simulate a single epidemic on a network. However, for each graph we produce several graph perturbations (up to 35 variants). 
    For each perturbation we simulate a number of epidemic parameters. 
    We allow $\beta\in [0.001,0.002,\dots,0.009,0.01,0.02,\dots,0.1]$ while the rolling quarantine capacity (qcap in Table~\ref{tab:epidemic-data-structure-table}) ranges from $0$ to $15\%$ of nodes in the network. 
    For most graphs listed in Table~\ref{table:graph-stats}, we perform 
    
    \[\text{50 nodes} \times \text{25 graphs} \times \text{19 values of }\beta\times \text{16 quarantining levels}=380,000\text{ epidemic simulations.}\]

    An exception is made Flickr as this network has over 1 million nodes. For Flickr, we only perform 11 rewirings. Without quarantining, each epidemic simulation can be done in $O(BFS)$, making this an expensive computation. To speed up the computation, we make use of a simple parallel algorithm, and provide details below. 

    We partition the computation so that a graph is loaded in only once on all workers, while all relevant epidemic simulations are generated across parameters. After each worker loads in the graph, we form the EventEpidemic data structure (see Table~\ref{tab:epidemic-data-structure-table}) locally on each worker. Each worker then simulates the epidemic using the parameters given and returns epidemic information (new infections and net infections at each time step) to the parent. Additionally, each worker initializes a matrix of results (nnodes $\times$ 16) representing the total number of times a node was susceptible at the end of the simulations. This is stored for each value of qcap (0 to 15$\%$ of nnodes).
    This information is aggregated across workers and saved.

\section{Synthetic Network Constructions}

There are three synthetic network models we use in the experiments: 

\begin{itemize}
    \item a purely geometric network where edges follow a planar topology;
    \item a geometric community model that seeks to blend local structure, large scale randomness, and community structure, where each node has a coordinate in a 2d space; and
    \item a purely network based model that uses local random walks to introduce local structure on top of a very sparse LFR network~\cite{lancichinetti2008benchmark}, without any underlying geometry.
\end{itemize}

\subsection{Local Geometric}
\label{sec:local-geometric-alldetails}
We introduce one purely geometric model where each node is connected to nearest neighbors in a two-dimensional space. This model includes no long range edges. The nodes are sampled randomly from the unit 2d square. Each node picks a degree from the uniform distribution between 1 and 20 neighbors and connects to that number of nearest neighbors. These connections are then made into undirected edges. This model has local structure, but not the large scale randomness that exists in empirically measured networks. Consequently, it is a limit for what might be true of local networks. We sample one such network with 100000 nodes. This is labeled ``geometric''.

    \subsection{GeometricCommunities: Synthetic Networks with Emergent Local Structure}
    The goal of the GeometricCommunities model 
    is to create a scenario where local pockets of community structure emerge but the network has large scale randomness. To do this, we simulate an evolving process for a number of steps. The network is initialized where every node is positioned in a unit 2d square, uniformly at random (we start with $50,000$ nodes). Each node is assigned a target degree from a log-normal distribution (the parameters can vary, the normal distribution in the log-normal had a mean of $\log(4)$ and variance of $1$). Then each node adds edges to neighbors based on their degrees. We also allocate $0.25\sqrt{d[v]}$ random connections from each node to seed global randomness. These random connections are added via preferential attachment. 

    The evolution process models a number of behaviors: the strength of weak / long ties~\cite{onnela2007structure,eckles2018long,mercier2022effective} along with social influence~\cite{Degroot1974} and preferential attachments~\cite{Price-1976-preferential-attachment,barabasi1999-scaling}. At each time step: nodes keep edges to any neighbor if the toroidal distance to that neighbor is at least the average distance among all nodes in the network. Then we move nodes around in space. They move towards their highest degree neighbor in toroidal space to model a combination of preferential attachment and social influence. If the node position is $p$ and its highest degree neighbor position is $q$, then we move node to $\gamma p + (1-\gamma) q + 0.001\cdot N(0,1)$, where the random variable is sampled independently in each coordinate. The value of $\gamma$ used is $0.95$. After the position update, nodes connect their residual degrees (those not used to keep long-distance connections) based on nearest neighbors in toroidal space. In the event that a node has lost connections over time, we also allow a correction back to the target degree. (The formula is $0.85$ times the residual degree ${}+\max(0, 0.25 \cdot (\text{target degree} - \text{residual degree}))$.)

    The idea here is that high degree vertices sampled from the log normal distribution form initial community centers that attract nodes around them. Yet nodes keep long distance connections, preserving global randomness. If the graph becomes disconnected, then we add about two edges to any connected component that isn't the largest connected component. Also, if there are any nodes in the largest connected component that are missing edges, we also try to add edges into the same component if a random single random draw works. We store the networks after every 10 steps. These networks are called ``study-25'' in the data tables. There is also a smaller version of the same network used for visualization called ``study-20-draft-150'' that was smaller (500 nodes) with a different log-normal degree distribution, but this network was only used for illustrations instead of analysis.

    \subsection{RandomWalkCommunities: Synthetic Networks with Planted Communities}
    The LFR parameterized network distribution is one of the better understood graph models \cite{lancichinetti2008benchmark} with planted community structure. It also features heterogeneity in degrees and community size. However, LFR lacks realistic local structure due to the fashion in which inter-community and intra-community edges are added. We attempt to add in local structure using sparse LFR graphs by augmenting these graphs via local random walks. The random walks should \emph{locally} explore a region to emulate the same type of structure. The base LFR graph we use is generated on 100,000 nodes with the following parameters: degree exponent - 3, community size distribution exponent - 2, minimum degree - 1, maximum degree - 2000, minimum community size - 5, maximum community size - 500. The python implementation via networkx~\cite{hagberg2008exploring} was used. Since this LFR graph was disconnected, the first step of our procedure is to ensure it is connected. We use a Chung-Lu-like process on that graph that adds in the edge if it decreases the number of connected components. The result after the Chung-Lu process is the graph we label as ``lfr-rand-walk-0'' in Table \ref{table:graph-stats}. For each random walk trial, we sample a seed node, $v$, and we start a random walk from that seed node. At each time step, we independently mark the current node $u$ with probability $0.5$ and attempt to stop the random walk. After the random walk has finished, edges are added between the seed $v$ and all marked nodes. This process of sampling a seed node and starting a random walk is performed up to 8000 times and we store the resulting graph every $1000$ random walks. The final graph with 8000 steps is denoted as ``lfr-rand-walk-8000'' in table~\ref{table:graph-stats}. The stopping probability was set to be 
    $$\dfrac{1}{\max(\text{maxdepth}-d[v]+1,3)} , $$ 
    so that low degree nodes take longer random walks while large degree nodes take shorter walks. We also choose $\text{maxdepth}=100$.

\section{Detailed Data Description}
    \label{sec:data-alldetails}
    
    The data in the main text uses semantic names for each network rather than the data source. The relationship between these two are given in Table~\ref{tab:semantic-names}, and the statistics underlying each network are given in Table~\ref{tab:data}. 
    \begin{table}[tp]
        \caption{The semantic names used in the main text correspond to the following data files and data sources. In the appendix, we tend to use the data names instead of the semantic names.
        } 
        \label{tab:semantic-names}
        \centering
        \begin{tabularx}{0.6\linewidth}{lX}
            \toprule
                Semantic Name & Data Name \\ 
            \midrule 
                US Commutes & commutes \\ 
                Mexico City Trace & modmexico-city \\ 
                Filtered US Flows & covidflows-filtered-20 \\ 
            \midrule 
                Sparsified College--Illinois & cn-modUIllinois20 \\ 
                Sparsified College--Penn & cn-Penn94 \\ 
                Sparsified College--Wisc & cn-modWisconsin87 \\ 
            \midrule 
                Collaboration & dblp-cc \\ 
                Email & email-Enron \\
                Facebook Interactions & anony-interactions \\
            \midrule 
                Citation & cit-HepPh\\
                Slashdot & Slashdot0811\\
                Flickr & flickr-links-sym \\ 
            \midrule 
                Local Geometric & geometric \\
                GeometricCommunities & study-25-150 \\ 
                RandomWalkCommunities & lfr-rand-walk-8000 \\
            \bottomrule 
        \end{tabularx} 
    \end{table}

    \begin{longtable}{lllllll}
        \caption{Key properties of the data used in the main text.}\label{tab:data}\label{table:graph-stats} \\ 
        \toprule 
        Graph & Nodes & Edges & Avg.~Deg. & max kcore & 2-core fraction & $\lambda_1(A)$ \\ 
        \midrule 
    \endfirsthead 
        \toprule 
        Graph & Nodes & Edges & Avg.~Deg. & max kcore & 2-core fraction & $\lambda_1(A)$ \\ 
        \midrule 
    \endhead 
        \bottomrule
        \multicolumn{7}{r}{{Continued on next page}} \\ 
    \endfoot
    \bottomrule 
    \endlastfoot
        commutes & 72580 & 7284448 & 100.4 & 97 & 0.0 & 261.73 \\
        covidflows-filtered-20 & 70930 & 1386068 & 19.5 & 29 & 0.06 & 59.9 \\ 
        modmexico-city & 129598 & 2747110 & 21.2 & 187 & 0.44 & 459.46 \\ 
        \midrule 
        cn-modUIllinois20 & 23165 & 69374 & 3.0 & 6 & 0.87 & 28.51 \\ 
        cn-Penn94 & 37794 & 108532 & 2.9 & 13 & 0.89 & 34.42 \\ 
        cn-modWisconsin87 & 18377 & 49368 & 2.7 & 8 & 0.93 & 22.81 \\ 
        \midrule 
        dblp-cc & 226413 & 1432920 & 6.3 & 74 & 0.37 & 75.13 \\ 
        email-Enron & 33696 & 361622 & 10.7 & 43 & 0.39 & 118.42 \\ 
        anony-interactions & 379500 & 1338248 & 3.5 & 9 & 0.76 & 21.03 \\ 
        \midrule 
        cit-HepPh & 34401 & 841568 & 24.5 & 30 & 0.08 & 76.58 \\ 
        Slashdot0811 & 77360 & 938360 & 12.1 & 54 & 0.54 & 131.34 \\ 
        flickr-links-sym & 1624992 & 30953670 & 19.0 & 568 & 0.68 & 1240.3 \\
        \midrule 
    
        geometric & 100000 & 1410362 & 14.1 & 12 & 0.0 & 17.66 \\ 
        study-25-1 & 50000 & 647044 & 12.9 & 12 & 0.0 & 37.36 \\
        study-25-2 & 50000 & 767730 & 15.4 & 12 & 0.0 & 38.38 \\
        study-25-150 & 50000 & 1595832 & 31.9 & 50 & 0.0 & 92.06 \\
        lfr-rand-walk-0 & 97714 & 207972 & 2.1 & 3 & 1.0 & 16.68 \\
        lfr-rand-walk-1000 & 97714 & 241718 & 2.5 & 6 & 0.95 & 17.89 \\ 
        lfr-rand-walk-2000 & 97714 & 272436 & 2.8 & 5 & 0.9 & 19.29 \\ 
        lfr-rand-walk-4000 & 97714 & 342412 & 3.5 & 6 & 0.76 & 20.44 \\ 
        lfr-rand-walk-6000 & 97714 & 407850 & 4.2 & 7 & 0.65 & 23.84 \\  
        lfr-rand-walk-8000 & 97714 & 468752 & 4.8 & 8 & 0.55 & 24.16 \\  
        \midrule 
        study-20-draft-150 & 500 & 5258 & 10.5 & 12 & 0.0 & 19.42 
    \end{longtable}  
    
    \subsection{Removing Large-Sized Conductance Bottlenecks}
    \label{sec:large-set-removal}
\begin{figure}
        \centering
        \includegraphics[width=\textwidth]{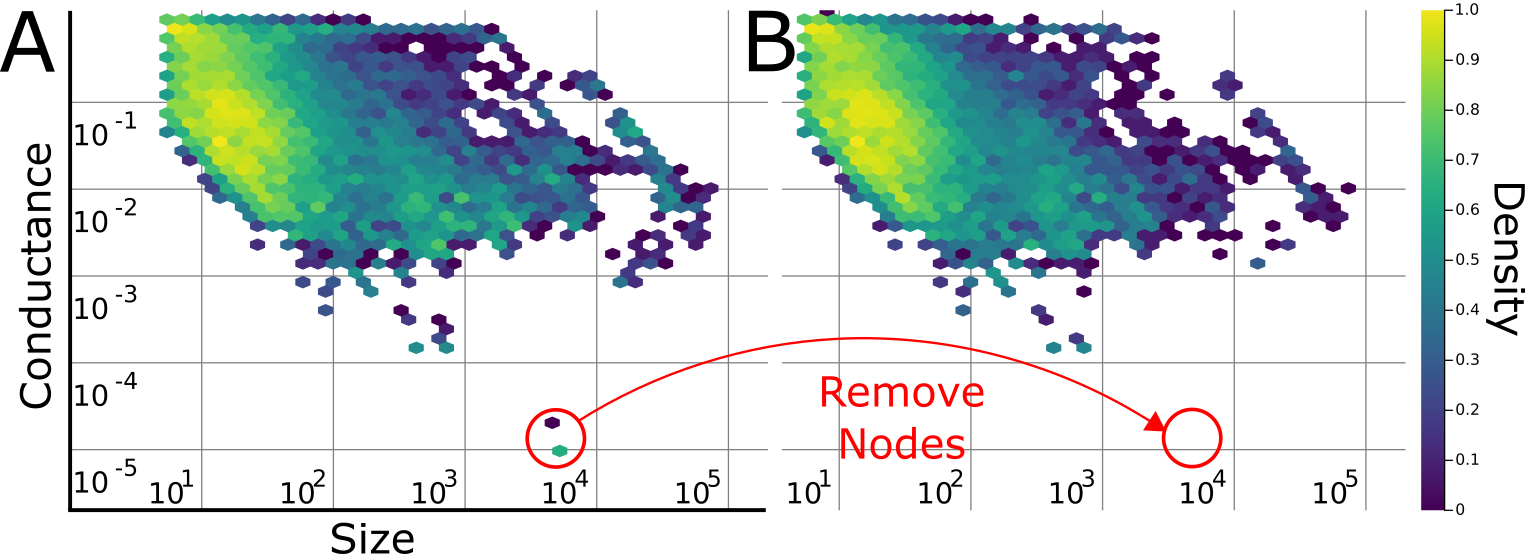}
        \caption{Personalized PageRank NCP for Mexico City Trace before and after removing a large-sized set of small conductance. The nodes in these deep conductance sets are well separated from the rest of the network.}
        \label{fig:large-set-removal-mexico-city}
    \end{figure}

    Many of the Facebook100 networks that we use as well as Mexico-City Trace were manually modified to remove artifacts in the seeded PageRank based NCP. In particular, small conductance bottlenecks at large size scales were removed. For Facebook networks, this corresponded to removing the incoming students for the year 2009 as those students tended to form connections primarily amongst themselves that produced a dip in the NCP at larger size scales. This is known to be an artifact of how the networks reflected social structure students required a college email to join Facebook~\cite{Veldt-2019-resolution}. \\
    For Mexico City Trace, there was a large set ($\approx$5000 nodes) with very small conductance $(\approx 10^{-5})$ that was removed. A very deep conductance set like this was an outlier that may have made results for Mexico City look more favorable. We opted to remove this set. The PPR NCP for Mexico City before and after the removal of these nodes in deep conductance sets is shown in Figure~\ref{fig:large-set-removal-mexico-city}. Due to the anonymized nature of the data, we cannot interpret what the removed nodes corresponded to. In Table\ref{table:graph-stats}, graphs that have been modified by removing nodes are labeled with the prefix ``mod''. For example, ``modmexico-city'' denotes the Mexico-City Trace graph after removing a large set with a deep conductance bottleneck.
    \subsection{Sparsified College Network Data}
    \label{sec:sparsified-data}
\begin{figure}
        \centering
        \includegraphics[width=\textwidth]{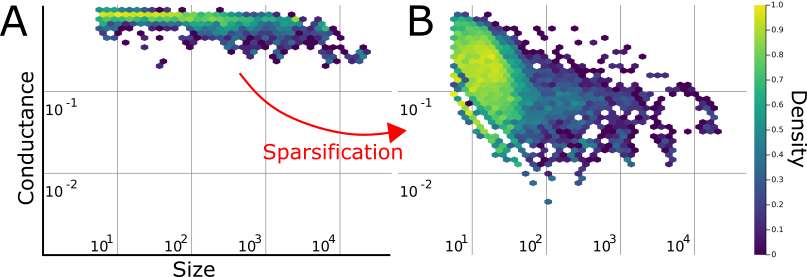}
        \caption{The PPR NCP looks characteristically different after sparsifying a network. Panel (A) shows the PPR NCP for Penn94 while (B) shows the PPR NCP after sparsifying. The average degree before and after sparsification is $\approx 65.6$ (A) and $\approx 2.9$ (B).}
        \label{fig:app-sparsification-penn}
    \end{figure}

    From the local structure perspective, when graphs are too dense or the epidemic diffusion is too strong, the impact of local structure is negligible in the observed diffusion. In order to study the impact of local network structure on epidemic diffusions, we need to make sure we are in a parameter regime where these effects can be observed. For this reason, we sparsify dense graphs to means highlight local structure. With this in mind, we sparsify all of the college Facebook networks that we used. The rationale is that their NCP's are very flat (no local heterogeneity) and don't have any good conductance sets~\cite{jeub2015think}. See Figure~\ref{fig:app-sparsification-penn} for example.
     
    We opted for an intuitive way to sparsify networks to focus on those edges most likely to transmit an infection. Essentially, each node ranks it's neighbors and elects to keep some fraction of it's neighbors based on that ranking. The metric that we used to rank neighbors is normalized common neighbors. If nodes $u,v$ are adjacent, then $u$ would score $v$ as $\frac{|N(u)\bigcap N(v)|}{d_u}$. Similarly, $v$ would score $u$ as $\frac{|N(u)\bigcap N(v)|}{d_v}$. This corresponds to choosing to interact with friends whom you have many joint friends with. The degree of sparsification used for each graph was decided based on the change in the NCP, the reduction in the average degree, and the size of the giant component in the graph (typically, this was at least 90\% of the original graph). All graphs and code is available in the accompanying code repository~\url{https://github.com/oeldaghar/network-epi}. 

\begin{figure}
        \centering
        \includegraphics[width=0.8\textwidth]{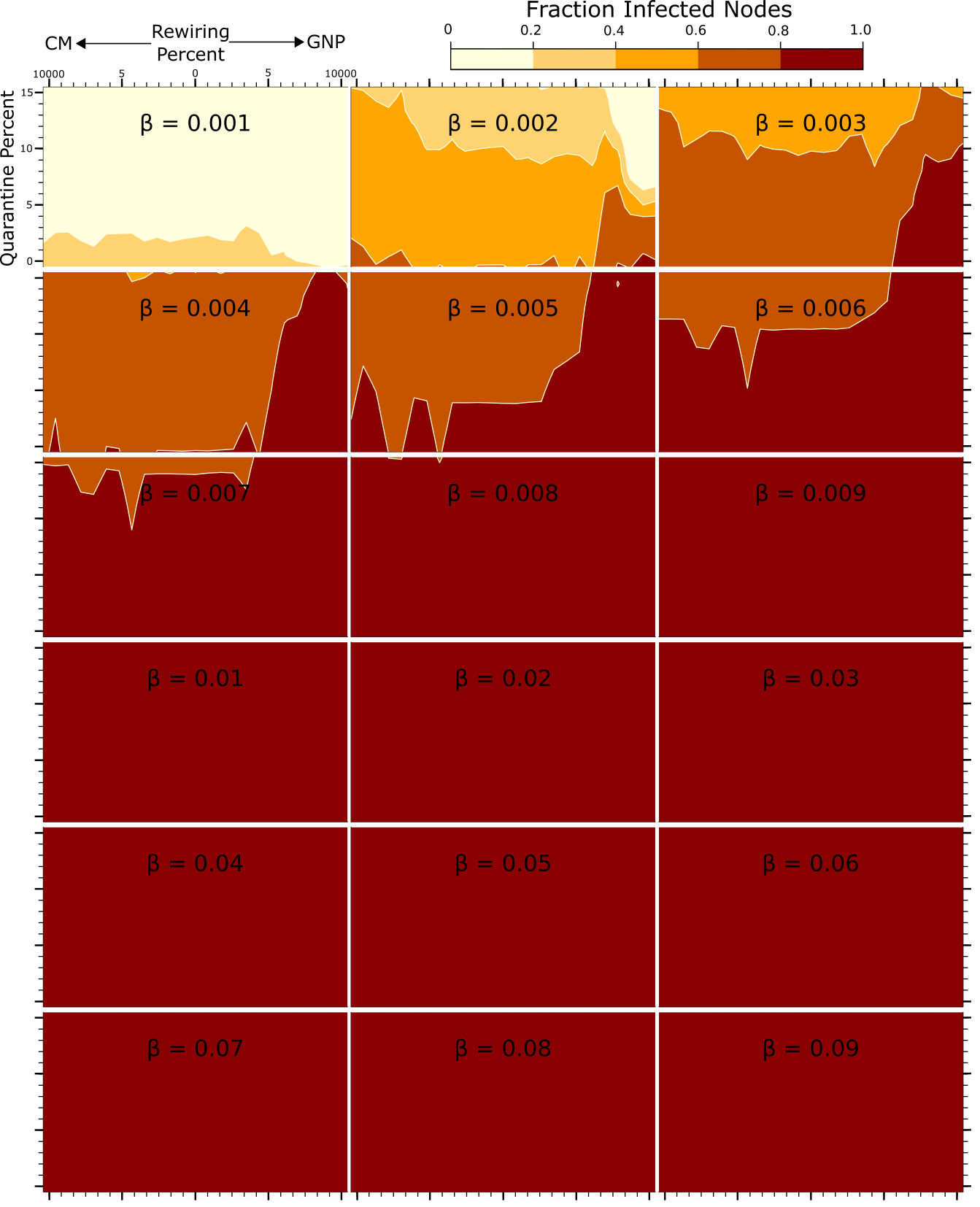}
        \caption{College-Penn diffusion data prior to performing sparsification. Epidemic ``strength'' (original network with no quarantining) varies from $s\approx 3.6$ to $s\approx 324.8$.}
        \label{fig:uplot-params-penn}
    \end{figure}

\begin{figure}
        \centering
        \includegraphics[width=0.8\textwidth]{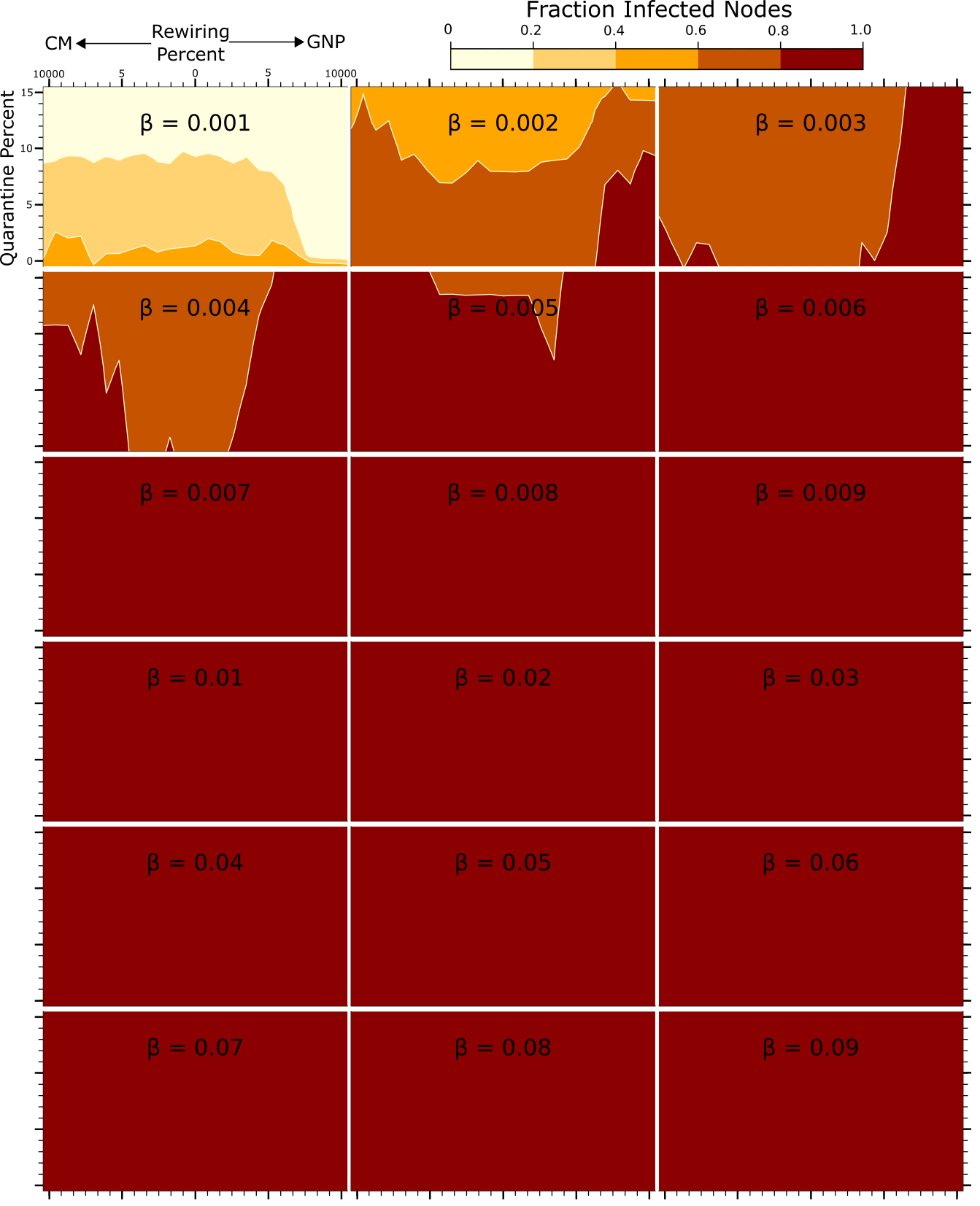}
        \caption{College-Illinois diffusion data prior to performing sparsification. Epidemic ``strength'' (original network with no quarantining) varies from $s\approx 3.8$ to $s\approx 340$.}
        \label{fig:uplot-params-uill}
    \end{figure}

\begin{figure}
        \centering
        \includegraphics[width=0.8\textwidth]{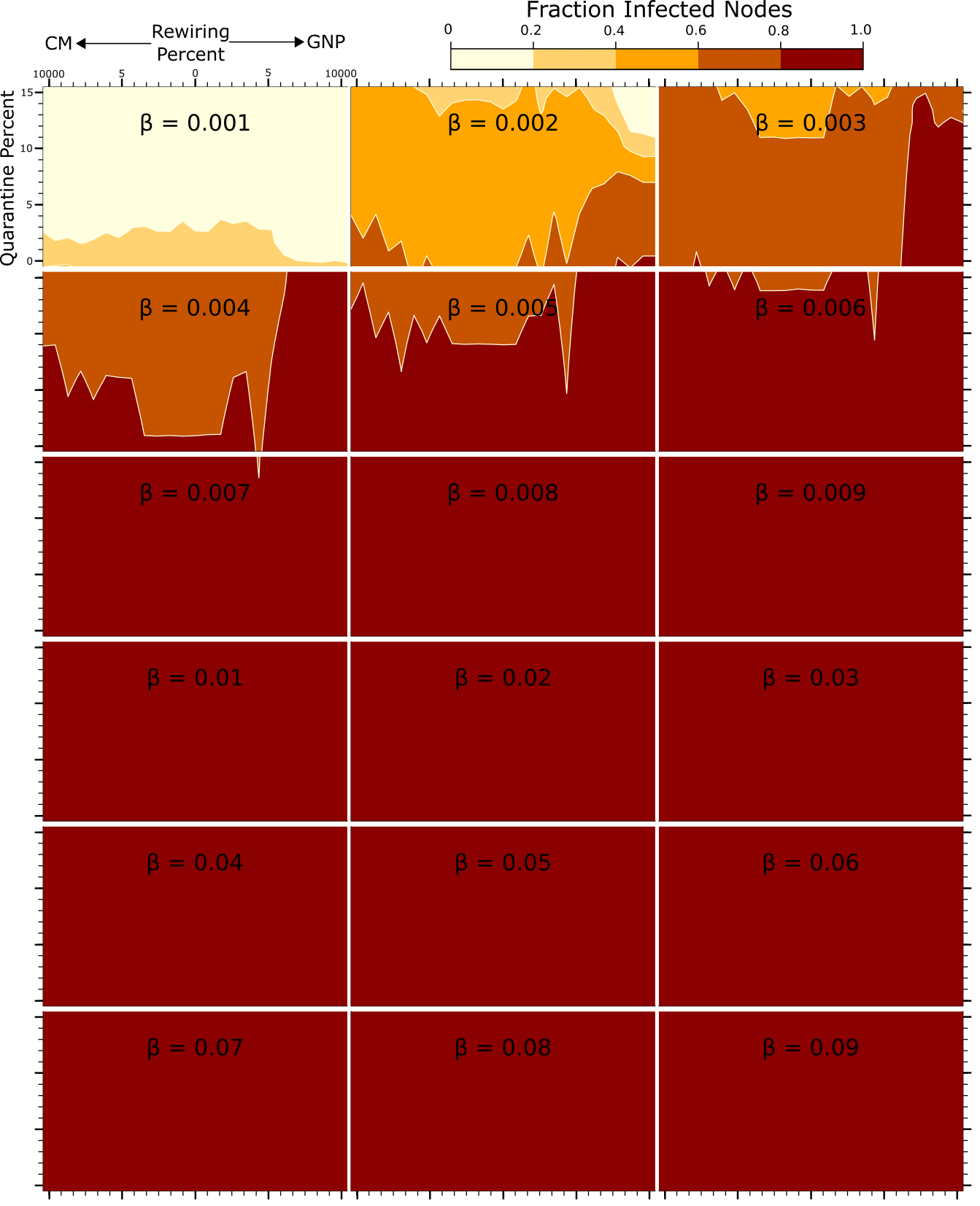}
        \caption{College-Wisc diffusion data prior to performing sparsification. Epidemic ``strength'' (original network with no quarantining) varies from $s\approx 2.9$ to $s\approx 265.3$.}
        \label{fig:uplot-params-wisc}
    \end{figure}

    Figures~\ref{fig:uplot-params-penn}--\ref{fig:uplot-params-wisc} show the diffusion data for the sparsified FaceBook networks prior to sparsification. Note that epidemics on these networks become very difficult to contain. The seeded PageRank NCP for College-Penn before and after sparsification is shown in Figure~\ref{fig:app-sparsification-penn}. Note the prior to sparsification, College-Penn lacks variability in local structure but after sparsification there is much more variability in local structure. Since the seeded PageRank method of computation is better at finding extremal conductance sets, these networks were sparsified to highlight local structure.
    
\section{Rewiring} 

The characteristic plot (Figure~\ref{fig:uplot-explanation}) involves rewiring from an original graph to either a uniform random graph $G_{n,p}$ or a Chung-Lu or Configuration Model random sample. We give details about how this was accomplished (which we believe is a standard process). Statistics about these graphs do not appear in Table~\ref{table:graph-stats} since the full set of graphs (including rewired variants) are too numerous.

\subsection{Rewiring to preserve average degree ($G_{n,p}$)}
In the first, we rewire to preserve average degree. To do so, we sequentially sample an edge and randomly choose it's new endpoints. This is done for an increasing number of rewiring steps. The number of edges rewired range from $10^{-4}|E|$ to $100|E|$, depending on the size of the graph. For each proportion of rewiring, all rewirings start from the base graph. This is done so that we do not introduce any unintended correlations in graph structure among different rewirings.

\subsection{Rewiring to preserve individual degrees (Configuration Model)} 
In the second, we use a degree preserving rewiring by sampling two edges at random and swapping their endpoints. This is done for an increasing number of rewiring steps. The number of edges rewired range from $10^{-4}|E|$ to $100|E|$, depending on the size of the graph. For each proportion of rewiring, all rewirings start from the base graph. This is done so that we do not introduce any unintended correlations in graph structure among different rewirings.

\subsection{Triangle Rewiring for Hypergraph Epidemic Models}
For triangle weighted epidemics, triangles in the network are shuffled to randomize the placement of triangles. For each base network, we iteratively sample a random triangle, and we randomize the endpoints of that triangle. 
For all networks apart from Flickr, we produce 10 triangle rewired variants (shuffling of 100k,200k,$\dots$, 1 million triangles). 
For Flickr, we produce 3 triangle rewired variants (shuffling of 100k,500k, and 1 million triangles).

\section{Full details on individual figures}

\subsection{Epidemic Network Community Profile (NCP) Plots}
\label{sec:epidemic-ncps-alldetails}

\paragraph{The original NCP from Leskovec et al.}
When originally introduced \cite{leskovec2009community}, the NCP was defined as the minimal conductance versus set-size as a function of set size. Formally, the minimum conductance, $\Phi(k)$, is a function of set size, $k$, where     
    \[\Phi(k) = \min_{S\subset V,\text{ }|S|=k}\text{cond}(S).\]
Since this is intractable to compute, they employ approximation algorithms to estimate this quantity. We note that recent advances have made it tractable to compute a lower bound on the NCP to validate empirical observations based on the NCP~\cite{Huang-2023-mucond-lrsdp}. 

\paragraph{Differences from NCPs in Leskovec et al.}
    We use NCP somewhat differently. In this paper, the NCP refers to the \emph{distribution} of conductance versus set size, rather than the minimum envelope. Moreover, the node sets, $S\subset V$, are distributed according to the particular graph diffusion under consideration, in this case either epidemic spread or seeded PageRank vectors. Formally, we are concerned with 
    \[\Phi_\mu(\phi,k) = \text{Pr}_{S\sim \mu}(\text{cond}(S),|S|),\]
    where $\mu$ is a distribution over subsets of nodes. In this paper, $\mu$ is implicitly defined by the epidemic spreading model and phenomena.    

    \paragraph{Epidemic Details}
        To produce the epidemic NCPs, we sampled 50,000 nodes and for each node sampled, and we run 20 SEIR epidemics from that node with a very strong infection probability ($\beta=0.3$). The rationale for this choice of $\beta$ is that we need the epidemic to infect a large percentage of the graph so that we can rank nodes using epidemic information. This has resulted in epidemics with more than $90\%$ of nodes being infected.\\
        
        For each sampled node, if more than 5 epidemics are trivial (fail to infect at least one other node), then we discard that node sample. To rank nodes, we use the function \[\vs[v] = -\sum_{i=1}^{20} min(I^{(i)}[v],l_i+1),\]
        where $I^{(i)}[v]$ denotes the infection time for node $v$ during epidemic sample $i$ and $l_i$ denotes the time at which epidemic sample $i$ ended (all nodes in $S$ or $R$). So the seed node should have $\vs[v]=0$ and $\vs[u]<\vs[v]$ for all other nodes. In particular, nodes that weren't infected during an epidemic sample are all assigned the same value. \\ 
        
        The full epidemic parameters used for all graphs is $(\beta,\gamma,\text{qpercent})=(0.3,0.05,0)$ for all epidemic diffusions. However, this is only performed for each base graph and it's extremal ER and configuration model vairants. See Table \ref{table:graph-stats} for the complete list of base graphs used.        
        
    \paragraph{Sub-sampling Details}

        Given the above node ranking $\vs$, a sweep-cut procedure on this ranking will sequentially form the set of the top-k ranked nodes, $S_k$, and then compute $\phi(S_k)$ for these sets.
        This gives a vector $\vp[k] = \phi(S_k)$.
        Typically, one takes the $\min(\vp)$ as the minimum conductance bottleneck and computes whatever statistics are desired (size of set, number of edges, size of cut, etc).
        Empirically, we observed that simply taking the minimum is insufficient as this procedure is biased to a single size scale. 
        Where this occurs (small set size vs large set size), it is often an artefact of the particular network used. 
        For example, geometric networks are biased towards larger size scales (see the geometric graph in Figure~\ref{fig:fig-ncp-trajectory}). In order to produce the epidemic NCP at multiple size scales, we sub-sample at several size scales. We could try storing the full vector $\vp$ but this would amount to forming a dense matrix with $(\text{number of nodes})\times (\text{number of epidemic trials})$ entries. For larger graphs, this requires large amounts of memory, hence we sub-sample this information.\\
        
        In particular, we first segment the sampling up to 5 regimes with splits at $10^2, 10^3, 10^4$ and $10^5$ nodes. For each portion, we then sample a lowerbound and upperbound on a uniform log scale from that bin. After that, we finally sample the minimum conductance set with size in $(lowerbound,upperbound]$ along with several statistics (seed node, set size, set conductance, cut size, and set volume). The rationale for the uniform log scale is purely for visualization purposes. This somewhat ad-hoc sub-sampling procedure can be thought of correcting the structural/dynamical bias observed in the epidemics trajectories. 
        
    \paragraph{Implementation Details}
        To perform this computation, the epidemic NCPs are parallelized. Each worker maintains a local copy of the epidemic data structure and a local copy of recorded statistics. The computation is split by seed nodes that are sampled. So each worker will perform a number of epidemic samples for each seed node and compute statistics. Afterwards, epidemic information is aggregated across workers. Also, after we have formed the NCP, we normalize by the maximum value over bins with the same bounds along the x-axis. For sets at the same size-scale, we are normalizing over the different conductance values. Bins with density close to $1$, are explored the most at that size scale. Bins with a density close to $0$ are seldom explored. See Figure~\ref{fig:fig-ncp-trajectory}. 

\begin{figure}
        \centering
        \includegraphics[width=0.8\textwidth]{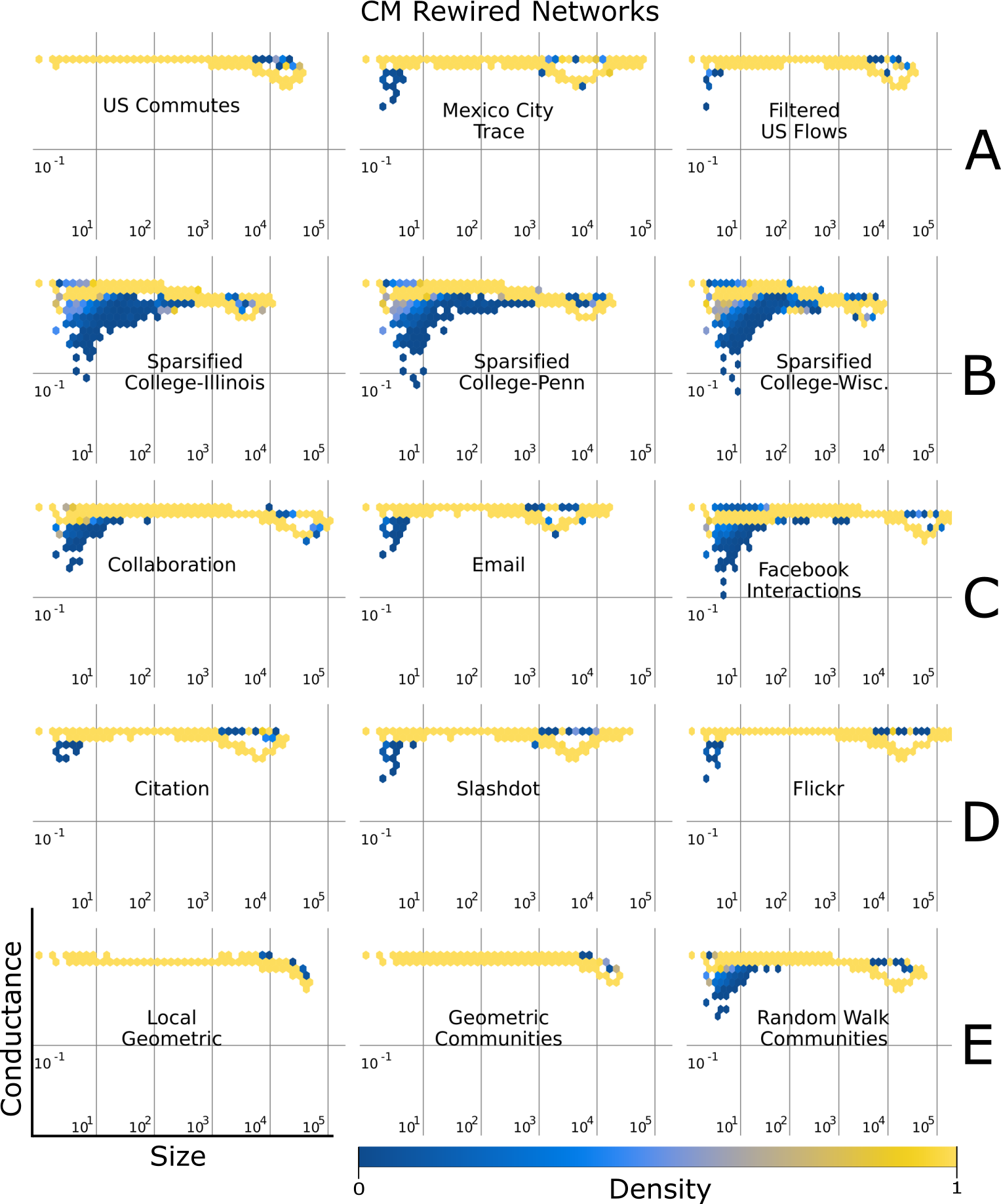}
        \caption{Epidemic NCP plots for extremal CM network variants.}
        \label{fig:app-cm-epidemic-ncp}
    \end{figure}

\begin{figure}
        \centering
        \includegraphics[width=0.8\textwidth]{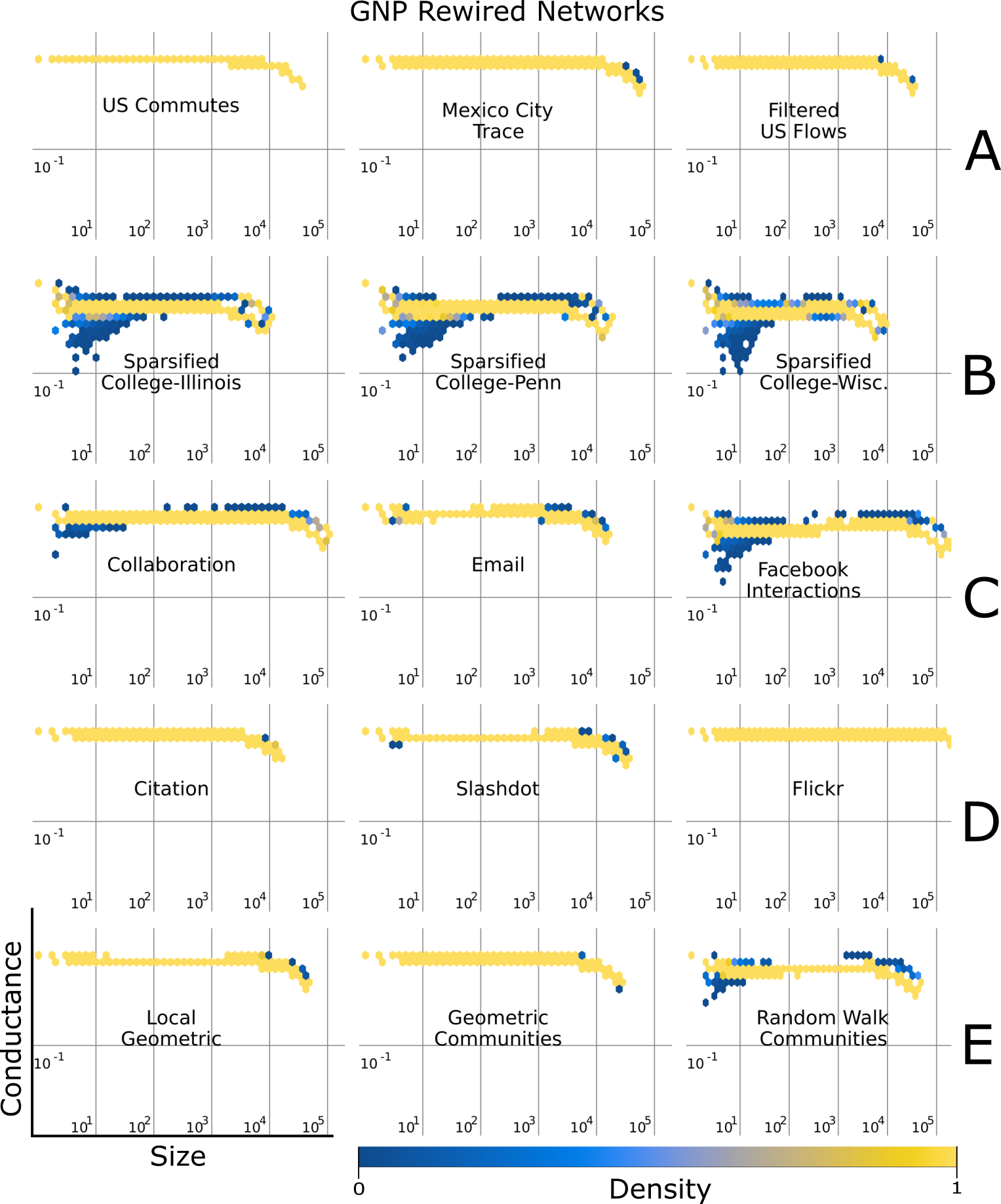}
        \caption{Epidemic NCP plots for extremal GNP network variants.}
        \label{fig:app-gnp-epidemic-ncp}
    \end{figure}

    \paragraph{Epidemic NCPs for Rewired Networks - Figure~\ref{fig:fig-ncp-trajectory}}
    Figures~\ref{fig:app-cm-epidemic-ncp} and~\ref{fig:app-gnp-epidemic-ncp} show the epidemic NCPs for the CM rewired and GNP rewired networks respectively. Networks are located in the same positions as Figure~\ref{fig:fig-ncp-trajectory} and Figure~\ref{fig:uplots-all}.  The rewiring percent for each network in Figures~\ref{fig:app-cm-epidemic-ncp} and~\ref{fig:app-gnp-epidemic-ncp} is 10000 so that we perform $100|E|$ edge rewirings before sampling the epidemic NCP.
    
\subsection{Details on Figure~\ref{fig:uplots-all}}
\label{sec:app-uplots-all}

Infection probabilities ($\beta$) in SEIR epidemics are: row 1 - 0.1, 0.06, 0.008, row 2 - 0.1, 0.04, 0.04, row 3 - 0.1, 0.1, 0.1, row 4 - 0.03, 0.05, 0.003, row 5 - 0.06, 0.02, 0.175.

\subsection{Figure~\ref{fig:area-vs-local-structure}: Details on Area of Epidemic NCP vs. Quarantine Impact}
\label{sec:area-vs-local-structure-details}

The data in this plot give a positive association between the simple area of the NCP measure and the impact of quarantine. We present a number of statistical details about the choices here in order to disclose all analyses done in preparing this figure. 

\paragraph{The intention of the measure.} Our goals in designing a single number to characterize the \emph{diversity} of an epidemic NCP is to quantify our qualitative observation that higher diversity in the NCP plot is associated with a large impact of quarantine. Consequently, we use a measure closely inspired by the original NCP that captures this notion of area.
In particular, given $\vx,\vy$ as vectors of approximate minimums, with $\vx$ normalized by half the total vertices (maximum value on x-axis for set-size), we compute
\[\text{AANCP}(\vx,\vy) = \text{AUC}(\log_{10}(\vx, \, -\log_{10}(\vy)).\]
This is akin to a normalized area above the NCP. In appropriate circumstances, useful lower-bounds on this envelope can be estimated~\cite{Huang-2023-mucond-lrsdp}.

\paragraph{Spearman Rho Coefficient.} 
The data sampled are non-normal with outliers, so we make use of Spearman's $\rho$ as a measure of correlation.

\subsection{Similar Effects for Seeded PageRank NCP}

\begin{figure}
    \centering
    \includegraphics[width=\textwidth]{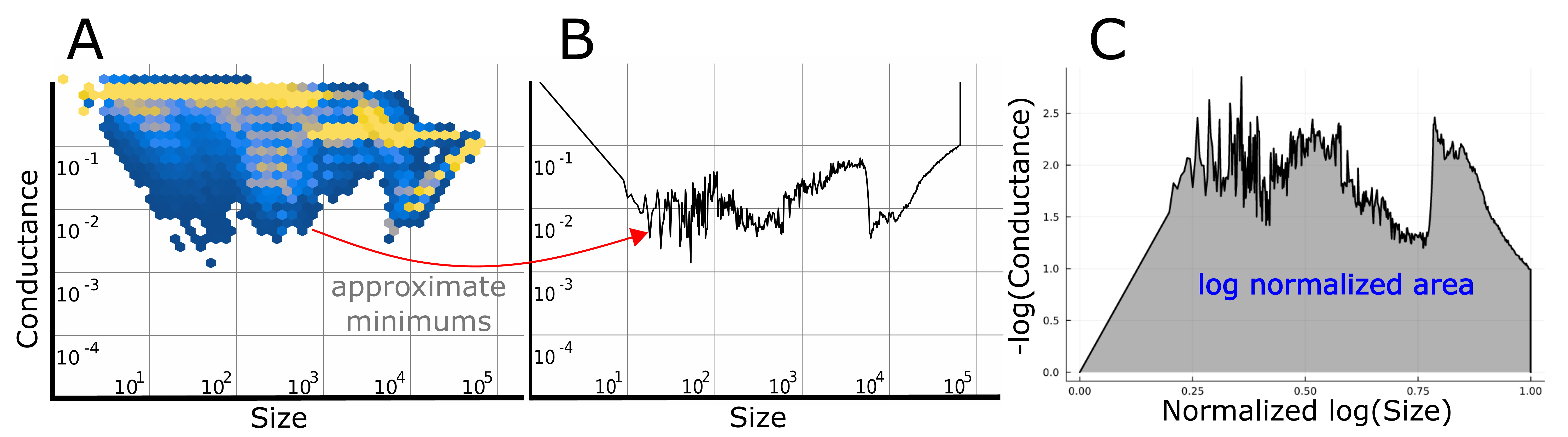}
    \caption{An example of how the AANCP is compute Mexico City Trace. (A) shows the epidemic NCP for Mexico City Trace. (B) computes approximate minimums for this distribution. (C) is a normalized area that is computed from these minimums. When normalizing the size of the set, half the vertices in the graph is set to 1 to handle issues of symmetry.}
    \label{fig:aancp-computation}
\end{figure}

\begin{figure}
    \centering
    \includegraphics[width=\textwidth]{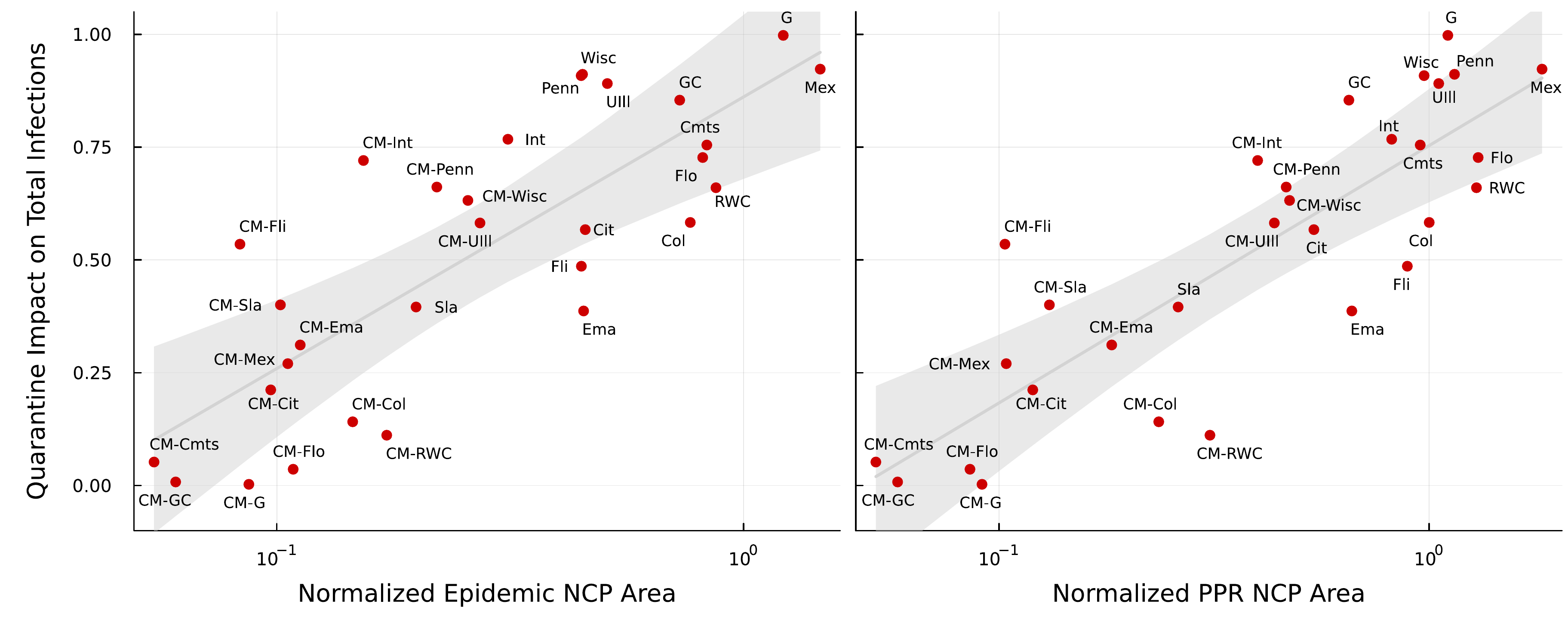}
    \caption{The main text uses the Epidemic NCP to display trends between local structure and quarantine impact. Here we display the same trend as in Figure~\ref{fig:area-vs-local-structure} using the Epidemic NCP (left) and seeded PageRank NCP (right). The same broad trend holds. The Spearman $\rho$ correlation coefficient for the seeded PageRank NCP (right) is 0.84 with 95\% CI $[0.69,0.92]$. While this shows a stronger correlation with quarantine impact compared to epidemic NCP, the epidemic NCP is more relevant for epidemic spreading.}
    \label{fig:ppr-area-local-structure-qimpact}
\end{figure}

Figure~\ref{fig:aancp-computation} shows how we compute the AANCP measure. 
While we use the epidemic NCP in Figure~\ref{fig:area-vs-local-structure} when quantifying local structure, one can use other dynamics. 
If instead, one uses the seeded PageRank NCP, we can get qualitatively similar results as shown in Figure~\ref{fig:ppr-area-local-structure-qimpact}.

\label{sec:ppr-area-local-structure-qimpact}

\subsection{Details on Epidemic Trajectory Experiments}
    
    \subsubsection{Missed Sets}
    The experimental results concerning which sets were missed were obtained by combing extremal conductance information with epidemic information about individual nodes. The epidemic information comes from simulating epidemics and recording the final state of nodes (susceptible or not). For our purposes, this comes from our experiments on total infections. During those epidemic simulations, one of the recorded statistics was the final epidemic state of each node. For information about the bins, we need not only aggregate statistics about the minimal conductance set for each sample, but we also want the extremal set itself. To obtain this, we perform many seeded PageRank samples for each graph and record set size, set volume, cut size, and the nodes in the set. 

    The seeded PageRank system corresponds to the linear system
    \[(\mI-\alpha\mP)\vx = (1-\alpha)\ve_i,\]
    where $\ve_i$ is the indicator vector where $\ve_i[k]=1$ if $k=i$ and zero otherwise. This system is approximately solved using the push-relabel algorithm~\cite{goldberg1998beyond}. We choose parameters based on the size of the graph. 
    For the largest graph sizes ($>45000$ nodes) we choose $\alpha=0.995$ so that we are predominantly exploring the graph structure in the seeded PageRank procedure.
    
    Epidemic information and extremal conductance bottlenecks are combined by using weighing the bottlenecks by fraction of susceptible nodes. 
    This is done by binning size and conductance information for sets found via the seeded PageRank system, and then weighing the bins by fraction of susceptible nodes in each bin. 
    We could have chosen to take the average over sets instead of nodes. 
    We tested both of these options but noticed minor differences and opted to use nodes instead due to ease of interpretability.

    \subsubsection{Supplementary Missed Sets Figures}
\begin{figure}
        \centering
        \includegraphics[width=0.95\textwidth]{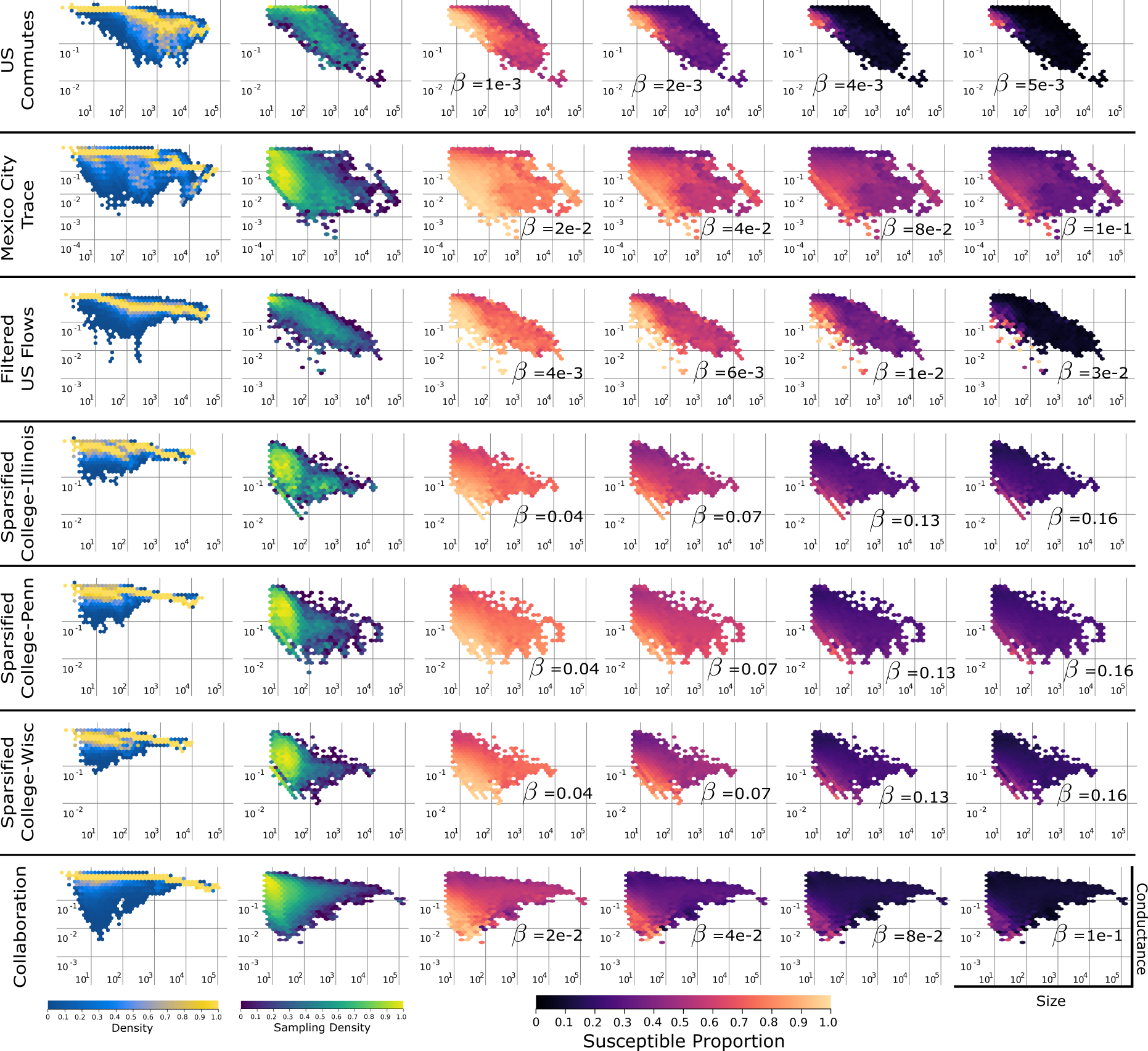}
        \caption{The above supplementary figure shows (leftmost) the epidemic NCP, the Personalized-PageRank NCP (second from the left) and finally how the susceptible proportion of sets evolves with larger infection probabilities (4 rightmost plots). }
        \label{fig:app-missed-sets1}
    \end{figure}

\begin{figure}
        \centering
        \includegraphics[width=0.95\textwidth]{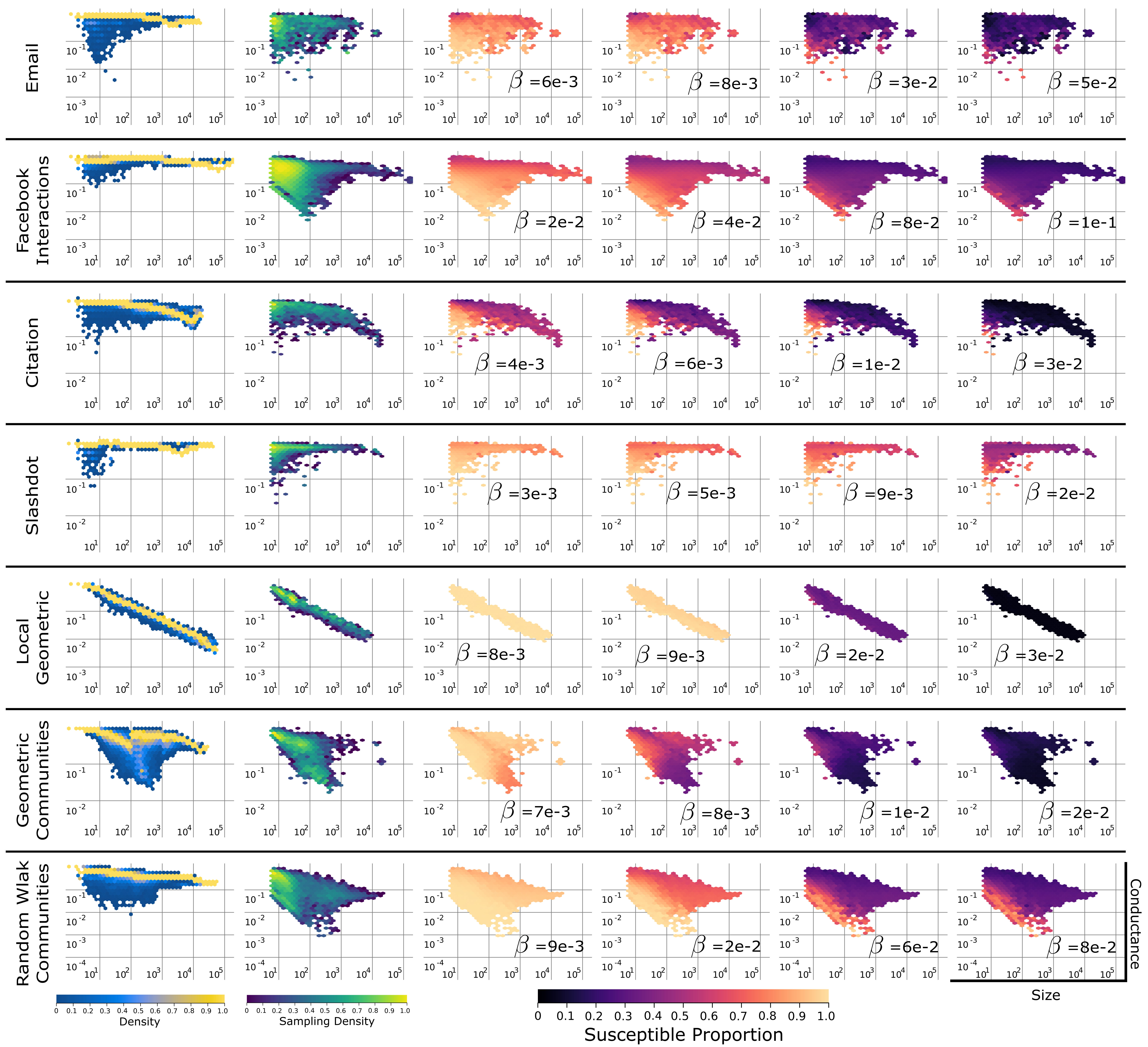}
        \caption{The above supplementary figure shows (leftmost) the epidemic NCP, the Personalized-PageRank NCP (second from the left) and finally how the susceptible proportion of sets evolves with larger infection probabilities (4 rightmost plots).}
        \label{fig:app-missed-sets2}
    \end{figure}

    In Figure~\ref{fig:missed-sets-fig}, we show the evolution of set susceptibility for a limited number of networks. In Figures~\ref{fig:app-missed-sets1} and~\ref{fig:app-missed-sets2}, we display this figure for all other networks (with the exception of Flickr, due to memory requirements). Values of $\beta$ are displayed in each subplot. The same general trend appears in most networks with the exception of Local Geometric.    
    
\subsection{Details on Eigenvalue Experiments}
    For eigenvalue experiments, we compute the dominant eigenvalue in Julia using Arpack. We compute dominate eigenvalues, the corresponding eigenvectors, and the residuals but only store the dominant eigenvalue and residuals. The largest residual across all networks and their rewired variants is no more than $6\text{e}^{-12}$.

\subsection{Details on Figure~\ref{fig:spatial-graph-fig}}
\label{sec:app-spatial-graph-fig}

Epidemics on these networks are generated for $s=\lambda_1(\mA)\frac{\beta}{\gamma} = 28$ with $\gamma=5 \cdot 10^{-2}$ (same $\gamma$ as all experiments). This is due to differences in average degree across the three networks (left to right these are $12.9,\quad 15.4,\quad 31.9$). 

\subsection{Details on Figure~\ref{fig:uplot-triangles}}
\label{sec:app-uplots-triangles}

Similar to Figure~\ref{fig:uplots-all} and the accompanying details in Appendix~\ref{sec:epidemic-params-app}, parameters for Figure~\ref{fig:uplot-triangles} were chosen in order to more readily observe notable effects. The infection probabilities used in Figure~\ref{fig:uplot-triangles} are as follows: row 1 - 0.0001, 0.003, 0.001, row 2 - 0.18, 0.18, 0.18, row 3 - 0.01, 0.001, 0.15, row 4 - 0.001, 0.006, 0.03, row 5 - 0.005, 0.0005, 0.06.

\section{Example of Heterogeneity in Community Structure}
\label{sec:app-community-heterogeneity}
\begin{figure}
    \centering
    \includegraphics[width=\textwidth]{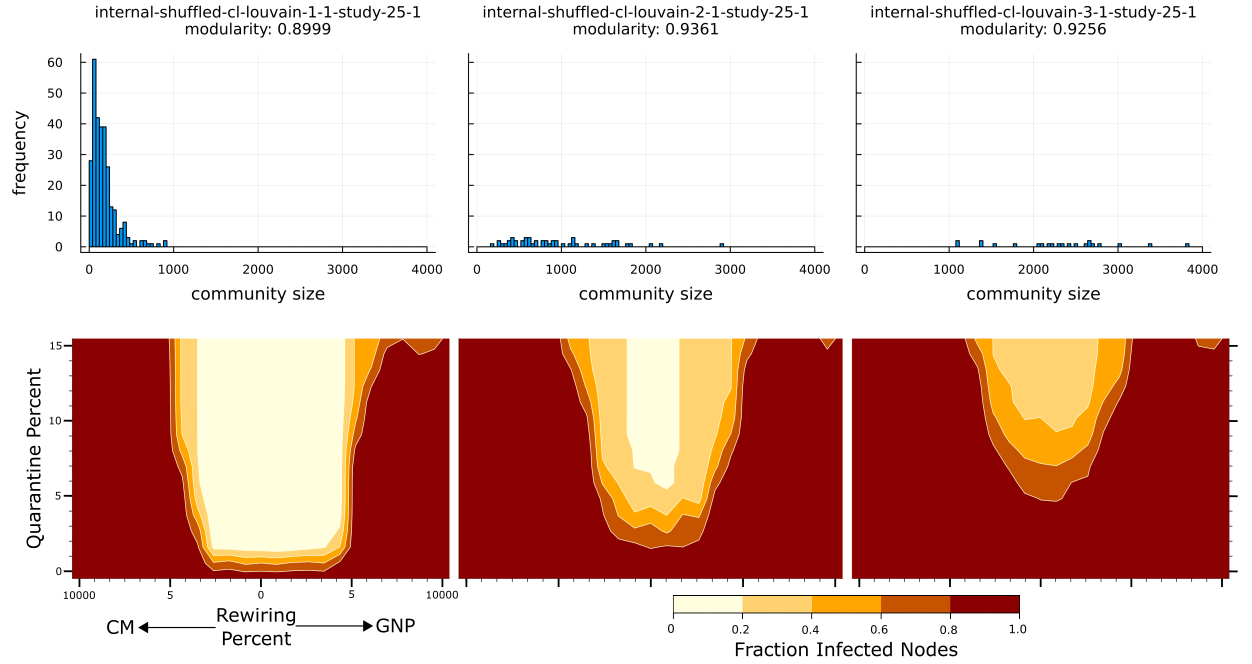}
    \caption{The level of granularity composing community structure is relevant for local interventions on epidemic spreading. As we move left to right, the larger communities are composed of aggregated communities from the more granular community structure. As communities become larger, local interventions become less effective. Simulations are SEIR epidemics with $\beta=0.06$ and $\gamma=0.05$.}
    \label{fig:mesoscale-vs-local}
\end{figure}

Similar networks with different scales of community sizes can behave very differently so that the internal community structures matter. In Figure~\ref{fig:mesoscale-vs-local}, we take a spatial network with 50,000 nodes with edges based on k-nearest neighbors in toroidal space, and we examine epidemics at several community granularity levels. This is the network `study-25-1'. Three levels of community structure were found via Louvain's algorithm, so the larger communities are a superset of smaller communities. For each partition, internal community edges are randomized while preserving node degrees, while edges between communities are left untouched. This destroys small-scale local structure while preserving mesoscale structure in these networks. As the communities become larger, reshuffling them causes quarantine interventions to be less effective because we destroy smaller-scale local structure. 

\section{Code and Data Availability}
\label{sec:app-code-data-availability}
All the experiment codes are available online at \url{https://github.com/oeldaghar/network-epi}. 
The full set of experiment data will be released prior to publication.

\end{document}